\documentclass[11pt]{JHEP3}

\JHEPspecialurl{http://jhep.sissa.it/JOURNAL/JHEP3.tar.gz}

\usepackage{epsfig,multicol}
\usepackage{amsmath,amssymb}
\usepackage{mathrsfs}

\newcommand\fverb{\setbox\fverbbox=\hbox\bgroup\verb}
\newcommand\fverbdo{\egroup\medskip\noindent%
			\fbox{\unhbox\fverbbox}\ }
\newcommand\fverbit{\egroup\item[\fbox{\unhbox\fverbbox}]}
\newbox\fverbbox

\newcommand{\pslash}{p\kern-1ex /}
\newcommand{\qslash}{q\kern-1ex /}
\newcommand{\lslash}{l\kern-1ex /}
\newcommand{\sslash}{s\kern-1ex /}
\newcommand{\kaslash}{k_a\kern-2ex /}
\newcommand{\kbslash}{k_b\kern-2ex /}
\newcommand{\Dslash}{\mathcal{D}\kern-1.5ex /}

\newcommand{\beqa}{\begin{eqnarray}}
\newcommand{\eeqa}{\end{eqnarray}}

\newcommand{\ba}{\begin{eqnarray}}
\newcommand{\ea}{\end{eqnarray}}
\newcommand{\be}{\begin{equation}}

\newcommand{\alg}[1]{\mathfrak{#1}}

\title{$AdS_5\times S^5$ mirror TBA equations from Y-system and discontinuity 
relations}

\author{J\'anos Balog and \'Arp\'ad Heged\H us\\
Research Institute for Particle and Nuclear Physics, 
Hungarian Academy of Sciences,
H-1525 Budapest 114, P.O.B. 49, Hungary\\}

\received{\today} 		
\accepted{\today}	

\abstract{
Using the recently proposed set of discontinuity relations we translate the
AdS/CFT Y-system to TBA integral equations and quantization conditions for a 
large subset of excited states from the ${\alg{sl}(2)}$ sector of the
$AdS_5 \times S^5$ string $\sigma$-model.
Our derivation provides an analytic proof of the fact that the exact Bethe
equations reduce to the Beisert-Staudacher equations in the asymptotic limit.
We also construct the corresponding T-system and show that in the language of 
T-functions the energy formula reduces to a single term which depends on a 
single T-function.
}

\begin{document}

\section{Introduction}

An important problem in the AdS/CFT correspondence \cite{adscft} is the calculation of
anomalous dimensions in the planar ${\cal {N}}=4$ super Yang-Mills theory (SYM)
or equivalently the energies of the dual $AdS_5 \times S^5$ string sigma-model.

The integrability discovered on both sides of the correspondence\footnote{For a comprehensive recent collection 
of review papers, 
see \cite{Beisert:2010jr}.} provided us with
an efficient mathematical apparatus to compute the exact spectrum of the planar
AdS/CFT models. As the central object of integrability the 2-particle S-matrix of the
model \cite{S,dressing,dresscross} plays a prominent {role}{\footnote{For a recent review see \cite{S1,S2} and
references therein.}} and is indispensable for the methods which enable us to compute the exact spectrum.

First, the spectrum of long operators or equivalently states with large R-charge
$J$ was determined by the asymptotic Bethe Ansatz (ABA) \cite{BS,S1}, which describes 
all polynomial corrections in $1/J$. 

Later, the leading exponentially small {corrections}{\footnote{These are the so 
called wrapping corrections.}} in $J$ could be taken into account by means of
the generalized L\"uscher formulae \cite{Luscher85,JL07,BJ08}. For the Konishi field their small coupling
expansion led to beautiful agreement with direct 4-loop field theoretical computations \cite{Sieg,Vel}.
For a certain class of states in the ${\alg{sl}(2)}$ sector the 4- and 5-loop expansion of these formulae 
\cite{Bajnok:2008qj,BJ09,Lukowski:2009ce} also satisfy
nontrivial consistency checks dictated by perturbation theory considerations in the
planar ${\cal {N}}=4$ SYM \cite{KL02,40_5}.

The exact energies, which resum all wrapping corrections in $J$, can be obtained by
the application of the Thermodynamic Bethe Ansatz method \cite{Zami} to the doubly
Wick rotated $AdS_5 \times S^5$ string sigma-model called the mirror model \cite{AJK,AF07}.

Strictly speaking the TBA method provides only the ground state energy of the model in finite
volume and its extension to excited states is only a conjecture even if
in most of the cases it rests on solid grounds. Based on the corresponding 
string-hypothesis \cite{AF09a} the ground state TBA equations of the $AdS_5 \times S^5$
mirror model were constructed in \cite{Bombardelli:2009ns,GKKV09,AF09b}
and simplified in \cite{AF09d}.
In AdS/CFT the ground state TBA equations do not give much information about the
spectrum since the ground state is protected by supersymmetry, i.e. $E_0(L)=0$.

The ground state equations are important nevertheless because they serve as
starting point for the excited state equations. An important further discovery is that
the Y-{functions}{\footnote{I.e. unknown functions of the TBA equations}} of the
mirror TBA equations satisfy the so called Y-system functional equations \cite{GKV09}.
The Y-functions can be rephrased in the language of T-functions  satisfying
the so-called T-system. The T-system of AdS/CFT lives on a T-hook \cite{GKV09}.
The discovery of the Y- and T-systems (see Figs. 1, 2) made it possible to determine the asymptotic
(large $J$) solutions of the excited state TBA problem. The asymptotic solution
is constructed so that the ABA and the generalized L\"uscher formulae are reproduced. 


Based on previously elaborated examples in lattice models \cite{KPS,33B} and in relativistic quantum field 
theory (QFT)
\cite{DT,BLZ,BH-SG,BH34}, the common experience is that excited state TBA equations differ from
the ground state ones only in source {terms}
and in quantization conditions imposed on objects appearing in the arguments of these terms.
These source terms can be found by various methods like analytic continuation
in some parameter of the model  \cite{DT}, deforming the integration contour of the ground
state equations \cite{KPS,Arutyunov:2009ax} or transforming the Y-system functional relations to integral 
equations \cite{KPS,33B,BH-SG,BH34}.
In AdS/CFT the analytic continuation \cite{GKKV09,GKV09b} and contour deformation \cite{Arutyunov:2009ax} 
methods together with requiring consistency with the large $J$ asymptotic solution were successfully applied 
to find the excited state TBA equations for certain states of the ${\alg{sl}(2)}$ sector of the model.
In \cite{Arutyunov:2011uz} a general strategy to construct the TBA equations for all states of the model
by the contour deformation method is outlined.

Since the TBA equations in AdS/CFT still cannot be derived from first principles it is
important to test them carefully. In the strong coupling limit it was shown  \cite{Gromov,ujKazi} 
that the TBA equations reproduce the 1-loop string energies in the semi-classical limit and
the strong coupling expansion of the energy of the Konishi state fitted from numerical TBA
computations  \cite{GKV09b,Frolnum} was found to be consistent with direct string theory computations 
\cite{StrongKonishi}.
At weak coupling 5-loop TBA results for the twist-2 states 
agree \cite{AFS,BHxxx,BH-BJ} with those based on the generalized  L\"uscher formula.

 

Though the analytic continuation and contour deformation methods provide the TBA equations
for excited states it is important to derive the TBA equations from the Y-system functional relations
as well. The main advantage of the Y-system based method is that the infinite Y-system can be solved
via the T-system and the related T-Q relations, thus it opens the way towards the NLIE formulation
of the AdS/CFT spectral problem. 

However, until ref. \cite{Tateo1} appeared, the Y-system based derivation of the TBA equations 
was impossible because the Y-functions are not meromorphic functions like in the relativistic models and
have nontrivial discontinuity structure.
The main discovery of ref.~\cite{Tateo1} was that this extra difficulty can be overcome for the 
ground state if the Y-system equations are supplemented by appropriate functional relations for
the square-root discontinuities of the Y-functions. The Y-system equations supplemented by the discontinuity 
functional
relations plus some analyticity assumption on the distribution of zeroes and poles of the Y-functions
are sufficient to transform the Y-system to TBA integral equations. 
In this spirit the ground state TBA equations (including the dressing kernel)
were derived \cite{Tateo1} assuming that none of the Y-functions have 
{local}{\footnote{Here the word \lq\lq local" means zeroes or poles.}} singularities in the entire complex 
plane.

It was conjectured in ref. \cite{Tateo1} that the  form of the discontinuity functional relations is 
state independent. If true, this allows us to derive excited state TBA equations from the 
Y-system as well. In this paper we carry out the Y-system based derivation of the excited state TBA equations.
We consider states consisting of fundamental particles only, i.e. we assume all particle rapidities are real.
As a first step we have checked that the conjecture is true in the asymptotic limit
and found that the asymptotic solutions given in appendix C satisfy the discontinuity relations
nontrivially.

Furthermore we  show that we do not need to  know the \lq\lq local" analyticity properties of the Y-functions
in the entire complex plane, it is sufficient to know their behaviour in certain regions near the real line.
This means that the assumption of ref. \cite{Tateo1} concerning the \lq\lq local" singularities of the 
ground state Y-functions is too restrictive. Indeed, in \cite{Tateo2}   
it has been shown numerically that the ground state Y-functions do have local singularities in the
complex plane. Reconsidering the derivation it turned out that the local singularities are arranged
in \lq\lq complexes" (see ref. \cite{Tateo2}), lie outside the physical strip and
do not modify the form of the TBA equations. 
 
Throughout the paper we assume that within certain regions of the complex plane the excited state
Y-functions are smooth deformations of the asymptotic solution given in \cite{GKV09}.
More precisely, we only discuss that part of the parameter space where the exact solution has
(qualitatively) the same analytic properties as the corresponding asymptotic solution. In practice,
if the size of the system ($J$) together with other quantum numbers are fixed, this is realized for
small enough coupling $g$. For larger $g$ the TBA equations for the given state may undergo phase transitions
similarly to what was found in \cite{Arutyunov:2009ax,Frolnum}. In this paper we restrict our attention to the 
form
of the TBA equations valid in the vicinity of the asymptotic solution. 
Then assuming that the solution of the Y-system is found we construct the T-functions in an appropriate
gauge. These T-functions are also smooth deformations of their asymptotic forms.

The benefit of introducing the T-functions is 2-fold. On the one hand they serve as internal variables
in terms of which the discontinuity relations and the derivation of the TBA equations
simplifies drastically. On the other hand T-functions seem to be more fundamental objects from the
point of view of the AdS/CFT spectral problem. For example we show that the complicated TBA 
energy formula becomes a very simple expression containing the single T-function $T_{1,0}$.
This fact indicates that there is an integrable $psu(2,2|4)$ spin-chain in the background
such that the Hamiltonian is related to a transfer matrix of the model similarly to the cases of
lattice models and lattice regularizations \cite{devegasum} of integrable QFTs \cite{FR,LC} studied previuosly.

Our final equations agree with previous results of \cite{Arutyunov:2009ax}. 
Our derivation is based on functional relations and analyticity assumptions which we know
are satisfied exactly in the asymptotic limit. Hence as a by-product our results prove analytically that the 
asymptotic solution 
satisfies the large $J$ limit of the TBA equations and also that the large $J$ limit of the exact Bethe
equations coincide with the ABA equations. This fact, although it 
played an important role in the extraction of the 5-loop L\"uscher terms from TBA \cite{AFS,BHxxx,BH-BJ},
has not been proven analytically so far. 

The organization of the paper is as follows. In the next section we introduce our starting relations
and assumptions. Section 3 contains important lemmas and the transformation of the Y-system to TBA integral
equations leaving the necessary discontinuities temporarily unspecified.
In section 4 we construct the T-system in a partially fixed special gauge 
which turns out to be very useful to simplify the derivation of the TBA equations. 
In section 5 and 6 we compute the so far undetermined discontinuities $Y_{-}^{(\alpha)}/Y_{+}^{(\alpha)}$
and $\Delta$ from dispersion relations and we fix completely our gauge choice for the T-functions.
In section 7 we derive the simplified version of the TBA equations, while section 8 contains their
canonical and hybrid form. In section 9 we discuss the quantization conditions and the
exact Bethe equations. Finally, in section 10 we discuss a simplified energy
formula. The paper is closed by our conclusions.
Appendix A contains the definitions of frequently used objects and kernels, appendix B the discussion of the 
branch cut 
discontinuities of the dressing part of the discontinuity function $\Delta$.
In appendix C 
the asymptotic solution of the Y- and T-system equations are presented,
in appendix D the precise definition of the discontinuity relations is given
and finally in appendix E we discuss the meaning of the exact Bethe
quantization conditions.

\section{Starting relations and assumptions}

Our starting point in this paper is the hypothesis that the same Y-system describes all excited
states in the planar AdS/CFT spectral problem \cite{GKV09}. The Y-system of AdS/CFT takes the standard form
\begin{equation} \label{Y}
Y_{a,s}^+ \, Y_{a,s}^-=\frac{(1+Y_{a,s-1}) \, (1+Y_{a,s+1})}{ (1+1/{Y_{a-1,s}}) \, (1+1/{Y_{a+1,s}})},    
\end{equation}  
but it lives on an $(a,s)$ lattice represented in Fig. 1. This is equivalent to imposing the
boundary conditions $Y_{0,s} \to \infty$, $Y_{2,|s|>2} \to \infty$ and $Y_{a>2,\pm 2} \to 0$ 
while the product $Y_{3,\pm 2} \, Y_{2,\pm 3}$ should be kept finite in order to $Y_{2,\pm 2}$ be finite. This
last requirement shows that the Y-system defined in the domain of Fig. 1
cannot form a closed set of equations because the new functions $Y_{3,\pm 2} \, Y_{2,\pm 3}$ enter 
the problem and we need additional equations (independent of the Y-system) to complete the system.

The Y-system is equivalent to a T-system defined on a T-hook of Fig. 2:
\begin{equation} \label{T}
T_{a,s}^+ \, T_{a,s}^-=T_{a+1,s} \, T_{a-1,s}+ T_{a,s+1} \, T_{a,s-1},
\end{equation}
where the relation to the $Y_{a,s}$ functions is given by
\begin{equation} \label{YT}
Y_{a,s}=\frac{T_{a,s+1} \, T_{a,s-1}}{T_{a+1,s} \, T_{a-1,s}}, \qquad 
1+Y_{a,s}=\frac{T_{a,s}^+ \, T_{a,s}^-}{T_{a+1,s} \, T_{a-1,s}}, \qquad
1+1/Y_{a,s}=\frac{T_{a,s}^+ \, T_{a,s}^-}{T_{a,s-1} \, T_{a,s+1}}.
\end{equation}
The T-equations can be extended to the infinite $(a,s)$ lattice by imposing the boundary conditions
that all $T_{a,s}=0$ outside the T-hook. The T-system is a more fundamental set of equations
than the Y-system because it forms a closed set of functional {equations}{\footnote{Here the word
\lq\lq closed" means that there are as many functional equations as T-functions.}} 
which determines all the Y-functions together with the supplementary combination 
$Y_{3,\pm 2} \, Y_{2,\pm 3}$.
The T-equations and the Y-functions given by (\ref{YT}) are invariant with respect to the 
gauge transformations\footnote{Our notations and conventions are explained in appendix A.}
\begin{equation} \label{gauge0}
T_{a,s} \rightarrow g_1^{[a+s]} \, g_2^{[a-s]} \, g_3^{[s-a]} \, g_4^{[-s-a]} \, T_{a,s},
\end{equation}
with $g_{1,2,3,4}$ being arbitrary functions. In this paper we will choose a gauge where
$T_{0,s}=1$. This fixes $g_1 g_3=g_2 g_4=1$. 

It is known that in AdS/CFT the Y-functions have square root branch cuts and live on an 
infinite genus Riemann surface \cite{AF09b,AF09d,Tateo1,Tquant}. The different sheets of the Riemann surface
are connected through square-root branch cuts starting from the branch points with real parts $\pm 2$
and run to infinity along the lines with integer (in $1/g$ units) imaginary parts.
To avoid the complications connected to using different sheets we will use a convention where
all our functions are assumed to be defined on the first Riemann sheet, defined as the entire complex plane 
excluding the
above cuts. In our convention if we analytically continue a function through one of the cuts, this becomes a new 
function,
which can also be continued to the entire first Riemann sheet. We will often use the operation $f(u)\rightarrow 
f_*(u)$ corresponding to
analytic continuation of the function $f(u)$ through the real cut $\vert u\vert\geq2$. The construction of 
$f_*(u)$ consists of the two steps
of analytic continuation of $f(u)$ through the real cut followed by analytic extension of the new function to 
the entire first Riemann sheet.
Our functions are assumed to be meromorphic in the first Riemann sheet: they have discontinuities along some 
(but not necessarily all) 
of the cuts of the first Riemann sheet and cannot have other discontinuities but may have local singularities 
(zeroes and/or poles) 
in the sheet (and in some cases even on the cuts).  

Restricting the Y-functions to the first Riemann sheet
the Y-system equations supplemented by some analyticity information on the local singularities 
are not enough to derive the TBA integral equations. Some additional information on the discontinuities
is needed as well. According to the proposal of ref. \cite{Tateo1} this missing piece of information
is a set of functional equations relating the discontinuities in a state independent way. These discontinuity
equations translated to dispersion relations determine the extra $Y_{3,2} \, Y_{2,3}$ functions as well.

We will use the conventions of ref. \cite{Arutyunov:2009ax} throughout the paper. 
The Y-functions in these conventions are related to the $Y_{a,s}$ variables as follows:
\begin{equation}
Y_{-}^{(\alpha)}=-1/Y_{1,\alpha  1}, \qquad 
Y_{+}^{(\alpha)}=-Y_{2,\alpha  2}, \quad  \alpha=\pm \qquad Y_Q=Y_{Q,0} \quad  \quad Q=1,2,... \label{Ydef1}
\end{equation}
\begin{equation}
Y_{m|vw}^{(\alpha)}=1/Y_{m+1,\alpha 1} \qquad Y_{m|w}^{(\alpha)}=Y_{1,\alpha(m+1)} \qquad \alpha=\pm \quad 
m=1,2,...
\label{Ydef2}
\end{equation}
From the ground state equations we obtain the following structure for the locations of the branch cuts:
\begin{itemize}
\item $Y_{\pm}^{(\alpha)}(u): \qquad u+i \,2\, m/g \quad  \quad
m \in \mathbb{Z}$, 
\item $Y_m(u), \quad Y_{m|vw}^{(\alpha)}(u), \quad Y_{m|w}^{(\alpha)}(u): \qquad u\pm i \,(m+2 \,j)/g, \qquad
j=1,2,...,\infty $, 
\end{itemize}
where $u \in (-\infty,-2) \cup (2,\infty)$.
One of our main assumptions is that this structure remains valid for the excited states as well.
For later convenience we rewrite the discontinuity relations proposed in \cite{Tateo1} for the square-root 
branch cuts 
in the conventions we are using in this paper. 
We introduce the symbol $[f]_Z$ with $Z \in \mathbb{Z}$ to denote the discontinuity of the function $f(u)$:
\begin{equation} \label{[f]_Z}
[f(u)]_Z=\lim_{\epsilon \to 0^+} \left( f(u+i \, Z/g+i \epsilon)- f(u+i \, Z/g-i \epsilon) \right), \qquad
u \in (-\infty,-2) \cup (2,\infty)
\end{equation}
and define
\begin{equation} \label{Y1cut}
\Delta=[\ln Y_1]_{+1}.
\end{equation}
The discontinuity relations relate the \lq\lq jumps" of this function and those
of some other Y-functions. They take the form
\begin{equation}   \label{Deltacut}
[\Delta]_{\pm 2N}= \pm \sum_{\alpha} \left(
\left[\ln \!  \left(1-\frac{1}{Y_{\mp}^{(\alpha)}}   \right)\right]_{\pm 2N}   \!
+\sum\limits_{m=1}^N \, \left[\ln  \! \left(1+\frac{1}{Y_{m|vw}^{(\alpha)}}\right)\right]_{\pm (2N-m)} 
\! \! \! \! \! \! \!  \!+ \! \ln \!  \left( \frac{Y_{-}^{(\alpha)}}{Y_{+}^{(\alpha)}} \right)  
\right)
\end{equation}
\begin{equation} \label{Ypmcut}
\left[ \ln \! \left( \frac{Y_{-}^{(\alpha)}}{Y_{+}^{(\alpha)}} \right) \right]_{\pm 2N}=
-\sum\limits_{Q=1}^N \, \left[ \ln \! \left( 1+Y_Q \right) \right]_{\pm (2N-Q)}
\end{equation}
with $N=1,2,...$ and
\begin{equation} \label{Yvw1cut}
\left[ \ln \! Y_{1|vw}^{(\alpha)} \right]_{\pm 1}=\ln \! 
\left( \frac{1-Y_{-}^{(\alpha)}}{1-Y_{+}^{(\alpha)}}   \right),  \qquad
\left[ \ln \! Y_{1|w}^{(\alpha)} \right]_{\pm 1}=\ln \! 
\left( \frac{1-1/Y_{-}^{(\alpha)}}{1-1/Y_{+}^{(\alpha)}}   \right).
\end{equation}

In the sequel $[f(u)]_Z$ means the analytic extension of the discontinuity (\ref{[f]_Z}) to generic values
of $u$. Using the notation defined earlier in this section we can write 
$[f(u)]_Z=f^{[+Z]}(u)-(f^{[+Z]})_*(u)$ where $(f^{[+Z]})_*(u)$ is the 
{function}
obtained by analytic continuation of $(f^{[+Z]})(u)$ through the cut lying along the real line  
(crossing it from below) 
and then extended to the first Riemann sheet.

Some important remarks on the interpretation of the discontinuity relations (\ref{Y1cut}-\ref{Yvw1cut})
are in order. As they stand, (\ref{Y1cut}-\ref{Yvw1cut}) are valid if the Y-function
combinations appearing in them have no logarithmic discontinuities crossing the lines of the square-root
discontinuities. In order to get rid of the difficulties caused by the logarithmic discontinuities we use
the derivative of the discontinuity relations (\ref{Y1cut}-\ref{Yvw1cut}) as our starting point and apply
dispersion relations for the derivatives of $\Delta$ and $\ln Y_{-}^{(\alpha)}/Y_{+}^{(\alpha)}$.
Then a second subtlety appears if there are local singularities of the Y-function combinations lying
exactly on the lines of square-root discontinuities. Such local singularities do not modify the discontinuity
relations, but they contribute to the corresponding dispersion relation through the residue theorem. To cure 
this problem
a finer interpretation of the discontinuity relations is necessary in which the contribution of such local
singularities are taken into account as well. 

To find the correct interpretation we can invoke the asymptotic solution. 
It can be shown that only (\ref{Deltacut}) at $N=1$ and at the positions of real poles 
of $1-1/Y_{\pm}^{(\alpha)}$ must be refined.
The term which causes the trouble is $\left[ \ln \left(1+{1}/{Y_{1|vw}^{(\alpha)}} 
\right)\right]_{\pm 1}$ on the right hand side of (\ref{Deltacut}), because its derivative has poles sitting 
right on the
{discontinuities}{\footnote {We note that other Y-combinations for other values of $N$ can also
have local singularities along the cuts under consideration, but they mutually cancel each other's
contribution. }} with $\mbox{Im} u=\pm 1/g$. This refined interpretation of (\ref{Y1cut}-\ref{Yvw1cut}) is 
necessary
only when they are translated to dispersion relations.
In order for the dispersion relation applied for the derivative of (\ref{Deltacut}) give the correct formula for 
$\Delta$ the following replacement must be done on the right hand side of (\ref{Deltacut}) at $N=1$:
\begin{equation} \label{CutInterpret}
\left[ \ln \left(1+\frac{1}{Y_{1|vw}^{(\alpha)}(u)} 
\right)\right]_{\pm 1} \rightarrow \left[ \ln \left \{ \left(1+\frac{1}{Y_{1|vw}^{(\alpha)}(u)} 
\right)  \frac{1}{p_2^{(\alpha) \mp}(u)} \right \}   \right]_{\pm 1},
\end{equation} 
where $p_2^{(\alpha)}(u)$ is the polynomial having zeroes at the positions of the real zeroes 
of $Y_{-}^{(\alpha)}$ with absolute values larger than $2$.
In other words the poles corresponding to the zeroes of $p_2^{(\alpha)}(u)$ must be ignored.
For a proof of this formula for the most general state of the model see appendix D.

We note that as a consequence of the definition (\ref{Y1cut}), and the fact that going around twice a square 
root branch point gives back
the original function, we have the relation 
\begin{equation} \label{Deltacut0}
\Delta(u)=-\Delta_*(u). 
\end{equation}

The Y-system (\ref{Y}) and the discontinuity relations (\ref{Y1cut}-\ref{Yvw1cut}) are not 
sufficient to derive the TBA equations. Some more information is needed about the discontinuities
lying along the real axis. 
Based on the properties of the solution for the ground state TBA equations and the similar properties of the
asymptotic solution for excited states, we require that the \lq\lq fermionic" Y-functions $Y^{(\alpha)}_\pm$
are analytic continuations of each other:
\begin{equation} \label{Y-cut0}
 Y_{-}^{(\alpha)}(u)=(Y_{+}^{(\alpha)})_*(u).
\end{equation}

We will also assume that all Y-functions are real analytic. This assumption
is based on the observations that the Y-functions are real analytic functions for the ground state
solution of the TBA equations and also in the asymptotic limit for excited state solutions. 
Since we consider the AdS/CFT Y-functions as smooth deformations
of their asymptotic counterparts we restrict ourselves to deformations that preserve the property
of real analyticity. We note that in the absence of this assumption several TBA integral equations
could be set up (such that their asymptotic limit coincides). 

To summarize we assume that 
\begin{itemize}

\item Y-functions satisfy the Y-system equations (\ref{Y}). 

\item The discontinuities are related by the modified relations 
(\ref{Y1cut}-\ref{Yvw1cut}),(\ref{CutInterpret}).

\item $Y_{-}^{(\alpha)}(u)$ and $Y_{+}^{(\alpha)}(u)$
are related by (\ref{Y-cut0}).

\item The Y-functions are real analytic functions.

\item The Y-functions are smooth deformations of their asymptotic limit.

\end{itemize}

If one assumes that both $\Delta$ and $Y_{-}^{(\alpha)}/Y_{+}^{(\alpha)}$ are analytic and  
bounded near $\pm 2$ then their discontinuities must be zero at the branching points $\pm 2$. 
From this requirement and from (\ref{Deltacut0}), (\ref{Y-cut0}) it follows
that the combinations $\Delta(u)/\sqrt{4-u^2}$ and $\ln  (Y_{-}^{(\alpha)}/Y_{+}^{(\alpha)})/\sqrt{4-u^2}$ 
have no square root branch cuts along the real axis. These properties have been used in the derivation given
in ref. \cite{Tateo1} and we think it is important to emphasize them since they play very important
role also in our considerations.

The form of the discontinuity relations (\ref{Y1cut}-\ref{Yvw1cut}) and (\ref{Y-cut0}) have 
been conjectured to be independent of the particular excited state of the $AdS_5 \times S^5$ sigma-model under 
consideration.
Using the formulae and results of appendices A, B, and C this conjecture can be proven to be valid for the 
asymptotic 
solutions. Here we will {verify}{\footnote{During the verification the upper case index $(\alpha)$
is ignored since in the limit under consideration the two wings of the Y-system become independent.
Furthermore we forget about possible logarithmic discontinuities since (\ref{Y1cut}-\ref{Yvw1cut})
account for the contribution of the square-root discontinuities only. }}
the relations (\ref{Ypmcut}),(\ref{Yvw1cut}) and (\ref{Y-cut0}) in the asymptotic 
{limit}{\footnote{ Asymptotic limit: $J \to \infty$ or $g \to 0$.}}.
The (asymptotic) justification of (\ref{Deltacut}) will be presented in section 6.

With the help of the T-function representation of $Y_{-}^{(0)}/Y_{+}^{(0)}$ and the formulae
of appendix C it can be shown that asymptotically:
\begin{equation} \label{Ym/Yp(0)}
\ln \left( Y_{-}^{(0)}/Y_{+}^{(0)}\right)= \ln \frac{R_p \, B_m}{R_m \, B_p},
\end{equation}
from which $\left[\ln \left(Y_{-}^{(0)}/Y_{+}^{(0)}\right)\right]_{\pm 2N}=0$,
the asymptotic $(Y_Q \to 0)$ limit of (\ref{Ypmcut}), follows immediately.

Next we write
\begin{equation}
\left(1-\frac{1}{Y^{(0)}_-}\right)_*=(1+Y_{1,1}^{(0)})_*=\frac{{\cal F}^{(0)+}{\cal G}^{(0)-}}{T_{2,1}^{(0)}}
=\frac{T_{2,2}^{(0)+}T_{2,2}^{(0)-}}{T_{2,1}^{(0)}T_{2,3}^{(0)}}=1+\frac{1}{Y_{2,2}^{(0)}}=1-\frac{1}{Y_+^{(0)}}
,
\end{equation}
where we have used the identity
\begin{equation}
T_{2,2}^{(0)+}\,T_{2,2}^{(0)-}={\cal F}^{(0)+}\,{\cal G}^{(0)-}T_{2,3}^{(0)}
\end{equation}
and this proves (\ref{Y-cut0}) asymptotically.

The asymptotic verification of (\ref{Yvw1cut}) goes as follows. Using the T-representation:
\begin{equation} \label{Y1vw(0)cut}
\left[ \ln Y_{1|vw}^{(0)} \right]_{\pm 1}=-\left[ \ln Y_{2,1}^{(0)} \right]_{\pm 1}=
\left[ \ln \frac{T_{1,1}^{(0)}}{T_{2,2}^{(0)}} \right]_{\pm 1}.
\end{equation}
With the help of (\ref{T0sT0a0})-(\ref{G0}) it can be shown that:
\begin{equation} \label{T1122cut}
\left[ \ln \frac{T_{1,1}^{(0)}}{T_{2,2}^{(0)}} \right]_{\pm 1}=
\ln \frac{T_{1,1}^{(0)+} \, T_{1,1}^{(0)-}}{{\cal F}^{(0)+}  \, {\cal G}^{(0)-}} 
\frac{T_{3,2}^{(0)}}{T_{2,3}^{(0)}}=
\ln \frac{1+1/Y_{1,1}^{(0)}}{1+Y_{2,2}^{(0)}}=\ln \frac{1-Y_{-}^{(0)}}{1-Y_{+}^{(0)}}.
\end{equation}
Formulae (\ref{Y1vw(0)cut}) and (\ref{T1122cut}) together verify the first relation of 
(\ref{Yvw1cut}) while the
second relation follows from (\ref{T1122cut}) and the following two relations:
\begin{equation} \label{Y1w(0)cut}
\left[ \ln Y_{1|w}^{(0)} \right]_{\pm 1}=\left[ \ln Y_{1,2}^{(0)} \right]_{\pm 1}=
\left[ \ln \frac{T_{1,1}^{(0)}}{T_{2,2}^{(0)}} \right]_{\pm 1}+\left[ \ln T_{1,3}^{(0)} \right]_{\pm 1} ,
\end{equation}
\begin{equation}
\left[ \ln T_{1,3}^{(0)} \right]_{\pm 1}=\ln Y_{1,1}^{(0)}Y_{2,2}^{(0)}=\ln \frac{Y_{+}^{(0)}}{Y_{-}^{(0)}}.
\end{equation}

In the derivation of the excited state TBA equations we can avoid the technical problems coming 
from dealing with branch cuts of the $\log$ functions if we first derive equations 
for the derivatives of the $\log Y$-functions. 
To this end we use the logarithmic derivative of the Y-system (\ref{Y}), 
and the derivative of the discontinuity relations (\ref{Y1cut}-\ref{Yvw1cut}), (\ref{Deltacut0})
and (\ref{Y-cut0}).
Qualitative information about the local singularities of the Y-functions can be read off from
their asymptotic form. Finally we integrate the equations for the derivatives. 

In order to be able to extract the necessary analyticity information on the local singularities
of the Y-functions we will make use of the assumption that the exact Y-functions are smooth deformations
of the asymptotic ones. This however cannot be satisfied in the whole complex plane, but only in certain strips 
(similarity regions). 
We will show that analyticity information regarding the behavior of our functions in these strips are enough to 
derive the TBA equations 
and to determine the exact Y-functions.

The Y-functions are smooth deformations of their asymptotic counterparts as long as the $Y_Q^{(0)}$
functions are small. From (\ref{Ya00}) it follows that this condition is satisfied in the region 
${{\Omega}}_{\varepsilon_Q}=\{u \in \mathbb{C}: \quad | \mbox{Im} \, u|<Q/g, \quad |u-u_j^{[\pm(Q-1)]}|> 
\varepsilon_Q \}$,
where $\varepsilon_Q$ is a small but not infinitesimal positive parameter for $Q\geq 2$ and zero
for $Q=1$.
Then the Y-system equations (\ref{Y}) imply for the other Y-functions 
the following similarity regions: 
\begin{itemize}
\item  $Y_Q(u) \sim Y_{Q}^{(0)}(u)=Y_{Q,0}^{(0)}(u) \qquad \, u \in \Omega_{\varepsilon_Q}, \qquad Q=1,2,...,$
\item $Y_{m|vw}^{(\alpha)}(u) \sim  Y_{m|vw}^{(\alpha)(0)}(u)=1/Y_{m+1,\alpha 1}^{(0)}(u)
\qquad |\mbox{Im} \, u|<m/g, \qquad m=1,2,...,$
\item $Y_{m|w}^{(\alpha)}(u) \sim  Y_{m|w}^{(\alpha)(0)}(u)=Y_{1,\alpha(m+1)}^{(0)}(u)
\qquad |\mbox{Im} \, u|<m/g, \qquad m=1,2,...,$
\item $Y_{-}^{(\alpha)}(u) \sim  Y_{-}^{(\alpha)(0)}(u)=-1/Y_{1,\alpha 1}^{(0)}(u), \qquad
|\mbox{Im} \, u|<2/g, \qquad \alpha=\pm,$
\item $Y_{+}^{(\alpha)}(u) \sim  Y_{+}^{(\alpha)(0)}(u)=-Y_{2,\alpha 2}^{(0)}(u), \qquad
|\mbox{Im} \, u|<2/g, \qquad \alpha=\pm.$
\end{itemize}

The discontinuity relations (\ref{Y1cut}-\ref{Yvw1cut}) simplify considerably in the language
of T-functions. The complicated products of Y-functions in the argument of the $\log$ function
become a product of a few T-functions only. In section 4 we will show
that there is a particular choice for the gauge where the exact T-functions are smooth deformations
of the asymptotic ones and due to their simple square-root branch cut structure the right hand sides of
(\ref{Y1cut}-\ref{Yvw1cut}) simplify drastically.

\begin{figure}
\begin{center}
\psfig{figure=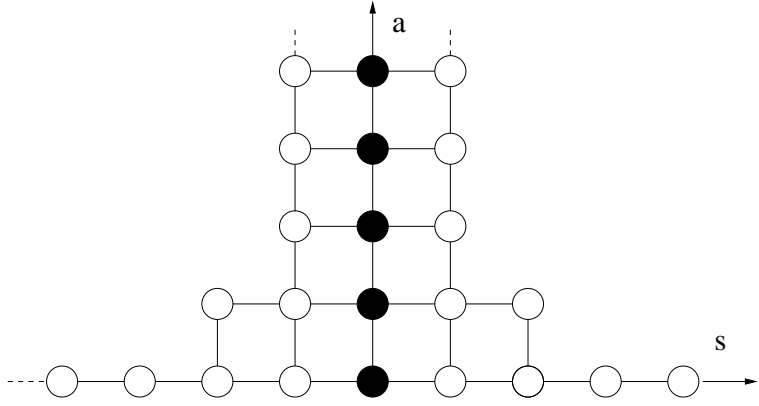,width=10cm}
\end{center}
\vspace{-0.5cm}
\caption{\footnotesize
The ${\rm AdS}_5{\rm /CFT}_4$ Y-system. Full circles on the $a$ axis correspond to
massive nodes with $s=0$. The $s$ axis corresponds to nodes with $a=1$.
}
\label{Yads}
\end{figure}

\begin{figure}
\begin{center}
\psfig{figure=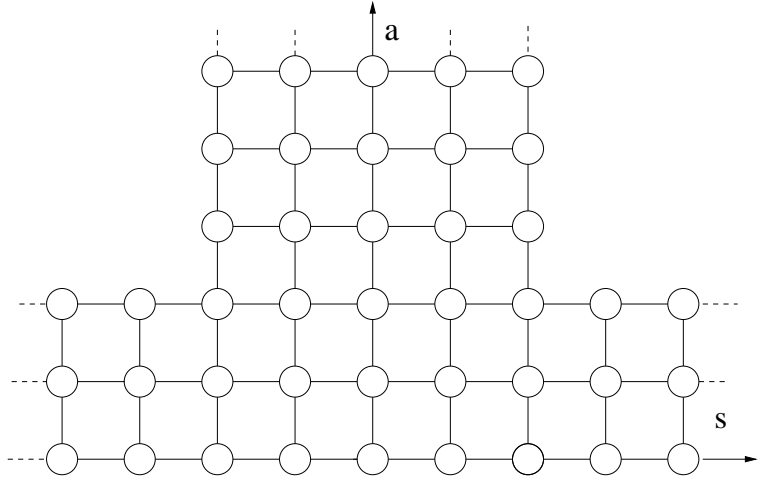,width=10cm}
\end{center}
\vspace{-0.5cm}
\caption{\footnotesize
The ${\rm AdS}_5{\rm /CFT}_4$ T-system.
The $s$ axis corresponds to nodes with $a=0$.
}
\label{Tads}
\end{figure}

\section{TBA equations with cuts}
In this section we transform the Y-system equations (\ref{Y}) into TBA integral
equations. This transformation is not as complete here as for the case of
integrable relativistic models because of the presence of cuts
(discontinuities) in the analytic extension of some of the Y-functions.
The Y-system equations are of the universal form
\begin{equation}
y^+y^-={\cal R},
\label{y}
\end{equation}
but the details of the corresponding integral equation depend on the
analytic properties of the \lq\lq unknown" function $y$.

\subsection{TBA lemma 1a}

Assume that the set of zeroes [poles] of $y(u)$ inside the physical strip
($-1/g<{\rm Im}\,u<1/g$) is $\{\xi_j\}$ $[\{\eta_k\}]$.
We also assume that there may be discontinuities along the cuts with imaginary
part~$\frac{i}{g}$:
\begin{equation}
\frac{y(u+\frac{i}{g}+i\epsilon)}{y(u+\frac{i}{g}-i\epsilon)}=
{\rm e}^{V(u+i\epsilon)},\qquad \vert u\vert\geq2, \qquad V(\pm2)=0.
\end{equation}
Further we assume the large $u$ asymptotic behaviour
\begin{equation}
y(u)\approx y_0\,u^{M_0},\qquad u\longrightarrow\infty
\end{equation}
below the cut and
\begin{equation}
y(u)\approx y_+\,u^{M_+},\qquad u\longrightarrow \infty+\frac{i}{g}+i\epsilon
\end{equation}
just above the cut. (If there is no discontinuity then of course
$y_0=y_+$, $M_0=M_+$.)
The Y-system equation (\ref{y}) implies that ${\cal R}(u)$ 
behaves asymptotically as
\begin{equation}
{\cal R}(u)\approx y_+y_0u^{M_0+M_+},\qquad u\longrightarrow \infty+i\epsilon.
\end{equation}

The function
\begin{equation}
\tau(u)=\frac{\prod_jt(u-\xi_j)}{\prod_kt(u-\eta_k)},
\end{equation}
where
\begin{equation}
t(u)=\tanh\frac{\pi gu}{4}
\end{equation}
has the same zeroes [poles] (in the physical strip) as $y(u)$ and 
satisfies $\tau^+\tau^-=1$.
Using this function and the universal TBA kernel function
\begin{equation}
s(u)=\frac{g}{4\cosh\frac{\pi gu}{2}}
\end{equation}
the solution of (\ref{y}) which has the analytic properties detailed above is
\begin{equation}
y={\rm sgn}(y_0)\tau\exp\{ 
\ln{\cal R}^\epsilon\star s-V^\epsilon \ \check\star\ s\}.
\label{lemma1a}
\end{equation}
Here ${\cal R}^\epsilon(u)={\cal R}(u+i\epsilon)$ and $V^\epsilon(u)=
V(u+i\epsilon)$ and this notation indicates that the integration
contour in (\ref{lemma1a}) goes just above the real line. 
Our definition of the log (ln) function is the standard one: we always
assume that the cut of this function is along the negative real axis.

The result (\ref{lemma1a}) can be proven directly by substituting this
expression into the functional equation (\ref{y}). 

\subsection{TBA lemma 1b}

In one of the TBA equations the Y-function has no cuts near the physical
strip but singular points right at its boundaries. In this case we
assume again that the set of zeroes [poles] of $y(u)$ inside the physical strip
is $\{\xi_j\}$ $[\{\eta_k\}]$ and further assume that the set of
zeroes [poles] on the boundary of the physical strip is $\{z_\alpha\pm
\frac{i}{g}\}$ $[\{w_\beta\pm\frac{i}{g}\}]$, where $z_\alpha$ $[w_\beta]$
are real. The large $u$ asymptotic behaviour is
\begin{equation}
y(u)\approx y_0\,u^{M_0},\qquad u\longrightarrow\infty.
\end{equation}

The Y-system equation (\ref{y}) implies that ${\cal R}(u)$ has double
zeroes [poles] at $\{z_\alpha\}$ $[\{w_\beta\}]$ and behaves asymptotically
as
\begin{equation}
{\cal R}(u)\approx y_0^2 u^{2M_0},\qquad u\longrightarrow \infty+i\epsilon.
\end{equation}

In this case the solution of (\ref{y}) 
with the right analytic properties is
\begin{equation}
y={\rm sgn}(y_0)\tau\exp\{ 
\ln{\cal R} \star s\}.
\label{lemma1b}
\end{equation}

The result (\ref{lemma1b}) can be proven directly because the logarithmic
singularities (at $\{z_\alpha\}$ $[\{w_\beta\}]$ or at $\infty$) are 
integrable and do not invalidate the result.

\subsection{TBA lemma 2}

Let us now assume that the \lq\lq unknown" function $y(u)$ has zeroes [poles]
inside the physical strip as before, and in addition has (multiplicative) 
discontinuities along the real cuts: 
\begin{equation}
\frac{y(u+i\epsilon)}{y(u-i\epsilon)}={\rm e}^{J(u+i\epsilon)},
\qquad \vert u\vert\geq2,\qquad J(\pm2)=0.
\end{equation}
We also assume $y(u)$ has constant asymptotics for large $u$: it approaches
the constants $y_+[y_-]$ when $u\longrightarrow\infty$ just above [below]
the real line. This implies
\begin{equation}
{\rm e}^{J(u+i\epsilon)}\approx J_\infty=\frac{y_+}{y_-},\qquad
{\cal R}(u)\approx y_+y_-, \qquad u\longrightarrow\infty.
\end{equation}

In this case the solution of (\ref{y}) is
\begin{equation}
y={\rm sgn}(y_+)\tau\exp\{ 
\ln{\cal R}\star s+J^\epsilon \ \check\star\ s_1\},
\label{lemma2}
\end{equation}
where
\begin{equation}
s_1(u)=\frac{-ig}{4\sinh\left(\frac{\pi gu}{2}-i\epsilon\right)}.
\end{equation}

\subsection{TBA integral equations}

Using the above two lemmas we now write down the set of TBA integral
equations corresponding to the Y-system (\ref{Y}). First we spell out
these equations using the notations of (\ref{Ydef1}-\ref{Ydef2}):
\begin{eqnarray}
Y^{(\alpha)+}_{m\vert vw}\,Y^{(\alpha)-}_{m\vert vw}&=&
\frac{(1+Y^{(\alpha)}_{m+1\vert vw})(1+Y^{(\alpha)}_{m-1\vert vw})}
{(1+Y_{m+1})},\qquad\qquad m\geq2,\label{Ymvw}\\
Y^{(\alpha)+}_{1\vert vw}\,Y^{(\alpha)-}_{1\vert vw}&=&
\frac{1-Y_-^{(\alpha)}}{1-Y_+^{(\alpha)}}\,
\frac{(1+Y^{(\alpha)}_{2\vert vw})}
{(1+Y_2)},\label{Y1vw}\\
Y^{(\alpha)+}_{m\vert w}\,Y^{(\alpha)-}_{m\vert w}&=&
(1+Y^{(\alpha)}_{m+1\vert w})(1+Y^{(\alpha)}_{m-1\vert w}),
\qquad\qquad m\geq2,\label{Ymw}\\
Y^{(\alpha)+}_{1\vert w}\,Y^{(\alpha)-}_{1\vert w}&=&
\frac{1-\frac{1}{Y_-^{(\alpha)}}}{1-\frac{1}{Y_+^{(\alpha)}}}\,
(1+Y^{(\alpha)}_{2\vert w}),\label{Y1w}\\
Y^+_Q\,Y^-_Q&=&\frac{Y_{Q+1}\,Y_{Q-1}}{Y^{(+)}_{Q-1\vert vw}\,
Y^{(-)}_{Q-1\vert vw}}\,\frac{(1+Y^{(+)}_{Q-1\vert vw})
(1+Y^{(-)}_{Q-1\vert vw})}{(1+Y_{Q+1})(1+Y_{Q-1})},\quad\quad 
Q\geq2,\label{YQ}\\
Y^+_1\,Y^-_1&=&\frac{Y_2}{Y^{(+)}_-\,Y^{(-)}_-}\,
\frac{(1-Y_-^{(+)})(1-Y_-^{(-)})}{1+Y_2},\label{Y1}\\
Y_-^{(\alpha)+}\,Y_-^{(\alpha)-}&=&\frac{1+Y_{1\vert vw}^{(\alpha)}}
{(1+Y_1)(1+Y^{(\alpha)}_{1\vert w})}.\label{Y-}
\end{eqnarray}

Using the notation $X$ for a general index, which can take the
values ${\phantom Y}_Q$, ${\phantom Y}^{(\alpha)}_{m\vert vw}$,
${\phantom Y}^{(\alpha)}_{m\vert w}$ or ${\phantom Y}^{(\alpha)}_\pm$,
and denoting the set of zeroes of $Y_X$ in the physical strip
by $\{\xi_{X,j}\}$ and the set of its poles by $\{\eta_{X,k}\}$, 
finally the sign of $Y_X(u)$ in the limit $u\longrightarrow\infty+i\epsilon$ 
by $({\rm sgn})_X$, we define
\begin{equation}
t_X(u)=({\rm sgn})_X\,\left\{\frac{\prod_jt(u-\xi_{X,j})}
{\prod_kt(u-\eta_{X,k})}\right\}.
\end{equation}

From the discontinuity relations we see that
we can use Lemma 1a with $V=0$ for the cases (\ref{Ymvw}), (\ref{Ymw}) and
if $Q\geq3$ also for (\ref{YQ}). Lemma 1b is used for (\ref{YQ}) if $Q=2$. 
Further we have to use Lemma 1a
\begin{eqnarray}
{\rm for\ the\ }Y_{1\vert vw}^{(\alpha)}{\rm \ equation\ (\ref{Y1vw})\ 
with\ }
V&=&\ln\left(\frac{1-Y_-^{(\alpha)}}{1-Y_+^{(\alpha)}}
\right)\,\nonumber\\
{\rm for\ the\ }Y_{1\vert w}^{(\alpha)}{\rm \ equation\ (\ref{Y1w})\ 
with\ }
V&=&\ln\left(\frac{1-\frac{1}{Y_-^{(\alpha)}}}{1-\frac{1}{Y_+^{(\alpha)}}}
\right)\,\nonumber\\ 
{\rm for\ the\ }Y_1{\rm \ equation\ (\ref{Y1})\ with\ }V&=&\Delta\,\nonumber
\end{eqnarray}
and finally Lemma 2 for the $Y_-^{(\alpha)}$ equation (\ref{Y-}) with
$J=J^{(\alpha)}$.

We find
\begin{eqnarray}
Y^{(\alpha)}_{m\vert vw}&=&t^{(\alpha)}_{m\vert vw}\exp\left\{
\ln\left[\frac{(1+Y^{(\alpha)}_{m+1\vert vw})(1+Y^{(\alpha)}_{m-1\vert vw})}
{(1+Y_{m+1})}\right]\star s\right\},\qquad m\geq2,\label{TBAmvw}\\
Y^{(\alpha)}_{1\vert vw}&=&t^{(\alpha)}_{1\vert vw}\exp\left\{
\ln\left[\frac{(1+Y^{(\alpha)}_{2\vert vw})}{(1+Y_2)}\right]\star s
+\ln\left[\frac{1-Y_-^{(\alpha)}}{1-Y_+^{(\alpha)}}\right]\ \hat\star\ s
\right\},\label{TBA1vw}\\
Y^{(\alpha)}_{m\vert w}&=&t^{(\alpha)}_{m\vert w}\exp\left\{
\ln\left[(1+Y^{(\alpha)}_{m+1\vert w})(1+Y^{(\alpha)}_{m-1\vert w})\right]
\star s\right\},\qquad m\geq2,\label{TBAmw}\\\
Y^{(\alpha)}_{1\vert w}&=&t^{(\alpha)}_{1\vert w}\exp\left\{
\ln\left[1+Y^{(\alpha)}_{2\vert w}\right]\star s+ 
\ln\left[\frac{1-\frac{1}{Y_-^{(\alpha)}}}{1-\frac{1}{Y_+^{(\alpha)}}}\right]
\ \hat\star\ s \right\}\,,\label{TBA1w}\\
Y_Q&=&t_Q\exp\left\{\ln\left[
\frac{Y_{Q+1}\,Y_{Q-1}
(1+Y^{(+)}_{Q-1\vert vw})(1+Y^{(-)}_{Q-1\vert vw})}
{Y^{(+)}_{Q-1\vert vw}Y^{(-)}_{Q-1\vert vw}
(1+Y_{Q+1})(1+Y_{Q-1})}\right]\star s \right\},\quad Q\geq2,\label{TBAQ}\\
Y_1&=&t_1\exp\left\{\ln\left[\frac{Y_2}{Y^{(+)}_-\,Y^{(-)}_-}\,
\frac{(1-Y_-^{(+)})(1-Y_-^{(-)})}{1+Y_2}\right]\star s
-\Delta\ \check\star\ s
\right\},\label{TBA1}\\
Y_-^{(\alpha)}&=&t_-^{(\alpha)}\exp\left\{\ln\left[
\frac{1+Y_{1\vert vw}^{(\alpha)}}{(1+Y_1)(1+Y^{(\alpha)}_{1\vert w})}
\right]\star s+J^{(\alpha)}\ \check\star\ s_1 \right\}.\label{TBA-}
\end{eqnarray}

There is no equation for $Y_+^{(\alpha)}$, but since it is the analytic
continuation of $Y_-^{(\alpha)}$, the very definition of the discontinuity
$J^{(\alpha)}$ can be written as
\begin{equation}
Y_+^{(\alpha)}=Y_-^{(\alpha)}\exp\left\{-J^{(\alpha)}\right\}.
\label{Y+}
\end{equation}
The above set of TBA integral equations is incomplete yet, since the
discontinuities $\Delta$ and $J^{(\alpha)}$ are undetermined. They
will be obtained using dispersion relations in Section 6 and Section 5,
respectively.

\section{Construction of the T-system}
In this section we start to construct a T-system from the Y-system. 
This is not unique because there is a gauge freedom in this transformation.
Since our main assumption is that the exact Y-functions are (at least
qualitatively) close to their asymptotic values given by the Bethe
Ansatz solution and in our subsequent considerations the T-system plays 
an important role, we will choose a gauge in which also the T-system is close
to the asymptotic solution. This will be achieved here only partially and
the final, complete gauge fixing will be given in Section 6.

\subsection{Chain lemma}

Let us assume that we want to find the solution of the infinite system
\begin{equation}
\frac{\sigma_a^+\sigma_a^-}{\sigma_{a+1}\sigma_{a-1}}=\xi_a,\qquad
a=1,2,\dots,\qquad \sigma_0\equiv1,
\label{chain}
\end{equation}
where the unknowns $\sigma_a(u)$ are assumed to have no zeroes/poles
near the physical strip (in the physical strip and a little beyond).
$\xi_a(u)$ on the right hand side are given functions not having zeroes/poles
near the real axis and taking complex values (excluding the 
negative real axis). Furthermore, (only) $\xi_1$ may have discontinuities
on the real axis:
\begin{equation}
\frac{\xi_1(u+i\epsilon)}{\xi_1(u-i\epsilon)}={\rm e}^{I(u+i\epsilon)},
\qquad I(\pm2)=0.
\end{equation}
In such a case we assume that (only) $\sigma_1$ has discontinuities along
the cuts with imaginary part~$\pm\frac{i}{g}$:
\begin{equation}
\frac{\sigma_1(u+\frac{i}{g}+i\epsilon)}
{\sigma_1(u+\frac{i}{g}-i\epsilon)}={\rm e}^{p(u+i\epsilon)},\qquad\quad
\frac{\sigma_1(u-\frac{i}{g}+i\epsilon)}
{\sigma_1(u-\frac{i}{g}-i\epsilon)}={\rm e}^{m(u+i\epsilon)},
\end{equation}
where
\begin{equation}
p+m=I,\qquad \quad p(\pm2)=m(\pm2)=0.
\end{equation}

The solution of (\ref{chain}) is given by
\begin{equation}
\sigma_a=\exp\left\{\sum_{A=1}^\infty \ln\xi_A^\epsilon
\star \ell^A_a -p^\epsilon \ \check\star\ K_a \right\}\,
\label{chain1}
\end{equation}
where
\begin{equation}
\ell^A_a=\sum_{j=0}^{A-1}K_{a+1-A+2j}.
\end{equation}
(Here we use the convention $K_{-a}=-K_a$, $K_0=0$.) Note that
\begin{equation}
\ell^A_a=\ell^a_A,\qquad\quad \ell^1_a=K_a.
\end{equation}
Because of the discontinuities, we have to integrate slightly above the real 
axis as indicated by
\begin{equation}
\xi_A^\epsilon(u)=\xi_A(u+i\epsilon),\qquad\quad
p^\epsilon(u)=p(u+i\epsilon),
\end{equation}
but this is only necessary for $\xi_1$.

\subsection{Constructing $T_{a,0}$}

To construct $T_{a,0}$ we can use the chain lemma with
\begin{equation}
\sigma_a=T_{a,0},\qquad\quad \xi_a=1+Y_a,\qquad a=1,2,\dots
\end{equation}
(We use the $T_{0,s}\equiv1$ gauge throughout this paper.)

From the asymptotic solution (and assuming similar behavior for the
exact solution) we see that for $a\geq2$ $1+Y_a$ has no zeroes/poles 
in the strip with imaginary part in the interval 
$(\frac{1-a}{g},\frac{a-1}{g})$ (the strip
$(1-a,a-1)$ for short) and no cuts in the strip $(-a,a)$. The first 
singularities are poles at $u_j^{[\pm(a-1)]}$ and the first cuts are
at $\pm\frac{ia}{g}$. $1+Y_1$ has no zeroes/poles/cuts in $(-1,1)$.

$\xi_1$ has no real cuts in this case and we can put $p=0$.
In the first step we construct $T_{a,0}$ by using the chain lemma
formula (\ref{chain1}). This gives the solution along the real axis
but it is easily extended to the whole physical strip (and a little
beyond). The next step is to use the defining relation
\begin{equation}
\frac{T_{a,0}^+T_{a,0}^-}{T_{a+1,0}T_{a-1,0}}=1+Y_a,\qquad
a=1,2,\dots,
\end{equation}
which (for $a\geq2$) extends the solution so that $T_{a,0}$ is free
of any zeroes/poles in the strip $(-a,a)$ and meromorphic in $(-1-a,a+1)$.
(The first poles are at $u_j^{[\pm a]}$ and the first cuts are
at $\pm\frac{i(a+1)}{g}$.) $T_{1,0}$ is more regular: it has no 
zeroes/poles/cuts in the strip $(-2,2)$ with first cuts at $\pm\frac{2i}{g}$.

\subsection{Constructing $T_{a,1}$}

Here we restrict our attention to the ${\alg{sl}(2)}$ (sub-)sector 
(see appendix C) only. In this
special case the two sides of the Y-system are identical:
\begin{equation}
Y^{(\alpha)}_{m\vert vw}=Y_{m\vert vw},\qquad
Y^{(\alpha)}_{m\vert w}=Y_{m\vert w},\qquad
Y^{(\alpha)}_\pm=Y_\pm,\qquad
m=1,2,\dots,\qquad \alpha=\pm.
\end{equation}

We can use here the chain lemma using the identifications
\begin{equation}
T_{a,1}=-\sigma_a\tau_a(-1)^a,\qquad a=1,2,\dots,\qquad\quad T_{0,1}\equiv1
\end{equation}
and
\begin{equation}
\xi_a=\tau_{a+1}\tau_{a-1}\left(1+\frac{1}{Y_{a-1\vert vw}}\right),\qquad
a=2,3,\dots,\qquad
\xi_1=-\tau_2\left(1-\frac{1}{Y_-}\right).
\end{equation}

In this case there are real cuts for $\xi_1$ and to specify the solution 
completely we also need $p(u)$. This function will be fixed later 
(in section 6). Again, we first use the formula (\ref{chain1}) to determine
$T_{a,1}$ near the physical strip with $T_{1,1}$ having discontinuities
along the $\pm\frac{i}{g}$ cuts. Then, using the defining relation
\begin{equation}
\frac{T_{a,1}^+T_{a,1}^-}{T_{a+1,1}T_{a-1,1}}=1+Y_{a,1},\qquad
a=1,2,\dots,
\label{Ya1}
\end{equation}
we can extend the solution. The $T_{a,1}$ constructed this way will
be meromorphic in the strip $(-a,a)$ and the first cuts occur at 
$\pm\frac{ia}{g}$.

\subsection{The complete T-system}

Having constructed the T-system elements $T_{a,0}$ and $T_{a,1}$, the rest of 
the T-system can simply be calculated from the relation between the T-system
and Y-system elements. For example, we have
\begin{equation}
T_{a,2}=\frac{T_{a-1,1}}{Y_{a-1\vert vw}}\,\frac{T_{a+1,1}}{T_{a,0}},
\qquad a=2,3,\dots
\end{equation}
From this representation we can see that $T_{a,2}$ is meromorphic in the strip
$(1-a,a-1)$. In the $a=1$ case we have
\begin{equation}
T_{1,2}=-\frac{T_{2,1}}{T_{1,0}}\,\frac{1}{Y_-}.
\end{equation}
This function has discontinuities along the real cuts inherited from $Y_-$.
The other factor is a meromorphic function in the $(-2,2)$ strip.

It is also possible to calculate $T_{1,s}$, $T_{2,s}$ for $s=3,4,\dots$
and $T_{a,s}$ for $s<0$. These functions will not be used in our 
considerations. We just note that $T_{a,s}\not=T_{a,-s}$ in general,
in spite of the fact that the two sides of the Y-system are identical
in the ${\alg{sl}(2)}$ (sub-)sector we are considering here. To illustrate
this, we calculate
\begin{equation}
T_{a,-1}=\frac{T_{a+1,0}\,T_{a-1,0}\,Y_{a,0}}
{T_{a,1}}
\qquad a=1,2,\dots
\label{Tam1}
\end{equation}
from the relation between the T-system and Y-system elements. Although it
is not at all obvious from the above formula, we know that by construction
the set $\{T_{a,-1}\}$ must also satisfy (\ref{Ya1}) with the same 
right hand side (assuming that we stay in the ${\alg{sl}(2)}$ \break
(sub-)sector).
This means that the T-functions $\{T_{a,-1}\}$ are gauge transforms of 
the functions $\{T_{a,1}\}$ in the sense discussed below. Indeed, using
the results given in appendix C, we can verify the above structure
by explicitly calculating $\{T_{a,-1}\}$ from (\ref{Tam1}) in the 
asymptotic limit.
We also find that the T-functions $\{T_{a,-1}\}$ are exponentially 
small asymptotically.

\subsection{Gauge transformations}

The relation between the Y-system and T-system is not unique, there is
a gauge freedom $T_{a,s}\rightarrow \hat T_{a,s}$. We restrict this gauge
freedom by demanding
\begin{equation}
\hat T_{0,s}=T_{0,s}\equiv1,\qquad\quad \hat T_{a,0}=T_{a,0},
\end{equation}
i.e. we work in the $T_{0,s}\equiv1$ gauge and also fix $T_{a,0}$ as 
constructed explicitly above using the chain lemma. The remaining
gauge freedom is of the form
\begin{equation}
\hat T_{a,s}=\frac{f^{[s-a]}f^{[a-s]}}{f^{[a+s]}f^{[-a-s]}}\,T_{a,s}.
\end{equation}
The gauge transformation is generated by a single function $\beta(u)$:
\begin{equation}
\hat T_{1,1}=\beta\, T_{1,1},\qquad \beta=\frac{f^2}{f^{++}f^{--}},
\end{equation}
\begin{equation}
\hat T_{2,1}=\beta^+\beta^- T_{2,1},\qquad \hat T_{2,2}=
\beta^{++}\beta^2 \beta^{--} T_{2,2},\qquad \dots
\end{equation}
We see that the $T_{a,1}$ solution constructed in this section with
some $p(u)$ is of the form as if it were
a gauge transform of the $p(u)=0$ solution with
the discontinuous gauge transformation
\begin{equation}
\beta=\frac{q^+}{q^-},\qquad q(u)=\exp\left\{\int \hspace{-3.9mm}\square
\frac{{\rm d}v}{2\pi i}\,\frac{p(v)}{u-v}\right\}.
\label{qdef}
\end{equation}
Since our Y-functions and T-functions have discontinuities, it is
natural to allow also gauge transformations that have discontinuities
along the same cut lines.

\subsection{Large $\vert u\vert$ asymptotics}

It is easy to show that if the function $f(u)$ has large $\vert u\vert$
asymptotics
\begin{equation}
f(u)\approx f_\pm+\kappa\ln\vert u\vert, \qquad\quad u\longrightarrow\pm\infty,
\end{equation}
then this is reproduced by the $f\star K_a$ convolution:
\begin{equation}
(f\star K_a)(u)\approx f_\pm+\kappa\ln\vert u\vert, 
\qquad\quad u\longrightarrow\pm\infty,
\end{equation}
and similarly for the modified convolution $f \ \check\star\ K_a$.
From the known asymptotics of $Y_{m\vert vw}$ and $Y_-$ we can determine
the large $\vert u\vert$ behavior of the functions on the right hand side
of (\ref{chain}) for the $T_{a,1}$ problem:
\begin{equation}
\xi_a(u)\approx \frac{a^2}{a^2-1},\quad a=2,3,\dots,\qquad\quad
\xi_1(u)\approx \frac{B}{u},\quad\quad 
\vert u\vert\longrightarrow\infty
\end{equation}
and the discontinuity behaves as
\begin{equation}
p(u)\approx \ln\frac{Bu}{\vert A\vert}.
\label{pasy}
\end{equation}
Thus we have
\begin{equation}
\ln\xi_1\star K_a
\approx\ln\frac{B}{u},\qquad\quad
p\ \check\star\ K_a \approx\ln\frac{Bu}{\vert A\vert}
\end{equation}
and further
\begin{equation}
\sigma_a(u)\approx\frac{\vert A\vert}{u^2}\exp\left\{
\sum_{m=1}^\infty\,{\rm min}(m+1,a)\ln\frac{(m+1)^2}{m(m+2)}\right\}=
\frac{2\vert A\vert a}{u^2},
\end{equation}
which reproduces the expected large $\vert u\vert$ behavior of $T_{a,1}$.

\section{Dispersion relation for $Y_-$}
In this section we determine the discontinuity $J^{(\alpha)}=
\ln\frac{Y_-^{(\alpha)}}{Y_+^{(\alpha)}}$ using the discontinuity relation
(\ref{Ypmcut}), which can be rewritten as
\begin{equation}
\left[J^{(\alpha)}\right]_{\pm 2N}=-\sum_{Q=1}^N\left[L_Q\right]_{\pm(2N-Q)},
\qquad N\geq1.
\label{Ypmcut1}
\end{equation}
Here we introduced the notation $L_Q=\ln(1+Y_Q)$. Using the notation
$t_Q=\ln T_{Q,0}$ ($t_0\equiv0$) and the relation
\begin{equation}
L_Q=t_Q^++t_Q^--t_{Q+1}-t_{Q-1},\qquad\quad Q=1,2,\dots,
\end{equation}
(\ref{Ypmcut1}) can be drastically simplified. In the language of
the $t_Q$ variables most of the terms cancel and we are left with
\begin{equation}
\left[J^{(\alpha)}\right]_{\pm 2N}=\left[t_{N+1}\right]_{\pm N}
-\left[t_1\right]_{\pm 2N}-\left[t_N\right]_{\pm (N-1)}=
-\left[t_1\right]_{\pm 2N}.
\label{Ypmcut2}
\end{equation}
The last equality follows from using the analytic properties of the
$T_{Q,0}$ elements constructed in the previous section by the chain lemma,
namely that $T_{Q,0}$ are analytic in the strip $(-Q,Q)$.
The last relation is equivalent to saying that the combination 
$\frac{Y_-^{(\alpha)}}{Y_+^{(\alpha)}}T_{1,0}$ has discontinuities along the
real axis only (in the gauge we are using) or equivalently that 
$H^{(\alpha)}$ has discontinuities only on the real axis, where
\begin{equation}
J^{(\alpha)}=-t_1-\ln H^{(\alpha)},\qquad
H^{(+)}=\frac{T_{2,3}}{T_{3,2}},\qquad
H^{(-)}=\frac{T_{2,-3}}{T_{3,-2}}.
\end{equation}

\subsection{Dispersion relation}

\begin{figure}
\begin{center}
\psfig{figure=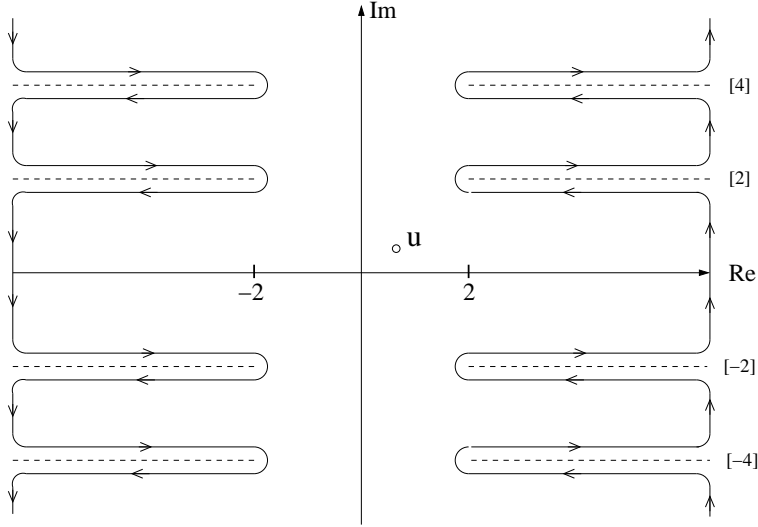,width=10cm}
\end{center}
\vspace{-0.5cm}
\caption{\footnotesize
The contour $\Gamma_0$. It goes around all positive and negative even cuts.
}
\label{gnull}
\end{figure}

Note that
\begin{equation}
J^{(\alpha)}(u+i\epsilon)=-J^{(\alpha)}(u-i\epsilon),\qquad
\vert u\vert\geq2,
\end{equation}
which follows from the fact that $(Y_\pm)_*=Y_\mp$, i. e. $Y_\pm$ are
analytic continuations of each other. Moreover, since we know that
$\exp\{J^{(\alpha)}\}$  
has no real zeroes/poles (this is true in the asymptotic limit
and according to our basic assumption it remains true also exactly) the 
combination
\begin{equation}
\frac{J^{(\alpha)\prime}(u)}{\sqrt{4-u^2}}
\label{Jpersqrt}
\end{equation}
is meromorphic near the real line (in the strip 
$\vert{\rm Im}u\vert<\frac{\gamma}{g}$, where $\gamma$ is small, 
but not infinitesimal) and has no other poles than those at $u=\pm2$. 
Note that the function (\ref{Jpersqrt}) 
goes like $1/u^2$ for large $\vert u\vert$, which is necessary 
for some of our integrals to converge. The above properties allow us to
define (for $0<{\rm Im}u<\frac{\gamma}{g}$) 
\begin{equation}
\Gamma^{(\alpha)}(u)=\oint_{\Gamma_0}{\rm d}vJ^{(\alpha)\prime}(v)K(v,u),
\label{Gammaalpha}
\end{equation} 
where the contour $\Gamma_0$ goes around all the even cuts $\pm 2N$ (see
Fig.~\ref{gnull}).

\begin{figure}
\begin{center}
\psfig{figure=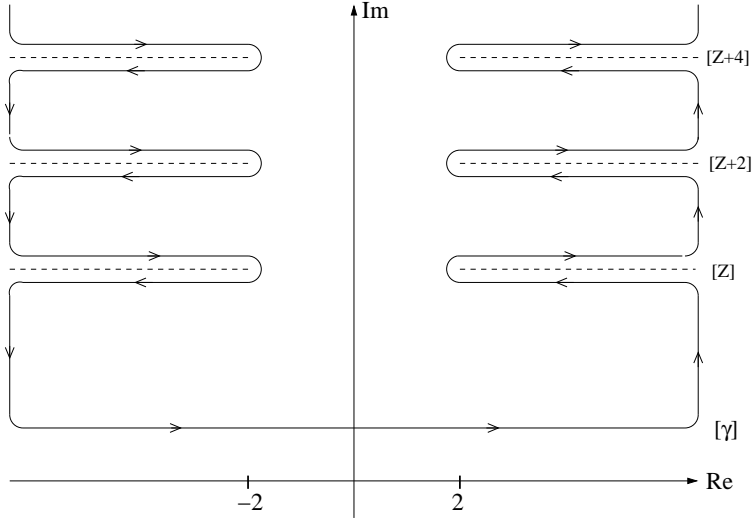,width=10cm}
\end{center}
\vspace{-0.5cm}
\caption{\footnotesize
This contour first goes parallel to the real axis and goes around
the cuts at $Z+2m$, $Z$ positive integer, $m=0,1,\dots$  
}
\label{gzinfty}
\end{figure}

For later use we define also the contour integral
$\oint_{[\gamma,Z,\infty]}{\rm d}v$, where the contour consists of a horizontal
line $v+\frac{i\gamma}{g}$ and goes around all the $Z+2m$ cuts ($m=0,1,\dots$).
Here $\gamma<Z$, $Z$ a positive integer. See Fig.~\ref{gzinfty}.
The contour integral $\oint_{[-\infty,-Z,-\gamma]}{\rm d}v$ is defined 
similarly. Here the contour goes around all the $-(Z+2m)$ cuts ($m=0,1,\dots$)
and comes back (from right to left) along the horizontal line 
$v-\frac{i\gamma}{g}$. Here $-Z<-\gamma$, $Z$ a positive integer.

Evaluating the integral (\ref{Gammaalpha}), it can be written as a sum of 
three terms, the first being the contribution of the narrow strip
$-\frac{\gamma}{g}<{\rm Im}v<\frac{\gamma}{g}$:
\begin{equation}
\Gamma^{(\alpha)}(u)=({\cal D}J^{(\alpha)})(u)+
\oint_{[\gamma,2,\infty]}{\rm d}v\,J^{(\alpha)\prime}(v)K(v,u)+
\oint_{[-\infty,-2,-\gamma]}{\rm d}v\,J^{(\alpha)\prime}(v)K(v,u).
\label{Gammaalpha1}
\end{equation}
Here we introduced the derivative operator ${\cal D}$ defined as
\begin{equation}
({\cal D}q)(u)=q^\prime(u)+\frac{1}{4\sqrt{4-u^2}}\left\{
(u-2)q_d(-2)-(u+2)q_d(2)\right\},
\end{equation}
where $q_d(u)=\sqrt{4-u^2}q^\prime(u)$. The above three terms correspond to the
three poles ($v=u$, $v=\pm2$) in the narrow strip, assuming that $q(u)$
has square root cuts for $\vert u\vert\geq2$, in which case $q_d(u)$ 
is analytic around the branch points $u=\pm2$.

The usefulness of this definition can be seen from the fact that the
${\cal D}$ derivative can be \lq\lq integrated" in the following 
sense. If the function $g(u)$ can be represented as
\begin{equation}
g(u)=\int_{-\infty}^\infty{\rm d}v\,f(v)K^{[a]}(v,u),
\label{calD1}
\end{equation}
then
\begin{equation}
({\cal D}g)(u)=\int_{-\infty}^\infty{\rm d}v\,f^\prime(v)K^{[a]}(v,u).
\label{calD2}
\end{equation}
If
\begin{equation}
h(u)=\frac{1}{2\pi i}\,\sigma(u,\xi),
\label{calD3}
\end{equation}
where $\xi$ is constant, then
\begin{equation}
({\cal D}h)(u)=-K(\xi,u)
\label{calD4}
\end{equation}
and finally if
\begin{equation}
{\cal A}(u)=2b\ln(-x(u)),
\label{zeromode}
\end{equation}
then 
\begin{equation}
({\cal D}{\cal A})(u)=0,
\label{calD6}
\end{equation}
i. e. (\ref{zeromode}) is the zero mode of the derivative ${\cal D}$.
Note that ${\rm e}^{{\cal A}(\pm2)}=1$ if $b$ is integer and note also that
\begin{equation}
{\cal A}(u)\approx -2b\ln\vert u\vert + {\rm const.},\qquad\quad
u\longrightarrow i\epsilon\pm\infty.
\end{equation}
This last property makes it possible to adjust the large $\vert u\vert$
behavior of the inverse of the operator ${\cal D}$, by adding a multiple 
of the zero mode, if necessary.

There exists an alternative way of calculating (\ref{Gammaalpha}) using the
(derivative of the) discontinuity relations (\ref{Ypmcut1}). The contribution
of the integrals along the cuts in the upper half plane is 
\begin{equation}
\sum_{Q=1}^\infty\left\{\int_{-\infty}^\infty{\rm d}v\,K^{[Q]}(v,u)
L^\prime_Q(v)-\oint_{[0,Q,\infty]}{\rm d}v\,K^{[Q]}(v,u)L^\prime_Q(v)
\right\}.
\end{equation}
Here the pole term ($\oint$ part) simplifies drastically in terms of $t_Q$
(using the same analyticity information as was already used above) and becomes 
\begin{equation}
-\oint_{[0,2,\infty]}{\rm d}v\,K(v,u)t^\prime_1(v).
\end{equation}
After similar considerations concerning the contribution of the cuts in the
lower half plane, we arrive at
\begin{equation}
\begin{split}
\Gamma^{(\alpha)}(u)=\sum_{Q=1}^\infty &\int_{-\infty}^\infty{\rm d}v\,
L^\prime_Q(v)\left\{K^{[Q]}(v,u)-K^{[-Q]}(v,u)\right\}\\
&-\oint_{[\gamma,2,\infty]}{\rm d}v\,t^\prime_1(v)K(v,u)
-\oint_{[-\infty,-2,-\gamma]}{\rm d}v\,t^\prime_1(v)K(v,u).
\end{split}
\label{Gammaalpha2}
\end{equation}
Comparing (\ref{Gammaalpha1}) and (\ref{Gammaalpha2}) we can
express ${\cal D}J^{(\alpha)}$ as
\begin{equation}
\begin{split}
({\cal D}J^{(\alpha)})(u)&=
\sum_{Q=1}^\infty \int_{-\infty}^\infty{\rm d}v\,
L^\prime_Q(v)\left\{K^{[Q]}(v,u)-K^{[-Q]}(v,u)\right\}\\
&+\oint_{[\gamma,2,\infty]}{\rm d}v\,\ln^\prime H^{(\alpha)}(v)K(v,u)
+\oint_{[-\infty,-2,-\gamma]}{\rm d}v\,\ln^\prime H^{(\alpha)}(v)K(v,u).
\end{split}
\label{derJalpha}
\end{equation}
We already know that $H^{(\alpha)}$ has no discontinuities in the
upper and lower half planes, but to be able to \lq\lq integrate"
(\ref{derJalpha}) we need to know the position of its zeroes/poles.
We assume that its set of zeroes is $\{x_a^{(\alpha)}\}$ and
the set of poles is $\{y_b^{(\alpha)}\}$ (and also assume that these
singular points are not real and are not on any of the even cuts).
Then, using (\ref{calD1}-\ref{calD4}), the \lq\lq integral" of 
(\ref{derJalpha}) can be written as
\begin{equation}
\begin{split}
J^{(\alpha)}(u)=\sum_b\sigma(u,& y_b^{(\alpha)})-
\sum_a\sigma(u,x_a^{(\alpha)})\\
&+\sum_{Q=1}^\infty \int_{-\infty}^\infty{\rm d}v\,
L_Q(v)\left[K^{[Q]}(v,u)-K^{[-Q]}(v,u)\right].
\end{split}
\label{Jalpha}
\end{equation}
Since for large $\vert u\vert$ $J^{(\alpha)}\longrightarrow{\rm const.}$, 
there was not necessary to add the zero mode (\ref{zeromode}) to this solution.

According to our basic assumption, $H^{(\alpha)}$ is close to its asymptotic
counterpart:
\begin{equation}
H^{(\alpha)}\approx H^{(0)}=\frac{T^{(0)}_{2,3}}{T^{(0)}_{3,2}}=
\frac{B_pR_m}{B_mR_p}.
\end{equation}
$H^{(0)}$ has no zeroes and its poles are at $u_j^\pm$. Being a smooth 
deformation, $H^{(\alpha)}$ must also be free of zeroes and must have the same
number of poles, which are close to the corresponding asymptotic positions.
Let us denote the positions of the poles by $w_j^{p(\alpha)}$ and
$w_j^{m(\alpha)}$, where
\begin{equation}
w_j^{p(\alpha)}\approx u_j^+,\qquad\quad
w_j^{m(\alpha)}\approx u_j^-.
\end{equation}
(These are indeed far from the real axis and all even cuts.) 

(\ref{Jalpha}) can now be written as:
\begin{equation}
\begin{split}
J^{(\alpha)}(u)=\sum_{j=1}^N &\left\{
\ln \frac
{x(u)-x_j^{m(\alpha)}}{\frac{1}{x(u)}-x_j^{m(\alpha)}}
+\ln\frac{\frac{1}{x(u)}-x_j^{p(\alpha)}}
{x(u)-x_j^{p(\alpha)}}\right\}\\
&+\sum_{Q=1}^\infty \int_{-\infty}^\infty{\rm d}v\,
L_Q(v)\left[K^{[Q]}(v,u)-K^{[-Q]}(v,u)\right],
\end{split}
\label{Jalpha1}
\end{equation}
where
\begin{equation}
x_j^{p(\alpha)}=\frac{1}{x(w_j^{p(\alpha)})}\approx x_j^+,\qquad\quad
x_j^{m(\alpha)}=x(w_j^{m(\alpha)})\approx x_j^-.
\end{equation}

\subsection{$u_j$ related singularities}

In the asymptotic solution the singularities of several Y-system
elements are given in terms of the set $\{u_j\}$, the set of
physical rapidities. In the exact solution, the positions of the
singular points are smoothly moving away from their asymptotic
values and it is possible that positions coinciding in the
asymptotic limit move away from each other. However, the Y-system
equations give very strong restrictions also for these positions.

In the asymptotic solution the set of real zeroes of 
$Y^{(\alpha)(0)}_{1\vert vw}$ coincides with the set of physical rapidities
for both values of $\alpha$. After the smooth deformation, the two
sets $\{u_j^{(\alpha)}\}$, $\alpha=\pm$ may be different, but since
$Y^{(\alpha)}_{1\vert vw}$ are real analytic functions, their zeroes remain
real. 

Next we use (\ref{YQ}) in the $Q=2$ case. This gives that $Y_2$ has
poles at $u_j^{(+)+}$ and $u_j^{(-)-}$ or $u_j^{(-)+}$ and $u_j^{(+)-}$  
(for all $j=1,2,\dots N$). Since $Y_2$ is real analytic, both options
imply that
\begin{equation}
u_j^{(+)}=u_j^{(-)}=u_j,\qquad\quad j=1,2,\dots,N.
\end{equation}

Now let us use the T-Y relation
\begin{equation}
T_{2,0}^+T_{2,0}^-=T_{1,0}T_{3,0}(1+Y_2).
\end{equation}
We conclude that $T_{2,0}$ has poles at $u_j^{++}$ and $u_j^{--}$. This is very
different from the asymptotic solution where $T_{2,0}^{(0)}\equiv1$.
These poles propagate also for higher $T_{a,0}$ functions: (for $a\geq2$)
the first cuts occur at $\pm\frac{i(a+1)}{g}$ but there are poles already at
$u_j^{[\pm a]}$.

From
\begin{equation}
Y^{(\alpha)}_{1\vert vw}=\frac{1}{Y_{2,\alpha 1}}=\frac{T_{1,\alpha1}
T_{3,\alpha1}}{T_{2,\alpha2}T_{2,0}}
\label{Y1vwzero}
\end{equation}
we see that $Y^{(\alpha)}_{1\vert vw}$ has zeroes at $u_j^{++}$ and
$u_j^{--}$. This is a new feature of the exact solution since 
$T_{2,0}$ is absent from the denominator of the asymptotic analog of
(\ref{Y1vwzero}).

Using this last piece of information in the Y-system equation (\ref{Y1vw})
we conclude that since the left hand side has double zeroes at $u_j^\pm$,
both $Y_2$ and $Y_+^{(\alpha)}$ must have poles at $u_j^\pm$.

Our final conclusion here is that
\begin{equation}
{\rm e}^{J^{(\alpha)}}=\frac{Y_-^{(\alpha)}}{Y_+^{(\alpha)}}
\end{equation}
has zeroes at $u_j^\pm$, hence $w_j^{p(\alpha)}=u_j^+$ and 
$w_j^{m(\alpha)}=u_j^-$ exactly and we arrive at the final result
\begin{equation}
\frac{Y_-^{(\alpha)}}{Y_+^{(\alpha)}}=
\frac{R_pB_m}{B_pR_m}\exp\left\{-\sum_{Q=1}^\infty L_Q\star K_{Qy}\right\}.
\label{YmperYp}
\end{equation}
Its logarithmic form is
\begin{equation}
J^{(\alpha)}=j+j_{{\rm red}},
\label{Jalpha2}
\end{equation}
where 
\begin{equation}
j=\ln\frac{R_pB_m}{B_pR_m}=\sum_{k=1}^N\left(
\ln\frac{\frac{1}{x}-x_k^+}{x-x_k^+}
-\ln\frac{\frac{1}{x}-x_k^-}{x-x_k^-}\right)
\label{j}
\end{equation}
and
\begin{equation}
j_{{\rm red}}=-\sum_{Q=1}^\infty L_Q\star K_{Qy}.
\label{jred}
\end{equation}

\subsection{Simplifying the equations for $Y_-^{(\alpha)}$ and 
$Y_+^{(\alpha)}$}

We can now substitute the results (\ref{j}) and (\ref{jred}) into the
TBA equation (\ref{TBA-}) and write
\begin{equation}
Y_-^{(\alpha)}=t^{(\alpha)}_-\exp\left\{
\ln\left[\frac{1+Y^{(\alpha)}_{1\vert vw}}{1+Y^{(\alpha)}_{1\vert w}}
\right]\star s
-L_1\star s+(j^\epsilon+j^\epsilon_{{\rm red}})\ \check\star\ s_1
\right\}.
\label{TBAY-1}
\end{equation}

We will simplify the $Y^{(\alpha)}_\pm$ TBA equations with the help of the
following two kernel identities.
\begin{equation}
\begin{split}
-2\int \hspace{-3.9mm} \square \,{\rm d}v\,s_1(v-u)K_{Qy}(w,v)
&+K_{Qy}(w,u)-2\delta_{Q,1}\,s(u-w)=\\
&-K_Q(w-u)+2\int_{-\infty}^\infty {\rm d}v\,s(u-v)K^{Q1}_{xv}(w,v)
\end{split}
\end{equation}
and
\begin{equation}
2\int \hspace{-3.9mm} \square\, {\rm d}v\, s_1(v-u)j^\epsilon(v)=
j(u)-\int_{-\infty}^\infty{\rm d}v\,s(u-v)\left(\ln\left[
\left(\frac{R_p^+R_p^-}{R_m^+R_m^-}\right)^2
\frac{Q^{--}}{Q^{++}}\right]\right)(v).
\end{equation}

Using the above identities and combining (\ref{YmperYp}) with
(\ref{TBAY-1}) we get
\begin{equation}
\begin{split}
Y^{(\alpha)}_+Y^{(\alpha)}_-=\left(t_-^{(\alpha)}\right)^2
&\exp\Bigg\{
2\ln\left[\frac{1+Y^{(\alpha)}_{1\vert vw}}{1+Y^{(\alpha)}_{1\vert w}}
\right]\star s+\sum_{Q=1}^\infty L_Q\star\left[
-K_Q+2K^{Q1}_{xv}\star s\right]\\
&-\ln\left[
\left(\frac{R_p^+R_p^-}{R_m^+R_m^-}\right)^2
\frac{Q^{--}}{Q^{++}}\right]\star s\Bigg\}.
\end{split}
\end{equation}

\section{Dispersion relation for $\Delta$}
In this section we determine the $\Delta$ discontinuity using the
relations (\ref{Deltacut}) and the corresponding dispersion relation.
At the same time we complete the construction of the T-system elements
we started in section 4. Our aim is to construct T-system elements that
are smooth deformations (similarly to the Y-system elements) of the 
corresponding asymptotic variables. As we have seen in section 4, this is
possible only partially. If, for example, $T_{a,1}$ is close to the
corresponding asymptotic $T_{a,1}^{(0)}$ solution for the right hand side of 
the diagram, then, in general, even for the left-right symmetric 
${\alg{sl}(2)}$ cases, $T_{a,-1}$ can be very different from the corresponding
asymptotic T-system elements on the left hand side of the diagram. We will 
work in this gauge which we call the R-gauge and when we want to emphasize 
this asymmetry we also use the notation $T^R_{a,s}$ for $T_{a,s}$. Of course, 
there also exists an analogous L-gauge, with corresponding T-system elements 
$T^L_{a,s}$, which are close to the asymptotic solution on the left hand side.

We start the calculation by writing
\begin{equation}
Y_1=Y_{1,0}=\frac{T_{1,1}T_{1,-1}}{T_{2,0}}=
\frac{T^R_{1,1}\,T^R_{1,-1}}{T_{2,0}}=Y_d\,T^R_{1,1}\,T^L_{1,-1}.
\label{dRL}
\end{equation}
The new object here is
\begin{equation}
Y_d=\frac{1}{T_{2,0}}\,\frac{T^R_{1,-1}}{T^L_{1,-1}}=\frac{t_{1,-1}}
{T_{2,0}},
\end{equation}
where $t_{a,s}=\frac{T^R_{a,s}}{T^L_{a,s}}$ is the gauge transformation
connecting the two gauges defined above. Corresponding to the three factors
in the last expression in (\ref{dRL}) we write
\begin{equation}
\Delta=\ln\frac{Y_1^+}{(Y^+_1)_*}=\Delta_R+\Delta_L+\Delta_3,
\end{equation}
where
\begin{equation}
\Delta_R=\ln\frac{(T^R_{1,1})^+}{((T^R_{1,1})^+)_*},\qquad
\Delta_L=\ln\frac{(T^L_{1,-1})^+}{((T^L_{1,-1})^+)_*},\qquad
\Delta_3=\ln\frac{Y_d^+}{(Y_d^+)_*}=\ln\frac{t_{1,-1}^+}{(t^+_{1,-1})_*}.
\end{equation}
We now rewrite the relation (\ref{Deltacut}), using also the results
of the previous section in the form
\begin{equation}
[\Delta]_{\pm 2N}=D^{(\pm2N)}_{12R}+D^{(\pm2N)}_{12L}\pm 
2(j^{[\epsilon]}+j^{[\epsilon]}_{{\rm red}}),
\label{Deltacut12RL}
\end{equation}
where
\begin{equation}
D^{(\pm2N)}_{12R}=\pm\left[\Lambda^{(+)}_\mp\right]_{\pm2N}
\pm\sum_{m=1}^N\left[\Lambda^{(+)}_m\right]_{\pm(2N-m)},\quad
D^{(\pm2N)}_{12L}=\pm\left[\Lambda^{(-)}_\mp\right]_{\pm2N}
\pm\sum_{m=1}^N\left[\Lambda^{(-)}_m\right]_{\pm(2N-m)}
\label{D12RL}
\end{equation}
and
\begin{equation}
\Lambda_\pm^{(\alpha)}=\ln\left(1-\frac{1}{Y^{(\alpha)}_\pm}\right),\qquad
\Lambda_m^{(\alpha)}=\ln\left(1+\frac{1}{Y^{(\alpha)}_{m\vert vw}}\right).
\end{equation}

So far (except the construction of the T-system elements in 
section~4) our considerations are valid for any state of the model.
From now on, for simplicity, we restrict our attention to states in the 
${\alg{sl}(2)}$ (sub-)sector defined in appendix C. We think that our methods
can straightforwardly be generalized to generic states, but some of the
subsequent formulae are considerably more complicated in the general case. 
Since the two sides are identical  in this special case, in the rest of
the paper we can omit the upper index $^{(\alpha)}$ of the Y-functions
$Y_\pm^{(\alpha)}$, $Y_{m\vert vw}^{(\alpha)}$ or $Y_{m\vert w}^{(\alpha)}$.

In the ${\alg{sl}(2)}$ (sub-)sector in the asymptotic limit we have
\begin{equation}
Y_d^{(0)}=y_d\,\qquad \Delta_3^{(0)}=\frac{y_d^+}{(y_d^+)_*}=\Delta_d
\end{equation}
and
\begin{equation}
[\Delta_3^{(0)}]_{\pm2N}=[\Delta_d]_{\pm2N}=\pm2j^{[\epsilon]},
\label{delt3}
\end{equation}
as discussed in appendix B.

The main part of the calculation is to determine (omitting the upper 
index $R$ of the T-functions)
\begin{equation} 
\Delta_R=\ln\frac{T^+_{1,1}}{{\cal F}^+},
\end{equation}
where ${\cal F}^+=(T_{1,1}^+)_*$ and similarly we define
${\cal G}^-=(T_{1,1}^-)_*$.
Note that $\Delta_R(u)=p(u)$, the discontinuity which was left undetermined 
in section 4. 

Considerable simplification occurs if we express $D^{(\pm2N)}_{12R}$
in terms of the T-system elements. For the upper sign we find
\begin{equation} \label{6.11}
D^{(2N)}_{12R}=[\ln T^+_{1,1}]_{2N}+[\ln T_{N+1,1}]_{N-1}
-[\ln T_{N+2,1}]_N.
\end{equation}
In the gauge we are using $T_{a,1}$ are meromorphic functions in the
strip $(-a,a)$, which means that this simplifies to
\begin{equation}
D^{(2N)}_{12R}=[\ln T^+_{1,1}]_{2N}.
\end{equation}
Next we further specify our gauge: we require that ${\cal F}^+$ has no
discontinuities in the upper half plane. This is satisfied by the
asymptotic function ${\cal F}^{(0)+}$. We then have
\begin{equation}
D^{(2N)}_{12R}=\left[\ln \frac{T^+_{1,1}}{{\cal F}^+}\right]_{2N}
=[\Delta_R]_{2N}.
\label{disc12R}
\end{equation}
This relation above is in the sense of a discontinuity relation. In this sense
one has for any function $f$ and $q$
\begin{equation}
\left[\ln\frac{f}{q}\right]_Z=[\ln f]_Z,
\end{equation}
provided $q$ is meromorphic around the cut $Z$. However, as explained in
section 2, in the sense of dispersion relation we have for example
\begin{equation}
\int_2^\infty{\rm d}v[\ln^\prime f]_Z(v)=
\int_2^\infty{\rm d}v[\ln^\prime f(v+\frac{iZ}{g}+i\epsilon)
-\ln^\prime f(v+\frac{iZ}{g}-i\epsilon)]
\end{equation}
and the result will be different if we use $f/q$ instead of $f$ when
the meromorphic function $q$ has zeroes/poles on the cut. The correct
interpretation of the original relation (\ref{Deltacut}) can be found
out from the asymptotic limit of the relation and we find that 
this subtlety is relevant for the $N=1$ case only and in that relation
we have to make the substitution $\Lambda_1\Rightarrow\frac{\Lambda_1}
{p_2^-}$, where $p_2$ is a polynomial whose roots are the real zeroes
of $T_{2,1}$. Actually, an equivalent (and simpler) way of taking into 
account this effect is to simply omit the contribution of these poles 
from the dispersion relation.

Repeating the same calculation for the lower sign we find
\begin{equation}
D^{(-2N)}_{12R}=-[\ln {\cal G}^-+\ln{\cal F}^+-\ln T^+_{1,1}]_{-2N}
-[\ln T_{N+1,1}]_{-(N-1)}
+[\ln T_{N+2,1}]_{-N}.
\end{equation}
Note that the first term is obtained from
\begin{equation}
1-\frac{1}{Y_+}=1-\frac{1}{(Y_-)_*}=
1+(Y_{1,1})_*=\frac{(T_{1,1}^+)_*\,(T^-_{1,1})_*}{(T_{2,1})_*}=
\frac{{\cal F}^+\,{\cal G}^-}{T_{2,1}}.
\end{equation}
Adding to the list of gauge fixing conditions the requirement that
${\cal G}^-$ has no discontinuities in the lower half plane (satisfied
by the asymptotic ${\cal G}^{(0)-}$) and using the properties of
the $T_{a,1}$ functions we have for both signs
\begin{equation}
D^{(\pm2N)}_{12R}=\left[\ln \frac{T^+_{1,1}}{{\cal F}^+}\right]_{\pm2N}
=[\Delta_R]_{\pm2N}.
\label{disc12Rpm}
\end{equation}
(With the understanding that the correct way of transforming the 
discontinuity relations into dispersion relations means that we have to 
omit the contributions of real zeroes in the $N=1$ case.)

At this point we see that the discontinuity relations (\ref{Deltacut})
(rewritten in the form (\ref{Deltacut12RL})) are satisfied by the asymptotic 
solution. Indeed, all the properties of the T-functions that we used in the 
derivation of (\ref{disc12Rpm}) are equally valid for the asymptotic 
T-functions and (\ref{disc12Rpm}), combined with (\ref{delt3}) and taking
into account the asymptotic relations (\ref{Ym/Yp(0)}) and the definition of
$j$ in (\ref{j}), prove the statement.

Since in the ${\alg{sl}(2)}$ (sub-)sector the two sides of the problem are identical
(in the language of Y-functions) we have
\begin{equation}
[\Delta]_{\pm2N}=2D^{(\pm2N)}_{12R}\pm2(j^{[\epsilon]}
+j_{{\rm red}}^{[\epsilon]})
\end{equation}
and using (\ref{disc12Rpm}) this simplifies to
\begin{equation}
[\Delta_3]_{\pm2N}=\pm2(j^{[\epsilon]}+j_{{\rm red}}^{[\epsilon]}).
\label{delt3red}
\end{equation}
We now define $Z_d$ and find the relations
\begin{equation}
Y_d=y_d\,Z_d,\qquad
\Delta_3=\Delta_d+\Delta_{{\rm red}},\qquad
\Delta_{{\rm red}}=\ln\frac{Z_d^+}{(Z_d^+)_*}.
\label{Zd}
\end{equation}

Since asymptotically $Z_d^{(0)}=1$, $\Delta_{{\rm red}}^{(0)}=0$ we conclude 
that in the exact solution ${\rm e}^{\Delta_{{\rm red}}}$ has no zeroes/poles
(and has constant asymptotics for large $\vert u\vert$). This means that if
we define (analogously to $\Gamma^{(\alpha)}(u)$ of section~5)
\begin{equation}
\Gamma_{{\rm red}}(u)=\oint_{\Gamma_0}{\rm d}v\,\Delta^\prime_{{\rm red}}(v)
K(v,u)
\label{Gammared}
\end{equation}
and write the corresponding dispersion relation we find
\begin{equation}
{\cal D}\Delta_{{\rm red}}=2\sum_{N=1}^\infty\,j_{{\rm red}}^{[\epsilon]\prime}
\ \check\star\ (K^{[2N]}-K^{[-2N]}).
\end{equation}
Here the left hand side is the contribution of the narrow strip only 
(the pole terms vanish) and the right hand side is simply the sum of 
contributions of the integrals along the even cuts.  Using the method
explained in section~5 we can \lq\lq integrate" this relation and get
\begin{equation}
\Delta_{{\rm red}}=2\sum_{N=1}^\infty\,j_{{\rm red}}^{[\epsilon]}
\ \check\star\ (K^{[2N]}-K^{[-2N]}).
\label{jredsol}
\end{equation}

\subsection{Determination of $\Delta_R$}

The dispersion relation for $\Delta_R$ can be written down using 
(\ref{D12RL}) and (\ref{disc12Rpm}). Proceeding as before we have
\begin{equation}
\begin{split}
\Gamma_R(u)=\oint_{\Gamma_0}{\rm d}v\,\Delta^\prime_R(v)&K(v,u)=
({\cal D}\Delta_R)(u)
+\oint_{[\gamma,2,\infty]}{\rm d}v\,
[\ln^\prime T_{1,1}^+(v)-\ln^\prime {\cal F}^+(v)]K(v,u)\\
&+\oint_{[-\infty,-2,-\gamma]}{\rm d}v\,
[\ln^\prime T_{1,1}^+(v)-\ln^\prime {\cal F}^+(v)]K(v,u).
\label{GammaR}
\end{split}
\end{equation}
The alternative calculation uses the explicit form of $D_{12R}^{(\pm2N)}$
and its direct integration along the cuts. The contribution of the upper
cuts is
\begin{equation}
\begin{split}
\Gamma^{{\rm upper}}_R(u)&=\sum_{N=1}^\infty\,\int \hspace{-3.9mm} \square
\,{\rm d}v\,K^{[2N]}(v,u)D^{(2N)\prime}_{12R}(v)\\
=&-\int_{-\infty}^\infty{\rm d}v\,K^{[\gamma]}(v,u)
\Lambda_-^{[\gamma]\prime}(v)
+\oint_{[\gamma,2,\infty]}{\rm d}v\,K(v,u)\Lambda_-^\prime(v)\\
&-\sum_{m=1}^\infty\int_{-\infty}^\infty{\rm d}v\,K^{[m+\gamma]}(v,u)
\Lambda_m^{[\gamma]\prime}(v)
+\sum_{m=1}^\infty
\oint_{[\gamma,m,\infty]}{\rm d}v\,K^{[m]}(v,u)\Lambda_m^\prime(v).
\end{split}
\end{equation}
Let us consider now the $\Lambda_m$ pole terms ($\oint\Lambda^\prime_m$ terms)
separately:
\begin{equation}
\begin{split}
&\sum_{m=1}^\infty\oint_{[\gamma+1,m+1,\infty]}{\rm d}v\,
K^{[m-1]}(v,u)\ln^\prime T_{m+1,1}(v)\\
+&\sum_{m=1}^\infty\oint_{[\gamma-1,m-1,\infty]}{\rm d}v\,
K^{[m+1]}(v,u)\ln^\prime T_{m+1,1}(v)\\
-&\sum_{m=0}^\infty\oint_{[\gamma,m+1,\infty]}{\rm d}v\,
K^{[m+1]}(v,u)\ln^\prime T_{m+1,1}(v)\\
-&\sum_{m=2}^\infty\oint_{[\gamma,m-1,\infty]}{\rm d}v\,
K^{[m-1]}(v,u)\ln^\prime T_{m+1,1}(v)
\end{split}
\end{equation}
We first note that in the integration contours $m-1$ can be changed
to $m+1$. This is possible because $T_{m,1}$ has no 
zeroes/poles/discontinuities on the $m-2$ cut for $m=3,4,\dots$ and
$T_{2,1}$ has no poles/discontinuities on the real cut while the
contribution of its real zeroes, as we have seen, has to be omitted.
We can also change all integrals to have contours of the type
$[\gamma,m+1,\infty]$ but here the change of $\gamma\pm1$ to $\gamma$
requires to add to this sum the corrections
\begin{equation}
-\sum_{m=1}^\infty\oint_{[\gamma,\gamma+1]}{\rm d}v\,
K^{[m-1]}(v,u)\ln^\prime T_{m+1,1}(v)\\
+\sum_{m=1}^\infty\oint_{[\gamma-1,\gamma]}{\rm d}v\,
K^{[m+1]}(v,u)\ln^\prime T_{m+1,1}(v).
\end{equation}
The rest of the sum (after taking into account that most terms cancel)
simplifies to:
\begin{equation}
\oint_{[\gamma,2,\infty]}{\rm d}v\,K(v,u)\ln^\prime T_{2,1}(v)
-\oint_{[\gamma,2,\infty]}{\rm d}v\,K(v,u)\ln^\prime T_{1,1}^-(v)
+\oint_{[\gamma-1,\gamma]}{\rm d}v\,K^+(v,u)\ln^\prime T_{1,1}(v).
\end{equation}
Adding also the pole terms corresponding to $\Lambda_-$
\begin{equation}
\oint_{[\gamma,2,\infty]}{\rm d}v\,K(v,u)\left[\ln^\prime T^+_{1,1}(v)
+\ln^\prime T^-_{1,1}(v)-\ln^\prime T_{2,1}(v)\right]
\end{equation}
the sum of all pole terms becomes:
\begin{equation}
\begin{split}
&-\sum_{m=1}^\infty\oint_{[\gamma,\gamma+1]}{\rm d}v\,
K^{[m-1]}(v,u)\ln^\prime T_{m+1,1}(v)\\
&+\sum_{m=0}^\infty\oint_{[\gamma-1,\gamma]}{\rm d}v\,
K^{[m+1]}(v,u)\ln^\prime T_{m+1,1}(v)
+\oint_{[\gamma,2,\infty]}{\rm d}v\,
K(v,u)\ln^\prime T^+_{1,1}(v).
\end{split}
\end{equation}
This can be simplified further using the relation $T_{m,1}=(-1)^{m-1}
\tau_m\sigma_m$ and the facts that $\sigma_m$ has no zeroes/poles near the
physical strip and $\tau_1^+$ has no zeroes/poles on the even cuts:
\begin{equation}
\begin{split}
-&\sum_{m=2}^\infty\oint_{[\gamma,\gamma+1]}{\rm d}v\,K^{[m-2]}(v,u)
\ln^\prime\tau_m(v)
+\sum_{m=1}^\infty\oint_{[\gamma-1,\gamma]}{\rm d}v\,K^{[m]}(v,u)
\ln^\prime\tau_m(v)\\
+&\int_{-\infty}^\infty{\rm d}v\,K^{[\gamma]}(v,u)
\ln^\prime\tau_1^{[1+\gamma]}(v)
+\oint_{[\gamma,2,\infty]}{\rm d}v\,
K(v,u)\ln^\prime \sigma^+_1(v).
\end{split}
\end{equation}
It remains to calculate the $\int_{-\infty}^\infty$ parts. We write these
contributions using the combinations (already used in section~4 for the
construction of the T-system elements)
\begin{equation}
{\mathscr L}_\pm=\ln\left[-\tau_2\left(1-\frac{1}{Y_\pm}\right)\right],\qquad
{\mathscr L}_m=\ln\left[\tau_m\tau_{m+2}
\left(1+\frac{1}{Y_{m\vert vw}}\right)\right]
\end{equation}
and find
\begin{equation}
\begin{split}
-&\int_{-\infty}^\infty{\rm d}v\,
K^{[\gamma]}(v,u)\Lambda_-^{[\gamma]\prime}(v)
-\sum_{m=1}^\infty\int_{-\infty}^\infty{\rm d}v\,
K^{[m+\gamma]}(v,u)\Lambda_m^{[\gamma]\prime}(v)\\
=-&\int_{-\infty}^\infty{\rm d}v\,
K^{[\gamma]}(v,u){\mathscr L}_-^{[\gamma]\prime}(v)
-\sum_{m=1}^\infty\int_{-\infty}^\infty{\rm d}v\,
K^{[m+\gamma]}(v,u){\mathscr L}_m^{[\gamma]\prime}(v)\\
+&\int_{-\infty}^\infty{\rm d}v\,
K^{[\gamma]}(v,u)\ln^\prime\tau_2^{[\gamma]}(v)
+\sum_{m=1}^\infty\int_{-\infty}^\infty{\rm d}v\,
K^{[m+\gamma]}(v,u)\left[\ln^\prime\tau_m^{[\gamma]}(v)
+\ln^\prime\tau_{m+2}^{[\gamma]}(v)\right].
\end{split}
\end{equation}
Adding everything together and using the identity $\tau_m^+\tau_m^-\equiv1$
in the form 
$\ln^\prime\tau_m^{[\gamma-1]}+\ln^\prime\tau_m^{[\gamma+1]}\equiv0$
we finally have
\begin{equation}
\begin{split}
\Gamma^{{\rm upper}}_R(u)&=\oint_{[\gamma,2,\infty]}{\rm d}v\,
K(v,u)\ln^\prime\sigma_1^+(v)\\
&-\int_{-\infty}^\infty{\rm d}v\,K^{[\epsilon]}(v,u)
{\mathscr L}_-^{[\epsilon]\prime}(v)
-\sum_{m=1}^\infty\int_{-\infty}^\infty{\rm d}v\,K^{[m]}(v,u)
{\mathscr L}_m^\prime(v).
\label{GammaRup}
\end{split}
\end{equation}
Similarly for the contribution of the lower half plane cuts we have
\begin{equation}
\begin{split}
\Gamma^{{\rm lower}}_R(u)&=\oint_{[-\infty,-2,-\gamma]}{\rm d}v\,
K(v,u)\left[\ln^\prime\sigma_1^+(v)
-\ln^\prime{\cal F}^+(v)-\ln^\prime{\cal G}^-(v)\right]\\
&-\int_{-\infty}^\infty{\rm d}v\,K^{[-\epsilon]}(v,u)
{\mathscr L}_+^{[-\epsilon]\prime}(v)
-\sum_{m=1}^\infty\int_{-\infty}^\infty{\rm d}v\,K^{[-m]}(v,u)
{\mathscr L}_m^\prime(v).
\label{GammaRlow}
\end{split}
\end{equation}

\subsection{The XXX gauge}

To calculate $\Delta_R$ we need to fix the gauge completely.
We will call this complete gauge fixing the XXX gauge and it will not be
our final gauge choice yet. Later we will have to perform an additional gauge 
transformation to arrive at the gauge in which all the T-functions are smooth 
deformation of the asymptotic solution. This \lq\lq intermediate" gauge,
which will be used in this subsection,
we call the XXX gauge because in the $g\rightarrow0$ limit the system
of functions $Y_{m\vert vw}$, $T_{m,1}$ decouples from the rest and they form
the Y-system and T-system of the XXX spin model (the latter in a gauge most 
natural in that model, namely in which the $T_{m,1}$ functions are polynomials
in the spectral parameter $u$). 

The XXX gauge is fixed by adding the requirements (in addition to those
already imposed in section~4 and this section):
\begin{itemize} 

\item ${\cal F}_{{\rm XXX}}^+$: no zeroes/poles in the upper 
half plane (no zeroes/poles on the real line) except poles at positions $u_j^+$
($j=1,2,\dots,N$).

\item ${\cal G}_{{\rm XXX}}^-$: no zeroes/poles in the lower 
half plane.

\end{itemize} 

Recalling that $T_{1,1}^{{\rm XXX}}=\tau_1\sigma_1$ we can express 
${\cal D}\Delta_R$ comparing (\ref{GammaR}) to the sum of
(\ref{GammaRup}) and (\ref{GammaRlow}). We find 
\begin{equation}
\begin{split}
{\cal D}&\Delta_R(u)=-\int_{-\infty}^\infty{\rm d}v\,
{\mathscr L}_-^{[\gamma]\prime}(v)K^{[\gamma]}(v,u)
-\int_{-\infty}^\infty{\rm d}v\,
{\mathscr L}_+^{[-\epsilon]\prime}(v)K^{[-\epsilon]}(v,u)\\
&-\sum_{m=1}^\infty\int_{-\infty}^\infty{\rm d}v\,
{\mathscr L}^\prime_m(v)k_m(v,u)+\oint_{[\gamma,2,\infty]}{\rm d}v\,\left[
\ln^\prime{\cal F}^+(v)-\ln^\prime\tau_1^+(v)\right]K(v,u)\\
&\qquad\qquad +\oint_{[-\infty,-2,-\gamma]}{\rm d}v\,
K(v,u)[-\ln^\prime\tau_1^+(v)].
\label{GammaRres}
\end{split}
\end{equation}
We can calculate the pole terms ($\oint$ terms) of this expression using
the residue theorem:
\begin{equation}
-2\pi i\sum_{j=1}^NK(u_j^+,u)+2\pi i\sum_{j=1}^N\sum_{\nu=-\infty}^\infty
\left\{K(u_j^++\frac{4\nu i}{g},u)-K(u_j^-+\frac{4\nu i}{g},u)\right\}.
\end{equation}
Their contribution, after \lq\lq integration", as explained in section~5,
becomes
\begin{equation}
{\cal W}(u)=\sum_{j=1}^N\sigma(u,u_j^+)
+\sum_{j=1}^N\sum_{\nu=-\infty}^\infty\left\{
\sigma(u,u_j^-+\frac{4\nu i}{g})-\sigma(u,,u_j^++\frac{4\nu i}{g})\right\}.
\end{equation}
Note that using the definition
\begin{equation}
\vartheta_1^+(u)=\prod_{j=1}^N\prod_{\nu=-\infty}^\infty\frac
{S(u,u_j^-+\frac{4\nu i}{g})}{S(u,u_j^++\frac{4\nu i}{g})},
\end{equation}
where the notation indicates that this function has exactly the same
zeroes/poles as $\tau_1^+$, we have
\begin{equation}
{\rm e}^{{\cal W}(u)}=\vartheta_1^+(u)\prod_{j=1}^NS(u,u_j^+).
\end{equation}
We can now write the \lq\lq integrated" version of (\ref{GammaRres}):
\begin{equation}
\begin{split}
\Delta_R(u)=-\int_{-\infty}^\infty{\rm d}v\,&{\mathscr L}_-^{[\gamma]}(v)
K^{[\gamma]}(v,u)
-\int_{-\infty}^\infty{\rm d}v\,{\mathscr L}_+^{[-\epsilon]}(v)
K^{[-\epsilon]}(v,u)\\
-&\sum_{m=1}^\infty\int_{-\infty}^\infty{\rm d}v\,
{\mathscr L}_m(v)k_m(v,u)+{\cal W}(u)-\ln x^2(u).
\label{DeltaRres0}
\end{split}
\end{equation}
Note that this formula is valid just above the real line in the strip 
$0<{\rm Im}u<\frac{\gamma}{g}$.

The last term in (\ref{DeltaRres0}) is necessary to balance the large
$\vert u\vert$ asymptotics of the equation. We know that in this limit
(just above the real line)
\begin{equation}
{\mathscr L}_+(u),\,{\mathscr L}_-(u)\approx-\ln\vert u\vert,\qquad
\Delta_R(u)\approx\ln\vert u\vert
\end{equation}
and we see that (\ref{DeltaRres0}) is asymptotically correct because
if a function $f$ behaves asymptotically as
$f(u)\approx\ln\vert u\vert$ then convolution with the kernel $K^{[\alpha]}$
gives
\begin{equation}
f\star K^{[\alpha]}\approx-\frac{1}{2}\ln\vert u\vert.
\end{equation}

Finally by manipulating the integrals containing 
${\mathscr L}_+,\,{\mathscr L}_-$ we write our final result for the 
discontinuity $\Delta_R$:
\begin{equation}
\begin{split}
\Delta_R(u)={\mathscr L}_-(u)-&\int_{-2}^2{\rm d}v\,
[{\mathscr L}_-(v)+{\mathscr L}_+(v)]K(v,u)\\
-&\sum_{m=1}^\infty\int_{-\infty}^\infty{\rm d}v\,
{\mathscr L}_m(v)k_m(v,u)+{\cal W}(u)-\ln x^2(u).
\label{DeltaRres}
\end{split}
\end{equation}
Note that the last term is chosen here so that $\Delta_R(\pm2)=0$ is
satisfied. The price we have to pay is that this term creates a cut
along the imaginary axis (but neither $\Delta^\prime_R(u)$ nor
${\rm e}^{\Delta_R(u)}$ have discontinuities there).

Our final result for the discontinuity $\Delta(u)$ in the ${\alg{sl}(2)}$
(sub-)sector is
\begin{equation}
\begin{split}
&\qquad\qquad\Delta=2\Delta_R+\Delta_d+\Delta_{{\rm red}}\\
&=2{\mathscr L}_--2({\mathscr L}_-+{\mathscr L}_+)\ \hat\star\ K-
2\sum_{m=1}^\infty{\mathscr L}_m\star k_m+
2{\cal W}-L\ln x^2+\ln\frac{{\cal D}_1^+}{({\cal D}_1^+)_*}+\Delta_{{\rm red}}.
\end{split}
\label{Deltares}
\end{equation}
Here we have used (\ref{yd}) to write
\begin{equation}
\Delta_d=-J\ln x^2+\ln\frac{{\cal D}_1^+}{({\cal D}_1^+)_*}.
\label{Deld}
\end{equation}
We note that the renormalization $J\rightarrow L=J+2$ is due to the addition
of the zero mode contribution in (\ref{DeltaRres0}) and it is universal
(we see this here for the case of states in the ${\alg{sl}(2)}$ (sub-)sector).
Indeed, this universality was shown for generic states in 
\cite{Arutyunov:2011uz}. The physical meaning of $L$ is that it is the
maximal value of the $J$-charge within the $psu(2,2\vert4)$ supermultiplet
to which the given state belongs. 

The result (\ref{Deltares}), together with $J^{(\alpha)}$ calculated in 
section~5, completes the set of TBA integral equations of section~3 for the
excited states in this sector of the model. This system of integral equations
is now closed. It still has to be supplemented by the quantization
conditions for the discrete parameters appearing in the source terms. This
will be discussed in section~9.

At this point we would like to emphasize that the Y-system and 
discontinuity relations (with some additional qualitative information 
on local singularities) determine the set of TBA equations completely.
We can summarize the logic of calculating the full $\Delta$ (including the
dressing phase part $\Delta_d$) as follows. First we
fix a gauge such that the T-factors in (\ref{dRL}) satisfy (\ref{disc12Rpm})
and thus the discontinuity relations simplify to (\ref{delt3red}). Next
we take (\ref{Zd}) as an Ansatz and use the (nontrivial) computations
presented in appendix B to show that (\ref{delt3red}) is now further
reduced to 
\begin{equation}
[\Delta_{red}]_{\pm 2N}= \pm 2 j_{red}^{[\epsilon]}, 
\end{equation}
which can be solved easily using (\ref{Gammared}-{\ref{jredsol}).
Of course, by writing (\ref{Zd}) we actually use the known formulae for the
dressing phase part and verify it satisfies the relations 
(\ref{Deltadu}-\ref{b24}). This simplifies our job here. However, even if 
we had not known the solution for $\Delta_d$ given by (\ref{Deld})
we could have calculated it from the discontinuity relation (\ref{Deltadu}) 
by transforming it into a dispersion relation as we did for the other building 
blocks.

In this section we obtained the result (\ref{DeltaRres}) by a long direct
calculation. An alternative logic could have been to simply postulate
the result (\ref{DeltaRres}) for $p(u)=\Delta_R(u)$ and show that the 
requirements on the upper/lower half plane behavior of the 
${\cal F}^+$/${\cal G}^-$ functions, which we used in the calculation, 
are indeed satisfied.

To show this, we start from
\begin{equation}
T^{{\rm XXX}}_{1,1}=\tau_1\sigma_1,\qquad\quad \sigma_1=\frac{q^+}{q^-}
\sigma_1^{(0)},
\end{equation}
where
\begin{equation}
\sigma_1^{(0)}=\exp\left\{{\mathscr L}_-^{[\epsilon]}\star(\kappa^--\kappa^+)
+\sum_{m=1}^\infty{\mathscr L}_m\star(\kappa^{[-m-1]}-\kappa^{[m+1]})
\right\}
\end{equation}
and $q$ is defined in (\ref{qdef}). Here we introduced the notation
\begin{equation}
\kappa(v,u)=\frac{1}{2\pi i}\,\frac{1}{v-u},\qquad\qquad
K_a=\kappa^{[-a]}-\kappa^{[a]}.
\end{equation}
Note that $K(v,u)-\kappa(v,u)$ is regular at $v=u$.

The discontinuity relation is
\begin{equation}
\frac{q(u+i\epsilon)}{q(u-i\epsilon)}={\rm e}^{-p(u+i\epsilon)},\qquad
\vert u\vert\geq2
\end{equation}
and just above the real line, in the strip $0<{\rm Im}u<\frac{\gamma}{g}$
we have
\begin{equation}
p={\mathscr L}_--({\mathscr L}_-+{\mathscr L}_+)\ \hat\star\ K-
\sum_{m=1}^\infty{\mathscr L}_m\star k_m+{\cal W}-\ln x^2
\end{equation}
and
\begin{equation}
\sigma_1^{(0)+}=\exp\left\{{\mathscr L}_-+{\mathscr L}_-^{[\epsilon]}\star
(\kappa^{--}-\kappa)+\sum_{m=1}^\infty{\mathscr L}_m\star(\kappa^{[-m-2]}
-\kappa^{[m]})\right\}
\end{equation}
and finally
\begin{equation}
\begin{split}
{\cal F}^+_{{\rm XXX}}&=\left(T_{1,1}^{{\rm XXX}}\right)^+{\rm e}^{-p}=
x^2\frac{q^{++}}{q}\,\frac{\tau_1^+}{{\rm e}^{\cal W}}\,\exp\Big\{
({\mathscr L}_-+{\mathscr L}_+)\ \hat\star\ K\\
+&{\mathscr L}_-^{[\epsilon]}\star(\kappa^{--}-\kappa)
+\sum_{m=1}^\infty{\mathscr L}_m\star(K^{[m]}-\kappa^{[m]}+
K^{[-m]}+\kappa^{[-m-2]})\Big\}.
\end{split}
\end{equation}
From this result we see that there are indeed no zeroes/poles/discontinuities
in the upper half plane, except for the poles at the positions $u_j^+$.

Similarly, just below the real line, in the strip 
$-\frac{\gamma}{g}<{\rm Im}u<0$ we have
\begin{equation}
\sigma_1^{(0)-}=\exp\left\{{\mathscr L}_-+{\mathscr L}_-^{[\epsilon]}\star
(\kappa-\kappa^{++})+\sum_{m=1}^\infty{\mathscr L}_m\star(\kappa^{[-m]}
-\kappa^{[m+2]})\right\},
\end{equation}
\begin{equation}
p=-{\mathscr L}_+-({\mathscr L}_-+{\mathscr L}_+)\ \hat\star\ K-
\sum_{m=1}^\infty{\mathscr L}_m\star k_m+{\cal W}-\ln x^2,
\end{equation}
and noting that from (\ref{Yvw1cut}) we have $I={\mathscr L}_--
{\mathscr L}_+$
\begin{equation}
m=I-p={\mathscr L}_-+({\mathscr L}_-+{\mathscr L}_+)\ \hat\star\ K+
\sum_{m=1}^\infty{\mathscr L}_m\star k_m-{\cal W}+\ln x^2,
\end{equation}
and finally
\begin{equation}
\begin{split}
{\cal G}^-_{{\rm XXX}}&=\left(T_{1,1}^{{\rm XXX}}\right)^-
{\rm e}^{-m}=\frac{1}{x^2}\,
\frac{q}{q^{--}}\,\frac{{\rm e}^{\cal W}}{\tau_1^+}\,\exp\Big\{
-({\mathscr L}_-+{\mathscr L}_+)\ \hat\star\ K\\
+&{\mathscr L}_-^{[\epsilon]}\star(\kappa-\kappa^{++})
+\sum_{m=1}^\infty{\mathscr L}_m\star(\kappa^{[-m]}-K^{[-m]}-
K^{[m]}-\kappa^{[m+2]})\Big\}.
\end{split}
\end{equation}
We see that there are indeed no zeroes/poles/discontinuities
in this expression in the lower half plane.

\subsection{Complete gauge fixing}

It is possible to show that the XXX gauge we have been using in this
section is a complete gauge fixing. This means that given an exact solution
of the Y-system equations, which also satisfies the reality conditions and
the relation between $Y_-$ and $Y_+$ and has the cut structure described
in section~2, we can always construct the corresponding XXX gauge T-system
elements $T^{{\rm XXX}}_{a,s}$, which satisfy all the requirements
listed below and this construction is unique.

The requirements are
\begin{itemize} 

\item $T^{{\rm XXX}}_{0,s}\equiv1$.

\item $T^{{\rm XXX}}_{a,0}$ are given as constructed explicitly in section~4
using the chain lemma.

\item $T^{{\rm XXX}}_{a,1}$ have discontinuities along the cuts 
$\pm(a+2m)$, $m=0,1,\dots$

\item The roots of $T^{{\rm XXX}}_{a,1}$ in the physical strip are the same 
as those of $\tau_a$ and there are no poles in this strip.

\item ${\cal F}_{{\rm XXX}}^+$: no zeroes/poles/discontinuities in the upper 
half plane (no zeroes/poles on the real line) except poles at positions $u_j^+$
($j=1,2,\dots,N$).

\item ${\cal G}_{{\rm XXX}}^-$: no zeroes/poles/discontinuities in the lower 
half plane.

\item The large $\vert u\vert$ behavior is the same as for the asymptotic
solution.

\end{itemize} 

We now perform the final gauge transformation which brings the T-system
solution in a gauge (we will call it the BA gauge) where the T-system
functions $T_{a,s}$ for $s\geq0$ are close to the asymptotic (Bethe Ansatz)
solution $T_{a,s}^{(0)}$. This is achieved by defining
\begin{equation}
T_{1,1}=\beta\,T^{{\rm XXX}}_{1,1},\qquad\quad
\beta=\frac{Q^{--}}{Q}.
\label{BAbeta}
\end{equation}
We see that $T_{1,1}$ has no zeroes/poles in the physical strip.
From
\begin{equation}
{\cal F}^+=\frac{Q^-}{Q^+}\,{\cal F}^+_{{\rm XXX}},\qquad\quad
{\cal G}^-=\frac{Q^{[-3]}}{Q^-}\,{\cal G}^-_{{\rm XXX}}
\end{equation}
we see that ${\cal F}^+$/${\cal G}^-$ has no zeroes/poles/discontinuities
in the upper/lower half plane. 

From the list of requirements above and the modification induced by
(\ref{BAbeta}) we see that the BA gauge T-functions are indeed
smooth deformations of the asymptotic solution, satisfying the same
requirements. Note that the final results for $p=\Delta_R$ and the full 
$\Delta$ are unchanged since the transformation (\ref{BAbeta}) is meromorphic.

\section{Simplified $Y_1$ equation}
In this section we want to simplify the TBA equation (\ref{TBA1}) in order to
be able to compare it with the results of \cite{Arutyunov:2009ax}. 
First we simplify $\Delta_R$ and $\Delta_{{\rm red}}$ and use these results
in the $Y_1$ TBA equation (\ref{TBA1}).

\subsection{Simplifying $\Delta_R$}

Let us introduce the new TBA variables $W_a$ with the definitions
\begin{equation}
Y_a=\tau_a^2\,W_a,\qquad a=2,3,\dots,\qquad Y_1=W_1.
\end{equation}
$W_a$ ($a=1,2,\dots$) have no zeroes/poles in the physical strip. In terms
of these variables we rewrite (\ref{TBAQ}) in the form
\begin{eqnarray}
\ln W_a&=&\left\{2{\mathscr L}_{a-1}+\ln W_{a+1}+\ln W_{a-1}-L_{a+1}
-L_{a-1}\right\}\star s,\qquad a=3,4,\dots\label{Wa}\\
\ln W_2&=&\left\{2{\mathscr L}_1+\ln W_3+\ln W_1-L_3-L_1-\ln\tau_1^2
\right\}\star s.\label{W2}
\end{eqnarray}
The simplification of (\ref{DeltaRres}) is based on the kernel identity
\cite{AF09d}
\begin{equation}
k_a+\delta_{a,1}\,s=s\star(k_{a+1}+k_{a-1}),\qquad a=1,2,\dots
\label{kernel1}
\end{equation}
where $k_0(w,v)=2\Theta(4-w^2)K(w,v)$.
The convolution of (\ref{Wa}) with $(k_a+k_{a-2})$ (from the right)
gives for $a=3,4\dots$, after using (\ref{kernel1})
\begin{equation}
\ln W_a*(k_a+k_{a-2})=(2{\mathscr L}_{a-1}+\ln W_{a+1}+\ln W_{a-1}
-L_{a+1}-L_{a-1})\star k_{a-1}
\label{Waconv}
\end{equation}
and from (\ref{W2}) we get
\begin{equation}
\ln W_2*(k_2+k_0)=\ln W_2+(2{\mathscr L}_1+\ln W_3+\ln W_1
-L_3-L_1)\star k_1-\ln\tau_1^2\star k_1.
\label{W2conv}
\end{equation}
Using the methods we employed in section~6 for the calculation of
pole terms we can evaluate the convolution $\ln\tau_1^2\star k_1$:
\begin{equation}
\ln\tau_1^2\star k_1=2{\cal W}-{\cal W}_0,
\qquad {\cal W}_0(u)=\sum_{j=1}^N[\sigma(u,u_j^+)
+\sigma(u,u_j^-)].
\end{equation}
The equations (\ref{Waconv}) and (\ref{W2conv}) can be used to express
the terms containing ${\mathscr L}_m$ in (\ref{DeltaRres}). Many terms
cancel and we find
\begin{equation}
\begin{split}
2\sum_{m=1}^\infty{\mathscr L}_m\star k_m=\sum_{Q=1}^\infty
(L_Q&+L_{Q+2})\star k_Q-\ln W_2\\
&+2\ln W_2 \ \hat\star\ K-\ln Y_1\star k_1+2{\cal W}-{\cal W}_0.
\end{split}
\end{equation}
This can be used to write the simplified $\Delta_R$ formula
\begin{equation}
\begin{split}
2\Delta_R&=-\sum_{Q=1}^\infty(L_Q+L_{Q+2})\star k_Q+\ln\left\{
Y_2\left(1-\frac{1}{Y_-}\right)^2\right\}\\
&-2\ln\left\{Y_2\left(1-\frac{1}{Y_-}\right)\left(1-\frac{1}{Y_+}\right)
\right\} \ \hat\star\ K+{\cal W}_0-2\ln x^2+\ln Y_1\star k_1.
\end{split}
\label{DeltaRsimpl}
\end{equation}

\subsection{Simplifying $\Delta_{{\rm red}}$}

We start from
\begin{eqnarray}
\Delta_{{\rm red}}(u)&=&2\sum_{N=1}^\infty \int \hspace{-3.9mm} \square\,
{\rm d}v\,j_{{\rm red}}(v+i\epsilon)\left\{K^{[2N]}(v,u)-K^{[-2N]}(v,u)
\right\}\\
&=&\sum_{Q=1}^\infty\int_{-\infty}^\infty{\rm d}w\, L_Q(w)\Sigma_Q(w,u),
\end{eqnarray}
where
\begin{equation}
\Sigma_Q(w,u)=2\sum_{N=1}^\infty \int \hspace{-3.9mm} \square\,
{\rm d}v\,\left\{K^{[Q]}(w,v+i\epsilon)-K^{[-Q]}(w,v+i\epsilon)\right\}
\left\{K^{[2N]}(v,u)-K^{[-2N]}(v,u)\right\}.
\end{equation}
The kernel identity we need here is
\begin{equation}
\int \hspace{-3.9mm} \square\, {\rm d}v\, K^{[\pm 2N]}(v,u)
K_{Qy}(w,v+i\epsilon)=\pm K^{[\pm(2N+Q)]}(w,u)\mp \int_{-2}^2{\rm d}v\,
K_{Qy}(w,v)K^{[\pm 2N]}(v,u).
\end{equation}
Using it we find
\begin{equation}
\begin{split}
\Sigma_Q(w,u)&=2\sum_{N=1}^\infty\left\{\int_{-2}^2{\rm d}v\,K_{Qy}(w,v)
k_{2N}(v,u)-k_{2N+Q}(w,u)\right\}\\
&=2I_Q(w,u)-2\int_{-2}^2{\rm d}v\,K_{Qy}(w,v)I_0(v,u)=2\check K^\Sigma_Q(w,u).
\end{split}
\end{equation}

\subsection{The $Y_1$ equation}

The results of the previous two subsections can be used to simplify
the full discontinuity $\Delta$. We get (with $L=J+2$)
\begin{equation}
\Delta=\ln\frac{{\cal D}_1^+}{({\cal D}_1^+)_*}
+\Delta^{(L)}+\Delta^{(12)}+{\cal W}_0+\ln\left\{
Y_2\left(1-\frac{1}{Y_-}\right)^2\right\}-L\ln x^2,
\end{equation}
where
\begin{equation}
\Delta^{(L)}=\sum_{Q=1}^\infty L_Q\star(2\check K^\Sigma_Q-k_Q)-
\sum_{Q=3}^\infty L_Q\star k_{Q-2}
\end{equation}
and
\begin{equation}
\Delta^{(12)}=\ln Y_1\star k_1-2\ln\left\{Y_2\left(1-\frac{1}{Y_-}\right)
\left(1-\frac{1}{Y_+}\right)\right\} \ \hat\star\ K.
\end{equation}
The simplified TBA equation for $Y_1$ becomes
\begin{eqnarray}
\ln Y_1&=&\ln\left\{Y_2\left(1-\frac{1}{Y_-}\right)^2\right\}
\star s-L_2\star s -\Delta \ \check\star\ s\\
&=&\ln\left\{Y_2\left(1-\frac{1}{Y_-}\right)^2\right\}
\ \hat\star\  s-L_2\star s -\Delta_{{\rm eff}} \ \check\star\ s,
\label{Y1simpl}
\end{eqnarray}
where 
\begin{equation}
\Delta_{{\rm eff}}=\Delta^{(L)}+\Delta^{(12)}+\Delta^{({\rm source})}
-L\ln x^2
\end{equation}
and
\begin{equation}
\Delta^{({\rm source})}={\cal W}_0+\ln\frac{{\cal D}_1^+}{({\cal D}_1^+)_*}.
\end{equation}

Using the results of appendix B we can write
\begin{equation}
\frac{{\cal D}_1^+}{({\cal D}_1^+)_*}=\left(\frac{B_p}{R_p}\right)^2\exp\left\{
2\sum_{j=1}^N[G(u,x_j^+)-G(u,x_j^-)]\right\},
\end{equation}
where
\begin{equation}
iG(u,\xi)=\Phi\left(\frac{1}{x(u)},\xi\right)-\Phi(x(u),\xi)+\psi(u,\xi)
\end{equation}
and the source term simplifies to
\begin{equation}
\Delta^{({\rm source})}=\ln\left(\frac{B_mB_p}{R_mR_p}\right)+
2\sum_{j=1}^N\left\{G(u,x_j^+)-G(u,x_j^-)\right\}.
\end{equation}
If we use for $G(u,\xi)$ the alternative representation 
\begin{equation}
G(u,\xi)=\int_{-2}^2{\rm d}v\,\ln\left(\frac{\frac{1}{x(v)}-\xi}
{x(v)-\xi}\right)\,I_0(v,u)
\end{equation}
we can check that our simplified equation (\ref{Y1simpl}) agrees with
eq. (4.10) of ref. \cite{Arutyunov:2009ax}.

\section{Canonical (and hybrid) equations for $Y_Q$}
Using the $W_a$ variables introduced in section~7 we can rewrite the
Y-system equations (\ref{YQ}-\ref{Y1}) in the form
\begin{equation}
\frac{W_a^+W_a^-}{W_{a+1}W_{a-1}}=\xi_a,\qquad a=1,2,\dots,\qquad W_0\equiv1,
\label{chainW}
\end{equation}
where
\begin{eqnarray}
\ln\xi_a&=& 2{\mathscr L}_{a-1}-L_{a+1}-L_{a-1}-\delta_{a,2}\ln\tau_1^2\,
\qquad a=2,3,\dots,\\
\ln\xi_1&=&2{\mathscr L}_--L_2.
\end{eqnarray}
Using the chain lemma of section~4 we can transform (\ref{chainW})
into the integral equations 
\begin{equation}
\ln W_a=\sum_{A=1}^\infty \ln\xi_A \star \ell^A_a
-\Delta \ \check\star\ K_a.
\label{can1}
\end{equation}
Here $\Delta$ is given by (\ref{Deltares}).

\subsection{Kernel identities}

We now list a number of kernel identities that are needed to simplify the 
integral equations (\ref{can1}).

We start by writing the kernel defined by (\ref{Svwx}) as
\begin{equation}
\begin{split}
K^{QM}_{vwx}(u,v)&=\frac{1}{2}K_{Q+M}(u-v)-\frac{1}{2}K_{Q-M}(u-v)+
\sum_{j=1}^{Q-1}K_{M-Q+2j}(u-v)\\
&+\frac{1}{2}k_Q(u,v^{[M]})-\frac{1}{2}k_Q(u,v^{[-M]}).
\end{split}
\end{equation}
Next we write the identity
\begin{equation}
K(y,v)K_a(v-u)=K(y,u^{[a]})K(v,u^{[a]})-
K(y,u^{[-a]})K(v,u^{[-a]})-K_a(y-u)K(v,y)
\end{equation}
and integrate with respect to $v$ just above the cuts using the result
\begin{equation}
\int \hspace{-3.9mm} \square{\rm d}v\,K(v+i\epsilon,\alpha)=\frac{1}{2}
\end{equation}
and obtain
\begin{equation}
K\ \check\star\ K_a=\frac{1}{2}K_{ya}-\frac{1}{2}K_a.
\end{equation}
This can be used to get the further identities
\begin{equation}
K^{Aa}_{vwx}=\ell^{A+1}_a+k_A\ \check\star\ K_a
\end{equation}
and for the \lq\lq fermionic" kernels defined by (\ref{KQypm}) 
\begin{eqnarray}
K^{ya}_-&=&\frac{1}{2}K_{ya}+\frac{1}{2}K_a=K_a+K\ \check\star\ K_a,\\
K^{ya}_+&=&\frac{1}{2}K_{ya}-\frac{1}{2}K_a=K\ \check\star\ K_a.\\
\nonumber
\end{eqnarray}
The most important identity is
\begin{equation}
K^\Sigma_{QQ^\prime}=\check K^\Sigma_Q\ \check\star\ K_{Q^\prime},
\end{equation}
which is proven in \cite{AF09d}. Using also
\begin{equation}
\ell^{Q+1}_a+\ell^{Q-1}_a=K_{Qa}
\end{equation}
we can write
\begin{equation}
\ell^{Q+1}_a+\ell^{Q-1}_a+\Sigma_Q\ \check\star\ K_a=K_{Qa}+
2\check K^\Sigma_Q\ \check\star\ K_a=K_{Qa}+2K^\Sigma_{Qa}=
-K^{Qa}_{{\alg{sl}(2)}}.
\end{equation}
Finally we write the chain lemma for $b_a=\frac{x^{[a]}}{x^{[-a]}}$.
Since it satisfies $b_a^+b_a^-=b_{a+1}b_{a-1}$ in this case $\xi_a\equiv1$
and the only nontrivial object is
\begin{equation}
p=\ln\frac{b_1^+}{(b_1^+)_*}=-\ln x^2
\end{equation}
and we find from the chain lemma
\begin{equation}
\tilde{\cal E}_a=-\ln b_a=-\ln x^2 \ \check\star\ K_a.
\end{equation}

\subsection{Canonical TBA equations}

Collecting all the terms proportional to ${\mathscr L}_m$, ${\mathscr L}_\pm$
or $L_Q$ and using the above identities and the simplifications of
subsection 7.2 we can rewrite (\ref{can1}). This
can be called the canonical $W_a$ TBA equation and is of the form
\begin{equation}
\ln W_a=-L\tilde{\cal E}_a+f_a+2\sum_{m=1}^\infty{\mathscr L}_m\star 
K^{ma}_{vwx}+2{\mathscr L}_-\ \hat\star\ K^{ya}_-
+2{\mathscr L}_+\ \hat\star\ K^{ya}_++\sum_{Q=1}^\infty L_Q\star 
K^{Qa}_{{\alg{sl}(2)}}.
\end{equation}
Here an alternative form of the ${\mathscr L}_\pm$ terms is
\begin{equation}
2{\mathscr L}_-\ \hat\star\ K^{ya}_-
+2{\mathscr L}_+\ \hat\star\ K^{ya}_+=\ln\left(\frac{
1-\frac{1}{Y_-}}{1-\frac{1}{Y_+}}\right)\ \hat\star\ K_a+
\ln\left(\tau_2^2\left(1-\frac{1}{Y_-}\right) \left(1-\frac{1}{Y_+}\right) 
\right)\ \hat\star\ K_{ya}
\end{equation}
and the source term $f_a$ can be written as
\begin{equation}
f_a=-\ln\tau_1^2\star(K_{a+1}+K_{a-1})-\left(2{\cal W}+\ln\frac{{\cal D}_1^+}{
({\cal D}_1^+)_*}\right)\ \check\star\ K_a.
\end{equation}

The source term can be simplified using the chain lemma for
\begin{equation}
{\cal D}_a(u)=\prod_{j=1}^N S^{a1*}_{{\alg{sl}(2)}}(u,u_j)=
\prod_{j=1}^N \left\{
S^{1*a}_{{\alg{sl}(2)}}(u_j,u)\right\}^{-1}.
\end{equation}
${\cal D}_a$ satisfy
\begin{equation}
{\cal D}_a^+{\cal D}_a^-={\cal D}_{a+1}{\cal D}_{a-1},\qquad
a=1,2,\dots,\qquad {\cal D}_0\equiv1.
\end{equation}
However, ${\cal D}_2$ has poles at $u_j^+$ and zeroes at $u_j^-$,
so the chain lemma does not directly apply for the ${\cal D}_a$ functions.
This problem is solved by introducing
\begin{equation}
d_a(u)=\frac{Q^{[a+1]}(u)Q^{[a-1]}(u)}{Q^{[-a-1]}(u)Q^{[1-a]}(u)}=
\prod_{j=1}^N\frac{1}{S_{a+1}(u-u_j)S_{a-1}(u-u_j)}
\end{equation}
and noting that $\tilde{\cal D}_a=\frac{{\cal D}_a}{d_a}$ also satisfy the
chain lemma equations with $\xi_a\equiv1$ and do not have any singularities
near the physical strip and thus can be represented as
\begin{equation}
\tilde{\cal D}_a=\exp\left\{-\ln\frac{{\cal D}_1^+}{({\cal D}_1^+)_*}\ 
\check\star\ K_a\right\}.
\end{equation}
Using this representation the source term becomes
\begin{equation}
\begin{split}
f_a(u)=&-\sum_{j=1}^N\ln S^{1*a}_{{\alg{sl}(2)}}(u_j,u)-\left[\ln\tau_1^2
\star(K_{a+1}+K_{a-1})\right](u)\\
&+\sum_{j=1}^N\left[\ln S_{a+1}(u-u_j)+\ln S_{a-1}(u-u_j)\right]
-2\left[T^\epsilon\star k_1\ \check\star\ K_a\right](u).
\end{split}
\end{equation}
Here we used the result
\begin{equation}
{\cal W}=T^\epsilon\star k_1,
\end{equation}
which can be proved by using the residue theorem. We have defined
\begin{equation}
T^{\pm\epsilon}(u)=\sum_{j=1}^N\ln t(u-u_j\pm i\epsilon).
\end{equation}

It is easy to prove that
\begin{eqnarray}
\left[(T^\epsilon-T^{-\epsilon})\star\ell^2_a\right](u)&=&-\sum_{j=1}^N
\left[\ln S_{a+1}(u-u_j)+\ln S_{a-1}(u-u_j)\right],\\
(T^\epsilon+T^{-\epsilon})\star\ell^2_a &=&\ln\tau_1^2\star\ell^2_a
\end{eqnarray}
and with the help of these relations and the results in the previous subsection
we can write the final form of the source terms: 
\begin{eqnarray}
f_a(u)&=&-\sum_{j=1}^N\ln S^{1*a}_{{\alg{sl}(2)}}(u_j,u)-2\left[T^\epsilon\star
(\ell^2_a+k_1\ \check\star\ K_a)\right](u)\\
&=&-\sum_{j=1}^N\ln S^{1*a}_{{\alg{sl}(2)}}(u_j,u)-2\left[T^\epsilon\star
K^{1a}_{vwx}\right](u).\label{fa}\\
\nonumber
\end{eqnarray}

\subsection{Hybrid TBA equations}

We can get rid of the infinite sums containing the functions ${\mathscr L}_m$
by the following trick \cite{Arutyunov:2009ax}. We write the (\ref{TBAmvw}) 
TBA equation as
\begin{equation}
r_m-{\mathscr L}_m=(r_{m+1}+r_{m-1}-L_{m+1})\star s +\delta_{m,1}
\ln\left[\frac{1-Y_-}{1-Y_+}\right]\ \hat\star\ s,\qquad
m=1,2,\dots,
\label{modTBAmvw}
\end{equation}
where we introduced the notation 
$r_m=\ln[1+Y_{m\vert vw}],\ m=1,2,\dots, r_0\equiv0$.
We now assume that the kernel functions ${\mathscr K}_m$ satisfy the relations
\begin{equation}
{\mathscr K}_m-s\star({\mathscr K}_{m+1}+{\mathscr K}_{m-1})=j_m,\qquad
m=1,2,\dots,\qquad{\mathscr K}_0\equiv0
\end{equation}
with some $j_m$. We take the convolution of (\ref{modTBAmvw}) with 
${\mathscr K}_m$ and sum over $m$. We find
\begin{equation}
\sum_{m=1}^\infty {\mathscr L}_m\star{\mathscr K}_m=\sum_{m=1}^\infty
L_m\star s\star{\mathscr K}_{m-1}+\sum_{m=1}^\infty r_m\star j_m-\ln\left[
\frac{1-Y_-}{1-Y_+}\right]\ \hat\star\ s\star {\mathscr K}_1.
\end{equation}
We now choose ${\mathscr K}_m=K^{ma}_{vwx}$. In this case
\begin{equation}
j_m=\delta_{m+1,a}s+\delta_{m,1} s\ \hat\star\ K_{ya}
\end{equation}
and using the above trick the TBA equations can be brought to the form
\begin{equation}
\begin{split}
\ln W_a=&-L\tilde{\cal E}_a+f_a+2r_{a-1}\star s+
2r_1\star s\ \hat\star\ K_{ya}\\
&-2\ln\left[\frac{1-Y_-}{1-Y_+}\right]\ \hat\star\ s\star K^{1a}_{vwx}
+2{\mathscr L}_-\ \hat\star\ K^{ya}_-+2{\mathscr L}_+\ \hat\star\ K^{ya}_+\\
&\qquad+\sum_{Q=1}^\infty L_Q\star\left[K^{Qa}_{{\alg{sl}(2)}}
+2s\star K^{Q-1\,a}_{vwx}
\right].
\end{split}
\label{hybrid}
\end{equation}
Here $K^{0a}_{vwx}\equiv0$ is understood. Our final result is in complete 
agreement with the corresponding results valid in the special
cases studied in \cite{Arutyunov:2009ax} and \cite{BH-BJ}.

\section{Quantization conditions and exact Bethe-Yang equations}
In this section we formulate the quantization conditions and the exact
Bethe equations which determine the discreet parameters $\xi_{m,j}$,
$\tilde\xi_{m,j}$ and $u_j$ occurring in the source terms of the TBA
integral equations. 

\subsection{Quantization conditions}

In this subsection we quantize the roots occurring in the functions
\begin{equation}
\tau_m(u)=\prod_{j=1}^{{\cal N}_m}t(u-\xi_{m,j}),\qquad
\tilde\tau_m(u)=\prod_{j=1}^{\tilde{\cal N}_m}t(u-\tilde\xi_{m,j}),\qquad
m=2,3,\dots
\end{equation}
(Note that $\tilde\tau_1(u)\equiv1$ and the physical rapidities $u_j$
occurring in $\tau_1(u)=\prod t(u-u_j)$ will be quantized by the exact Bethe
equations discussed in the next subsection.)
We assume that (as is the case in the asymptotic solution) the
functions $Y_{m\vert vw}^{[\pm2]}$ have no poles around the points $\xi_{m,j}$
or $\xi_{m+2,j}$ and therefore the zeroes on the left hand side of the
Y-system equations (\ref{Ymvw}) must be accompanied by corresponding
zeroes also on the right hand side. This leads to the quantization
conditions
\begin{equation}
1+Y^\pm_{m\vert vw}(\xi_{m+1,j})=0,\qquad m=1,2,\dots,\qquad
j=1,\dots,{\cal N}_{m+1}.
\end{equation}
These are well-defined even for $m=1$ since
\begin{equation}
\frac{Y_{1\vert vw}^+(\xi_{2,j}+i\epsilon)}
{Y_{1\vert vw}^+(\xi_{2,j}-i\epsilon)}=
\frac{1-Y_-(\xi_{2,j}+i\epsilon)}
{1-Y_+(\xi_{2,j}+i\epsilon)}=1.
\end{equation}

Similarly we assume that the functions $Y_{m\vert w}^{[\pm2]}$ have no 
poles around the points $\tilde\xi_{m,j}$ or $\tilde\xi_{m+2,j}$ and 
therefore we have the quantization conditions
\begin{equation}
1+Y^\pm_{m\vert w}(\tilde\xi_{m+1,j})=0,\qquad m=1,2,\dots,\qquad
j=1,\dots,\tilde{\cal N}_{m+1}.
\end{equation}
These are well-defined even for $m=1$.

We discuss an important special case in detail. In this special case
(which is relevant for example for the case of twist-two states in the
${\alg{sl}(2)}$ sector) we have no $\tilde\xi_{m,j}$ roots at all and all
$\xi_{m,j}$ roots are real. Moreover all ${\cal N}_m$ are even numbers.
Then we can make the following definitions. The (\ref{TBAmvw}-\ref{TBA1vw})
TBA equations can be written
\begin{equation}
Y_{m\vert vw}=\tau_m\tau_{m+2}\exp\left\{f_m\star s\right\},
\end{equation}
where
\begin{equation}
f_m(u)=\ln\left(\frac{[1+Y_{m-1\vert vw}(u)][1+Y_{m+1\vert vw}(u)]}
{1+Y_{m+1}(u)}\right)+\Theta(4-u^2)\delta_{m,1}\ln\left(\frac{
1-Y_-(u)}{1-Y_+(u)}\right)
\end{equation}
and we define
\begin{equation}
{\cal B}_m(u)=\frac{g}{4}\,{\cal P}\,\int_{-\infty}^\infty{\rm d}v\,
\frac{f_m(v)}{\sinh\frac{(u-v)g\pi}{2}}.
\end{equation}
Note that the principal value prescription ${\cal P}$ is actually
superfluous at the points we will need this function since
\begin{equation}
f_m(\xi_{m+1,j})=0,\qquad m=1,2,\dots,\qquad j=1,\dots,{\cal N}_{m+1}.
\end{equation}
We also define
\begin{equation}
R_m(u)=2\sum_{j=1}^{{\cal N}_m}\arctan\tanh\frac{(u-\xi_{m,j})g\pi}{4}.
\end{equation}
In the special case the quantization conditions take the form 
\begin{equation}
R_m(\xi_{m+1,k})+R_{m+2}(\xi_{m+1,k})+{\cal B}_m(\xi_{m+1,k})=
2\pi\nu_{m+1,k}\quad m=1,2,\dots,\quad k=1,\dots,{\cal N}_{m+1},
\label{QuantCond}
\end{equation}
where
\begin{equation}
\nu_{m+1,k}=\left\{
\begin{split}
{\rm integer\ \ \ \ \ \ } \quad {\rm if}\quad
 &\frac{{\cal N}_m+{\cal N}_{m+2}}{2} \quad {\rm is\ odd},\\
{\rm half-integer} \qquad {\rm if} \quad 
&\frac{{\cal N}_m+{\cal N}_{m+2}}{2} \quad {\rm is\ even}.
\end{split}
\right.
\end{equation}
We note that the above quantum numbers $\nu_{m+1,k}$ are not free parameters.
They are determined by the state under study and in principle can be calculated
by computing the left hand side of (\ref{QuantCond}) in the asymptotic limit.

\subsection{Exact Bethe equations}

The quantization conditions for the physical rapidities $u_k$ are
the exact Bethe equations:
\begin{equation}
Y_{1(*)}(u_k)=-1,\qquad   k=1,2,\dots,N,
\label{exactBethe}
\end{equation}
where the analytic continuation (denoted by the subscript $_{(*)}$) means that
we have to analytically continue the function from the real line just below the
cut line at $-\frac{i}{g}$ coming down between the branch points 
$\pm2-\frac{i}{g}$, and then going back to the real axis through the cut.
Using our previously introduced notation, for any function $F(u)$ we have
\begin{equation}
F_{(*)}(u)=\left(\left(F^-\right)_*\right)^+(u-i\epsilon).
\end{equation}
Our definition is in agreement with the transformation rules
$x^\pm_{(*)}(u)=x^\pm_s(u)$ and \hfill\break
$S^{1*\,1}_{{\alg{sl}(2)}(*)}(u_j,u)=S^{1*\,1*}_{{\alg{sl}(2)}}(u_j,u)$.
It is also easy to see that the following transformation rules hold:
\begin{eqnarray}
f=F\ \check\star\ K_1  
\quad\Rightarrow\quad f_{(*)}(u)&=&f(u)-F^+(u-i\epsilon),\\
\psi=F\ \hat\star\  K_1\quad\Rightarrow\quad 
\psi_{(*)}(u)&=&\psi(u)+F_*^+(u-i\epsilon).
\end{eqnarray}
Using the last formula we can write the following further identities
\begin{eqnarray}
\left(F\ \hat\star\ K^{y1}_+\right)_{(*)}&=&F\ \hat\star\ K^{y1}_{+(*)},\\
\left(F\ \hat\star\ K^{y1}_-\right)_{(*)}&=&F\ \hat\star\ K^{y1}_{-(*)}
+F^{[1-\epsilon]}_*,\\
\left(F\ \hat\star\ K_{y1}\right)_{(*)}&=&F\ \hat\star\ K_{y1(*)}
+F^{[1-\epsilon]}_*.
\end{eqnarray}

Using this set of identities we can rewrite the exact Bethe equations:
\begin{equation}
\ln Y_{1(*)}(u_k)=2\pi i\nu_k,\qquad \nu_k:{\rm \ half-integer},\qquad
k=1,\dots,N, 
\label{nuk}
\end{equation}
where
\begin{equation}
\begin{split}
\ln Y_{1(*)}=&-L\ln(x^+x^-)+f_{1(*)}+2r_1\star s^{[\epsilon-1]}+
2r_1\star s\ \hat\star\ K_{y1(*)}\\
&-2\ln\left[\frac{1-Y_-}{1-Y_+}\right]\ \hat\star\ s\star K^{11}_{vwx(*)}
+2{\mathscr L}_-\ \hat\star\ K^{y1}_{-(*)}
+2{\mathscr L}_+\ \hat\star\ K^{y1}_{+(*)}\\
&+2{\mathscr L}_+^{[1-\epsilon]}+\sum_{Q=1}^\infty L_Q\star\left\{
K^{Q1}_{{\alg{sl}(2)}(*)}+2s\star K^{Q-1\,1}_{vwx(*)}\right\}
\end{split}
\label{BY}
\end{equation}
and
\begin{equation}
f_{1(*)}(u)=-\sum_{j=1}^N\ln S^{1*\,1*}_{{\alg{sl}(2)}}(u_j,u)-2\left(T^{[\epsilon]}
\star K^{11[\epsilon]}_{vwx(*)}\right)(u).
\label{f1star}
\end{equation}
We note that since
\begin{equation}
2{\mathscr L}_+^+=-2iR_2+\ln\left(1-\frac{1}{Y_+^+}\right)^2
\end{equation}
and here the second term vanishes at $u_k$, the first term in the third
line in (\ref{BY}) can be substituted by $-2iR_2$. We also note that the
$\epsilon$ prescription in the third term in (\ref{BY}) is not really needed
since $r_1$ vanishes at $u=u_k$.

In (\ref{nuk}) $\nu_k$ are momentum quantum numbers and can be used
to label the state (instead of the particle momenta).

Finally we give an alternative variant of (\ref{f1star}) using the 
regularization introduced in~\cite{Arutyunov:2009ax}:
\begin{equation}
\begin{split}
f_{1(*)}(u)=&-\sum_{j=1}^N\ln S^{1*\,1*}_{{\alg{sl}(2)}}(u_j,u)+
2\int_{-\infty}^\infty{\rm d}v\,\ln\frac{Q(v)}{\tau_1(v)}\,K^{11}_{vwx(*)}
(v+i\epsilon,u)\\
&-2\sum_{j=1}^N\ln\left[(u_j-u^{++})\,\frac{x_j^--\frac{1}{x^-(u)}}
{x_j^--x^+(u)}\right].
\end{split}
\end{equation}
At $u=u_k$, where we need it, we have
\begin{equation}
\begin{split}
f_{1(*)}(u_k)=&-\sum_{j=1}^N\ln S^{1*\,1*}_{{\alg{sl}(2)}}(u_j,u_k)+
2\int_{-\infty}^\infty{\rm d}v\,\ln\frac{Q(v)}{\tau_1(v)}\,K^{11}_{vwx(*)}
(v+i\epsilon,u_k)\\
&-2\sum_{j=1}^N\ln\left[(u_j-u_k-\frac{2i}{g})\,\frac{x_j^--\frac{1}{x_k^-}}
{x_j^--\frac{1}{x_k^+}}\right].
\end{split}
\end{equation}

Again, the final formula for the exact Bethe-Yang equations agrees with
the results obtained in \cite{Arutyunov:2009ax} by contour deformation 
techniques.

\section{Simplifying the energy formula}
The energy of the multi-particle state we have been studying in this paper
is given by the formula
\begin{equation}
E=J+\sum_{j=1}^N{\cal E}(p_j)-\frac{1}{2\pi}\sum_{Q=1}^\infty
\int_{-\infty}^\infty {\rm d}u\,\frac{{\rm d}\tilde p^Q}{{\rm d} u}
\ln(1+Y_Q),
\label{energy1}
\end{equation}
where 
\begin{equation}
{\cal E}(p)=\sqrt{1+4g^2\sin^2\frac{p}{2}},\qquad
{\rm e}^{ip}=\frac{x_s^+(u)}{x_s^-(u)} 
\end{equation}
gives the energy of a physical particle with
momentum $p$ in the string model and $\tilde p^Q(u)$ is the momentum
of the bound state $Q$-particle with rapidity $u$ in the mirror theory.
Not only the above energy formula, but many other results in this paper
contain an infinite sum of convolutions of the form
\begin{equation}
\sum_{Q=1}^\infty L_Q\star f_Q.
\label{sumcon}
\end{equation}
The results that contain an infinite sum of this form include the
hybrid equation for $Y_1$, the exact Bethe equation and the results
for both $Y_-/Y_+$ and $Y_+Y_-$. The fact that allows the simplification
of the infinite sum is that in all these formulae the coefficient functions
$f_Q$ satisfy the functional relation
\begin{equation}
f_Q^++f_Q^-=f_{Q+1}+f_{Q-1}.
\label{fQ}
\end{equation}
We find that if we express $L_Q$ in the sum in terms of the T-system
functions $T_{Q,0}$ then using the above identity most of the terms cancel 
and we end up with a simple finite expression containing the single function 
$T_{1,0}$ only. 
(In some cases two functions remain, $T_{1,0}$ and $T_{2,0}$.)

For the case of the energy formula we have
\begin{equation}
f_Q(u)=-\frac{1}{2\pi}\,\frac{{\rm d}}{{\rm d}u}\tilde p^Q(u)=\frac{g}
{2\pi}\left(x^{[Q]\prime}(u)-x^{[-Q]\prime}(u)\right)
\end{equation}
and the full energy expression can be written as
\begin{equation}
E=J+\sum_{j=1}^N{\cal E}(p_j)+\frac{ig}{2\pi}
\int \hspace{-3.9mm} \square\ {\rm d}u\,\ln T_{1,0}(u)
\frac{u}{\sqrt{4-u^2}},
\label{energy2}
\end{equation}
where the contour of the $u$ integral has to lie a little above the real line.

We discuss just one more example here. In the case of the $Y_-/Y_+$ ratio
the relevant coefficient function is
\begin{equation}
f_Q=K^{[Q]}-K^{[-Q]}
\end{equation}
and the simplified final formula is
\begin{equation}
\frac{Y_-}{Y_+}=\frac{R_pB_m}{B_pR_mT_{1,0}}\exp\left\{
2\ln T_{1,0}\ \check\star\ 
K^{[\epsilon]}\right\}.
\end{equation}

In our opinion the fact that the energy formula can be rewritten such 
that it depends on the single variable $T_{1,0}$ only might indicate
that there is a transfer matrix formulation behind the exact TBA equations
in the model. This may also be an important step towards the NLIE description
of the system since $T_{1,0}$ can easily be expressed by
the elementary objects appearing in that approach \cite{KL10,GKLT}.

\section{Conclusions}

In this paper we derived the TBA equations for the ${\, \alg{sl}(2) \,}$ (sub-) sector of the
$AdS_5 \times S^5$ mirror model based on the Y-system and the discontinuity relations proposed 
in \cite{Tateo1}. 
The proposal for the discontinuity relations was based on the analyticity properties of the 
solutions
of the ground state TBA equations and was conjectured to be state independently valid for the 
excited
states as well.

In this paper we have shown that the discontinuity relations hold nontrivially for the 
asymptotic
solutions of the excited states. This corroborates the state independent nature of
the discontinuity relations.
In addition we studied the discontinuity relations carefully and concluded that a technical 
subtlety
requires the use of a refined interpretation when translating them to dispersion relations. 

Our derivation of the TBA equations is based on the fact that the Y-system equations,
the discontinuity relations plus some qualitative information on the local singularities
of the Y-functions and their asymptotics at infinity together make possible to transform the 
functional equations into TBA integral equations in a unique way.

In this derivation we assumed that the Y-functions for the excited states are smooth 
deformations
of their asymptotic form, so the qualitative information on their local singularities and on 
their
behavior at infinity can be read off from the explicitly known asymptotic solution.

Since as we have proven the asymptotic solutions satisfy the limiting functional equations 
exactly, by construction
they also satisfy (the $Y_Q\rightarrow0$ limit of) the TBA integral equations and the 
quantization conditions,
including the Bethe-Yang equations. An important consequence of this observation is that the
asymptotic limit of the exact Bethe-Yang equation (\ref{BY}) accounts for the Beisert-Staudacher 
equations (\ref{Beisert}).
This fact has not been proven analitically so far, though it was an important starting point in 
the
5-loop tests of the mirror TBA equations \cite{AFS,BHxxx,BH-BJ}. 

Beyond the derivation of the TBA equations we also constructed the T-system elements in a 
special gauge
in terms of the Y-functions. The benefit of this construction is 2-fold. 
On the one hand the discontinuity relations are a lot simpler in the language of T-functions and 
their
introduction makes the derivation of the TBA equations easier.
For example with their help we could show that to derive the TBA equations 
the knowledge of local singularities of the Y-functions lying only within certain finite strips 
around the real axis is needed.  
On the other hand we recognized that important infinite sums of the TBA problem simplify 
drastically
if they are expressed in terms of the T-functions. The most important such simplification 
appears in the energy
formula, which can be expressed as a simple expression of a single T-function $T_{1,0}$. 
This might indicate that there exists a hidden transfer matrix formulation of the model.

Independently of these speculations, we think that our construction of the T-system and the 
expression for
the energy in terms of $T_{1,0}$ gives an important step towards the NLIE formulation 
\cite{nlie1,nlie2} of the
AdS/CFT spectral problem. There the T-functions are more fundamental objects and can be 
expressed easier 
than the Y-functions since the NLIE construction is based on the T-Q relations of the model.
For example in the approach of \cite{KL10} and \cite{GKLT} the $T_{a,s}$ functions are expressed 
by 
the Wronskian determinants of certain fundamental Q-functions whose combinations  are
the unknowns of the NLIE. 

As a final remark we note that recently \cite{RyoHybrid} with the help of the T-Q relations the 
left and right $SU(2)$
wings of the TBA equations could be resummed by a hybrid-NLIE and so the number of unknown 
functions were
remarkably reduced.


\section*{Acknowledgments}

We would like to thank G. Arutyunov, S. Frolov and R. Tateo
for reading the manuscript and for useful suggestions.
We also thank an anonymous referee for useful comments and questions
and in particular for suggesting the analysis presented in appendix E.
This work was supported by the Hungarian
Scientific Research Fund (OTKA) under the grant K 77400.

\appendix

\section{Notations, kinematical variables, kernels}

In this paper we adopted the definitions and conventions of ref. \cite{Arutyunov:2009ax}.
For completeness, in this appendix we collect these definitions and give a list
of all kernel functions used  in the paper. 

We will use the notation $f^{\pm}(u)=f(u\pm \frac{i}{g})$ for any function $f$ and in general
$f^{[a]}(u)=f(u+\frac{i}{g} a)$. We will also use $w^\pm=w\pm\frac{i}{g}$ for $w$ some parameter.

Most of the kernels and also the asymptotic solution of the Y-system is expressed
in terms of the function $x(u)$:
\begin{equation}
x(u)=\frac12 (u-i\sqrt{4-u^2}), \qquad \mbox{Im}\, x(u)<0,
\end{equation} 
which maps the $u$-plane with cuts $[-\infty,-2] \cup [2,\infty]$ onto the physical region
of the mirror theory, and the function $x_s(u)$
\begin{equation}
x_s(u)=\frac{u}{2} \left(1+\sqrt{1-\frac{4}{u^2}}  \right), \qquad |x_s(u)|\geq 1,
\end{equation}
which maps the $u$-plane with the cut $[-2,2]$ onto the physical region of the string theory.
Both functions satisfy the identity $x(u)+\frac{1}{x(u)}=u$ and they are related by $x(u)=x_s(u),$ and 
$x(u)=1/x_s(u)$ in the lower and upper halves of the complex plane respectively.

The momentum $\tilde{p}^Q$ and the energy $\tilde{\cal{E}}_Q$ of a
mirror $Q$-particle are expressed in terms of $x(u)$ as follows
\begin{eqnarray}
 \tilde{p}_Q=g x\big(u-\frac{i}{g}Q\big)-g
x\big(u+\frac{i}{g}Q\big)+i Q\, ,
~~~~~\tilde{\cal{E}}_Q=\log\frac{x\big(u-\frac{i}{g}Q\big)}{x\big(u+\frac{i}{g}Q\big)}\,.
 \end{eqnarray}
Three different types of convolutions appear in the TBA equations. These are:
\begin{eqnarray}
\nonumber
&&f \star {\cal K}(v) \equiv \int_{-\infty}^\infty\, du\, f(u) \, { \cal K}(u,v)\,,
\quad f \, {\, \hat{\star} \,}{\cal K}(v) \equiv \int_{-2}^2\, du\, f(u) \, {\cal K}(u,v)\,,\\
&&f{\, \check{\star} \,} {\cal K}(v) \equiv \left(\int_{-\infty}^{-2} +\int_{2}^\infty\right)\, du\, 
f(u) \,{\cal  K}(u,v)\,. \label{convs}
\end{eqnarray}
The last operation $\left( \int_{-\infty}^{-2} +\int_{2}^\infty\right)\, du\,$ is often denoted by 
$\displaystyle{\int 
\hspace{-3.9mm}\square \, du}$ for short. If the kernel ${\cal K}$ depends on a single variable, then
the convolutions in (\ref{convs}) are understood as $ \int du \, f(u) \, {\cal K}(u-v)$. 
For a kernel ${\cal K}$  and parameter $a \in \mathbb{R}$ 
we often use the notation ${\cal K}^{[a]}(u,v)={\cal K}(u+{i \,a \over g},v)$.

The different but equivalent formulations of the mirror TBA {equations}{\footnote{Using the terminology of ref.
\cite{Arutyunov:2009ax} they are called canonical, simplified, hybrid etc.}} contain a number of kernels
which we list below.

We start with kernels depending on a single variable:
\begin{alignat}{2}
s (u) & = \frac{1}{2 \pi i} \, \frac{d}{du} \log t^-(u)= {g \over 4\cosh {\pi g u \over 2}}\,,\quad t(u)=\tanh[ 
\frac{\pi g}{4} u ]\,,
\nonumber \\
K_Q (u) &= \frac{1}{2\pi i} \, \frac{d}{du} \, \log S_Q(u) = \frac{1}{\pi} \, \frac{g\, Q}{Q^2 + g^2 
u^2}\,,\quad S_Q(u)= \frac{u - \frac{iQ}{g}}{u + \frac{i Q}{g}} \,, \nonumber\\
K_{MN}(u) &= \frac{1}{2\pi i} \, \frac{d}{du} \, \log 
S_{MN}(u)=K_{M+N}(u)+K_{N-M}(u)+2\sum_{j=1}^{M-1}K_{N-M+2j}(u)\,,\nonumber\\
S_{MN}(u) &=S_{M+N}(u)S_{N-M}(u)\prod_{j=1}^{M-1}S_{N-M+2j}(u)^2 =S_{NM}(u)\,. \label{sKQ}
\end{alignat}
The fundamental building block of kernels which are not of difference type is:
\begin{eqnarray}
 K(u,v) = \frac{1}{2 \pi i} \, \frac{d}{du} \, \log S (u,v) = \frac{1}{2 \pi i} \, 
\frac{ \sqrt{4-v^2}}{\sqrt{4-u^2}}\, {1\over u-v} \,,\ \  S(u,v)=\frac{x(u) - x(v)}{x(u) x(v) - 1}\,.~~~
\label{Kuv}
 \end{eqnarray}
An important function in the equations for $J^{(\alpha)}$ and $\Delta$ is $\sigma(u,v)$
which is the logarithm of $S(u,v)$ (i.e $e^{\sigma(u,v)}=S(u,v)$).
To treat the logarithmic discontinuities appropriately we define it more precisely. 
Let
\begin{equation}
\phi(u,\xi)=\frac{x(u)-\xi}{{1 \over x(u)}-\xi}, \qquad |\xi|>1, \qquad \mbox{Im}\, \xi \neq 0,
\end{equation}
and assume that $u$ lies close to the real axis. 
Then the definition of $\sigma(u,v)$ is as follows:
\begin{eqnarray}
\sigma(u,v)= \left\{ \begin{array}{rcl} 
\ln\left({-1 \over x(u)} \right)+\ln \phi (u,x(v)) \qquad & \mbox{Im}\, v<0,  \\
\ln\left({- x(u)} \right)-\ln \phi(u,{1/ x(v)}) \qquad & \mbox{Im}\, v>0.
\end{array} \right.
\end{eqnarray}
We list its most important properties below:
\begin{equation}
\sigma(u+i \, \epsilon,v)=-\sigma(u-i \, \epsilon,v), \qquad |u|>2,
\end{equation}
\begin{equation}
\sigma(u_0+i \, \epsilon,v) \rightarrow  \ln x(v)+2 \, \pi i \, \Theta(\mbox{Im}\, v) \,
\Theta(u_0), \qquad u_0 \rightarrow \pm \infty, \qquad \mbox{Im}\, v \neq 0,
\end{equation}
\begin{equation}
\sigma(-2,v)=0, \qquad \sigma(2+i \, \epsilon,v)=i  \pi \, {\rm sgn}(\mbox{Im}\,v).
\end{equation}
with $\Theta(u)$ being the unitstep function and $\epsilon$ is a positive infinitesimal parameter.

Using the kernels $K(u,v)$ and $K_Q(u-v)$ it is possible to define a series of kernels which are connected
to the fermionic $Y_{\pm}^{(\alpha)}$-functions.
They are:
\begin{eqnarray}
K_{Qy}(u,v)&=&K(u-\frac{i}{g}Q,v)-K(u+\frac{i}{g}Q,v)\,, \label{KQy}\\
K^{Qy}_\mp(u,v)&=&{1\over 2}\Big( K_Q(u-v) \pm  K_{Qy}(u,v)\Big) \label{KQypm}
\end{eqnarray}
and
\begin{eqnarray}
K_{yQ}(u,v)&=&K(u,v+{i\over g}Q)-K(u,v-{i\over g}Q), \label{KyQ}  \\ 
K^{yQ}_\pm(u,v) &=& {1\over 2}\Big(K_{yQ}(u,v)\mp K_Q(u-v)\Big)\, . \label{KyQpm}
\end{eqnarray}
The equation for the discontinuity function contains the kernel
\begin{equation}
k_m(u,v)=K^{[m]}(u,v)+K^{[-m]}(u,v), \qquad m=1,2,...
\end{equation} 
The kernels entering the infinite sums in the canonical equations are
\begin{eqnarray}
\nonumber
K_{xv}^{QM}(u,v) &=&{1\over 2\pi i}{d\over du}\log S_{xv}^{QM}(u,v)\,,\\
\nonumber S_{xv}^{QM}(u,v) &=&
\frac{x(u-i{Q \over g })-x(v+i{M \over g})}{x(u+i{Q \over g })-x(v+i{M
\over g})}\, \frac{x(u-i{Q \over g })-x(v-i{M \over g})}{x(u+i{Q \over g
})-x(v-i{M \over g})}\, \frac{x(u+i{Q \over g })}{x(u-i{Q \over g
})}~~~~\\ &\times
&\prod_{j=1}^{M-1}\frac{u-v-\frac{i}{g}(Q-M+2j)}{u-v+\frac{i}{g}(Q-M+2j)} \label{Sxv}
\end{eqnarray}
and
\begin{eqnarray}
\nonumber
K_{vwx}^{QM}(u,v) &=&{1\over 2\pi i}{d\over du}\log S_{vwx}^{QM}(u,v)\,,\\
\nonumber S_{vwx}^{QM}(u,v) &=&
\frac{x(u-i{Q \over g })-x(v+i{M \over g})}{x(u-i{Q \over g })-x(v-i{M\over g})}\,
 \frac{x(u+i{Q \over g })-x(v+i{M \over g})}{x(u+i{Q \over g})-x(v-i{M \over g})}\,
 \frac{x(v-i{M \over g })}{x(v+i{M \over g})}~~~~
\\ &\times
&\prod_{j=1}^{Q-1}\frac{u-v-\frac{i}{g}(M-Q+2j)}{u-v+\frac{i}{g}(M-Q+2j)}\,. \label{Svwx}
\end{eqnarray}

The equations for the momentum carrying nodes contain the dressing phase, an important building block
of the ${\alg{sl}(2)}$ S-matrix of the model \cite{AFrev}. It is of the form
\begin{eqnarray}
\label{Ssl2}
S_{{\alg{sl}(2)}}^{QM}(u,v)= S^{QM}(u-v)^{-1} \,
\Sigma_{QM}(u,v)^{-2}\,, 
\end{eqnarray}
where  $\Sigma^{QM}$ is the improved dressing factor \cite{dresscross}.
The corresponding ${\alg{sl}(2)}$ and dressing kernels are defined in the 
usual way
\begin{eqnarray}
K_{\alg{sl}(2)}^{QM}(u,v)= \frac{1}{2\pi i}\frac{d}{du}\log
S_{\alg{sl}(2)}^{QM}(u,v) \,,\quad K_{QM}^{\Sigma}(u,v)=\frac{1}{2\pi
i}\frac{d}{du}\log \Sigma_{QM}(u,v)\,.~~~~ 
\end{eqnarray}

The source terms in the equations for the mirror magnons involve the ${\alg{sl}(2)}$ S-matrix
analytically continued to the physical region in the first argument.
\begin{eqnarray}\nonumber
S_{\alg{sl}(2)}^{1_*M}(u,v)&=&{1\over
S_{1M}(u-v)\Sigma_{1_*M}(u,v)^2}\,.~~~~ 
\end{eqnarray}
Explicit expressions for the improved dressing factors $\Sigma_{QM}(u,v)$ and $\Sigma_{1_*M}(u,v)$ 
can be found in section 6 of ref. \cite{dresscross}. Their expressions contain two important functions
 $\Phi(x_1,x_2)$ and $\Psi(x_1,x_2)$  defined by 
\begin{eqnarray}\label{Phi}
\Phi(x_1,x_2)&=&\oint\frac{{\rm d}w_1}{2\pi i}\oint \frac{{\rm
d}w_2}{2\pi i}\frac{i}{(w_1-x_1)(w_2-x_2)} \ln{\Gamma\big[1+{i\over
2}g\big(w_1+\frac{1}{w_1}-w_2-{1\over w_2}\big)\big]\over
\Gamma\big[1-{i\over 2}g\big(w_1+\frac{1}{w_1}-w_2-{1\over
w_2}\big)\big]} , ~~~~  \\ 
 \label{Psi} 
\Psi({x_1},x_2)&=&\oint\frac{{\rm d
}w}{2\pi i} \frac{i}{w-x_2}\ln\frac{\Gamma\big[1+{i\over
2}g\big(x_1+\frac{1}{x_1}-w-{1\over w}\big)\big]}
{\Gamma\big[1-{i\over 2}g\big(x_1+\frac{1}{x_1}-w-{1\over
w}\big)\big]}\,,~~~~~~~ \end{eqnarray}
where the integrals run over the unit circles in anti-clockwise direction.

The simplified equations for the momentum carrying nodes involve the kernel 

\begin{eqnarray} \label{KQsig}
 {\check K}^\Sigma_{Q} = {1\over 2\pi i} {\partial \over \partial u} \log{\check \Sigma}_{Q}
= - K_{Qy}\,\hat{\star}\,  {I}_0 +{I}_Q
  \end{eqnarray}
where
\begin{eqnarray} \label{IQ}
 &&{I}_Q(u,v)=-  \sum_{n=1}^\infty
{k}_{2n+Q}(u,v+i\epsilon)=K_\Gamma^{[Q+2]}(u-v)-2 \,\int_{-2}^2 {\rm d}t
\, K_\Gamma^{[Q+2]}(u-t) {K}(t,v+i\epsilon) \,,  \nonumber \\\label{KG0}
&&\qquad\qquad\qquad K_\Gamma^{[Q]}(u)={1\over 2\pi i} {d\over d u} \log
\frac{\Gamma\big[{Q\over 2}-\frac{i}{2}g u\big]}{\Gamma\big[{Q\over
2}+\frac{i}{2}g u\big]}.
\end{eqnarray}

\section{The dressing phase discontinuities}
In this appendix we present the verification of the formula (\ref{delt3}),
which was used in the calculation of $\Delta$ in section 6. 
The calculation is entirely based on ref.
\cite{dresscross}, where the analytic continuation of the dressing phase 
to the mirror region and the corresponding integral representations were found.

We define the dressing part of the massive node $Y_1^{(0)}$ 
in the ABA limit as
\begin{equation}
(\ln Y_1^{(0)})_{\rm dress}=-2i\sum_j\theta(u,u_j)\,.
\label{Y1dress}
\end{equation}
This appears in (\ref{calD1C}).
$u_j$ are the physical particle rapidities and the second variable of the
dressing phase $\theta(u,u_j)$ lives on the string sheet which means that
the corresponding $x^\pm_{\rm s}(u_j)$ functions are evaluated in the physical 
(string) kinematics:
\begin{equation} \label{theta}
\theta(u,w)=f^{\rm m}(u,x^+_{\rm s}(w))-f^{\rm m}(u,x^-_{\rm s}(w))\,,
\end{equation}
where $f^{\rm m}(u,\xi)$ is the restriction to the mirror plane of
\begin{equation}
f(z,\xi)=\chi(x^+(z),\xi)-\chi(x^-(z),\xi)\,,\quad \xi=x^\pm_{\rm s}(w)\,,
\end{equation}
defined on the entire $z$ torus (see Figure 1 of \cite{dresscross}).

Actually, we will need this function only in three regions of the rapidity
torus: ${\cal R}_{k,0}$, $k=0,1,2$ \cite{dresscross}. 
We will denote the overlap of these regions with the mirror $u$ plane 
with $R^{\rm m}_k$, $k=0,1,2$. The regions are characterized by

$R^{\rm m}_0$:\ \ \ \ ${\rm Im}\, u<-\frac{1}{g}$;\ \ \  
\ \ \ \ \ \ \ \ \ \ \ \ \ $\vert x^\pm(u)\vert>1$,

\smallskip

$R^{\rm m}_1$:\ \ \ \ $-\frac{1}{g}< {\rm Im}\, u<\frac{1}{g}$;\ \ \
$\vert x^-(u)\vert>1\,,\, \, \vert x^+(u)\vert<1$,

\smallskip

$R^{\rm m}_2$:\ \ \ \ $\frac{1}{g}<{\rm Im}\, u$; \ \ \
\ \ \ \ \ \ \ \ \ \ \ \ \ \ $\vert x^\pm(u)\vert<1$.

\smallskip

The function $\chi(x,\xi)$ (and the complete dressing phase $\theta$) can
be expressed in terms of the functions $\Phi(x,\xi)$ and $\Psi(x,\xi)$
defined in ref. \cite{dresscross}. These definitions can be found
in appendix A. Since in our analysis the second argument of
these functions ($\xi=x^\pm_{\rm s}(u_j)>1$) plays no role in the analytic
continuation process, in the rest of this appendix we will suppress the
dependence of $\Phi,\Psi$ on $\xi$. The function $\Phi(x)$ is given by the 
double integral formula (\ref{Phi}) for all $\vert x\vert\not=1$ whereas for 
$\Psi(x)$ the single integral representation (\ref{Psi}) is valid for all 
$\vert x\vert\not=1$ except for an infinite number of cuts (see below). 
Both functions can be analytically
extended starting from a certain region but the analytic extensions in general
will differ from the integral formula. In this appendix $\Phi(x),\Psi(x)$
are always understood as given by the integral formulae. If we analytically
continue, for example, $\Phi(x)$ from the region $\vert x\vert>1$ to
values $\vert x\vert<1$, we have (for $\vert x\vert$ close to 1):
\begin{equation}
\Phi_{\rm cont}(x)=\Phi(x)-\Psi(x)\,,\qquad \vert x\vert 
{\rm \ \ (slightly)\ } < 1.
\label{cont1} 
\end{equation}

In the three regions we need the $f(z)$ function, analytically continued
from the physical ${\cal R}_{0,0}$ region, is given as \cite{dresscross}

\smallskip

${\cal R}_{0,0}:\qquad f(z)=\Phi(x^+)-\Phi(x^-)\,,$

\smallskip

${\cal R}_{1,0}:\qquad f(z)=\Phi(x^+)-\Psi(x^+)-\Phi(x^-)\,,$

\smallskip

${\cal R}_{2,0}:\qquad f(z)=\Phi(x^+)-\Psi(x^+)+
\frac{1}{i}\ln\frac{\frac{1}{x^-}-\xi}{x^--\xi}
-\Phi(x^-)+\Psi(x^-)\,.$

\smallskip

To calculate the discontinuity function, we also need the analytic continuation
of $f(z)$ from $R^{\rm m}_1$ to the region $R^{\rm m}_2$ through the upper cut
(${\rm Im}\,u=\frac{1}{g}$) in the mirror plane. 
Crossing this cut from below, we remain in ${\cal R}_{1,0}$ and 
correspondingly we have

\smallskip

$R^{\rm m}_2:\qquad f^*(u)=\left((f^+)_*\right)^-(u)=
\Phi(x^+)-\Psi(x^+)-\Phi(x^-_{\rm cont})\,,$

\smallskip

and since through this cut $x^-_{\rm cont}=1/x^-$ this can be written as

\smallskip  

$R^{\rm m}_2:\qquad f^*(u)=\Phi(x^+)-\Psi(x^+)-\Phi(\frac{1}{x^-})\,.$

\smallskip

Defining (in the region just above the cut) the discontinuity
\begin{equation}
D(u)=f(u+\frac{i}{g})-f^*(u+\frac{i}{g})
\end{equation}
we find
\begin{equation}
D(u)=\frac{1}{i}\ln\frac{\frac{1}{x}-\xi}{x-\xi}-\Phi(x)+\Psi(x)
+\Phi(\frac{1}{x})\,,\qquad {\rm Im\ }u>0\,.
\label{Dpos}
\end{equation}
To obtain the discontinuity just below the real line we have to
analytically continue it through the \lq\lq slot" $[-2,2]$. This means
that we have to continue from $\vert x\vert <1$ (above the real line)
to $\vert x\vert >1$ (below the real line). Using (\ref{cont1}) we have
\begin{equation}
-\Phi(x)+\Psi(x)\rightarrow -\Phi(x)
\end{equation}
and
\begin{equation}
\Phi(\frac{1}{x})\rightarrow \Phi(\frac{1}{x})-\Psi(\frac{1}{x})=
\Phi(\frac{1}{x})-\Psi(x)\,,
\end{equation}
since $\Psi(1/x)=\Psi(x)$. Just below the real line we thus have
\begin{equation}
D(u)=\frac{1}{i}\ln\frac{\frac{1}{x}-\xi}{x-\xi}-\Phi(x)
+\Phi(\frac{1}{x})-\Psi(x)\,,\qquad {\rm Im\ }u<0\,.
\label{Dneg}
\end{equation}
Comparing (\ref{Dpos}) and (\ref{Dneg}) we see that (since through the
cut $x\leftrightarrow\frac{1}{x}$) the jump of $D(u)$ through the cut
is equivalent to a sign change, as expected.

We now want to extend the discontinuity $D(u)$ to the first Riemann sheet,
i.e. the whole mirror
plane with its infinitely many cuts at ${\rm Re\ }u>2$, ${\rm Im\ }u=Z/g$,
for all integers $Z$. From (\ref{Dpos}) we see that
only $\Psi(u)$ has to be extended to the upper part of the mirror plane,
the rest is already unambiguously defined. Similarly from (\ref{Dneg})
we see that we have to extend $-\Psi(u)$ to the lower half of the mirror
plane. The point is that although $\Psi$ is already defined through
its integral representation but this representation has cuts
at the \lq\lq wrong" place ($\vert{\rm Re\ }u\vert<2$, 
${\rm Im \ }u=2Z/g$, $Z\not=0$).
Therefore we have to modify the analytic extension starting from
near the real line, where $D(u)$ is given by (\ref{Dpos}) and
(\ref{Dneg}).   

We now perform a partial integration in (\ref{Psi}) and get
\begin{equation}
\begin{split}
\psi(u)=\Psi(x(u))=-&\oint\frac{{\rm d}w}{2\pi}
\ln(w-\xi)\left(1-\frac{1}{w^2}\right)\\
&\frac{{\rm d}}{{\rm d}v}\left\{\ln\Gamma\left[1+\frac{ig}{2}(u-v)\right]
-\ln\Gamma\left[1-\frac{ig}{2}(u-v)\right]\right\}\,,
\end{split}
\end{equation}
where $v=w+1/w$. This can be rewritten as
\begin{equation}
\begin{split}
\psi(u)=\Psi(x(u))=&\oint_{\gamma_0}\frac{{\rm d}v}{2\pi}
\ln(x_{\rm s}(v)-\xi)\\
&\frac{{\rm d}}{{\rm d}v}\left\{\ln\Gamma\left[1+\frac{ig}{2}(u-v)\right]
-\ln\Gamma\left[1-\frac{ig}{2}(u-v)\right]\right\}\,.
\end{split}
\end{equation}
Here the overall sign has changed since the curve $\gamma_0$ is defined
as integration (just above the real line) from $-2$ to $2$, and then back
from $2$ to $-2$ just below the real line, and this is a {\it clockwise}
curve. From this representation we see that the cuts come from where
\begin{equation}
v=u-\frac{2iN}{g}\quad ({\rm first\ term})\qquad{\rm or}\qquad
v=u+\frac{2iN}{g}\quad ({\rm second\ term})\qquad N=1,2,\dots
\end{equation}
On the upper half of the $u$ plane the cuts are at
\begin{equation}
u=u_0+\frac{2iN}{g}\,,\qquad -2<u_0<2
\end{equation}
and come from the first term. Using the residue theorem we can calculate
the jumps and we find that for all $N>0$:
\begin{equation}
\begin{split}
J(u_0+\frac{2iN}{g})&=\psi(u_0+\frac{2iN}{g}+i\epsilon)-
\psi(u_0+\frac{2iN}{g}-i\epsilon)=\\
&-i\ln(x_{\rm s}(u_0+i\epsilon)-\xi)+
i\ln(x_{\rm s}(u_0-i\epsilon)-\xi)
=i\ln\frac{x(u_0)-\xi}{\frac{1}{x(u_0)}-\xi}\,.
\end{split}
\end{equation}
Introducing the notation
\begin{equation}
h(u)=\ln\frac{\frac{1}{x(u)}-\xi}{x(u)-\xi}
\end{equation}
we can write this jump as 
\begin{equation}
J(u)=-i\,h(u-\frac{2iN}{g})\,,\quad {\rm Im\ }u=\frac{2N}{g}\,,\quad
\vert{\rm Re\ }u\vert<2\,.
\end{equation}
Similarly on the lower half plane the jumps come from the second term
and we have ($N>0$)
\begin{equation}
J(u_0-\frac{2iN}{g})=\psi(u_0-\frac{2iN}{g}+i\epsilon)-
\psi(u_0-\frac{2iN}{g}-i\epsilon)=i\,h(u_0)=i\,h(u+\frac{2iN}{g})\,.
\end{equation}

We can now define the modified analytic extension $\psi^{{\rm (u)}}_{\rm m}(u)$
which has cuts at the right place ($\vert{\rm Re\ }u\vert>2$)
on the upper half of the mirror plane and which is defined as
\begin{equation}
\begin{split}
0<{\rm Im\ }u<\frac{2}{g}:\quad
\psi^{{\rm (u)}}_{\rm m}(u)&=\psi(u)\,,\\
\frac{2}{g}<{\rm Im\ }u<\frac{4}{g}:\quad
\psi^{{\rm (u)}}_{\rm m}(u)&=\psi(u)+i\,h(u-\frac{2i}{g})\,,\\
\frac{4}{g}<{\rm Im\ }u<\frac{6}{g}:\quad
\psi^{{\rm (u)}}_{\rm m}(u)&=\psi(u)+i\,h(u-\frac{2i}{g})
+i\,h(u-\frac{4i}{g})\,,\\
\end{split}
\end{equation}
and similarly for larger values of ${\rm Im\ }u$. For this function
we have for all $N>0$:
\begin{equation}
[\psi^{{\rm (u)}}_{\rm m}(u)]_{2N}=
\psi^{{\rm (u)}}_{\rm m}(u+\frac{2iN}{g}+i\epsilon)-
\psi^{{\rm (u)}}_{\rm m}(u+\frac{2iN}{g}-i\epsilon)=i\,h(u+i\epsilon)\,.
\end{equation}

Similarly on the lower part of the $u$ plane we define 
$\psi^{(\ell)}_{\rm m}(u)$ by
\begin{equation}
\begin{split}
-\frac{2}{g}<{\rm Im\ }u<0:\quad
\psi^{(\ell)}_{\rm m}(u)&=\psi(u)\,,\\
-\frac{4}{g}<{\rm Im\ }u<-\frac{2}{g}:\quad
\psi^{(\ell)}_{\rm m}(u)&=\psi(u)+i\,h(u+\frac{2i}{g})\,,\\
-\frac{6}{g}<{\rm Im\ }u<-\frac{4}{g}:\quad
\psi^{(\ell)}_{\rm m}(u)&=\psi(u)+i\,h(u+\frac{2i}{g})
+i\,h(u+\frac{4i}{g})\,,\\
\end{split}
\end{equation}
and so on. Again, the cuts are at $\vert{\rm Re \ }u\vert>2$ and
we have for all $N>0$
\begin{equation}
[\psi^{(\ell)}_{\rm m}(u)]_{-2N}
=\psi^{(\ell)}_{\rm m}(u-\frac{2iN}{g}+i\epsilon)-
\psi^{(\ell)}_{\rm m}(u-\frac{2iN}{g}-i\epsilon)=-i\,h(u-i\epsilon)\,.
\end{equation}

Recall that because of the sign change in (\ref{Dneg}) it is $-\psi(u)$
that has to be extended to the lower half plane and we have for all $N\not=0$
\begin{equation}
[D(u)]_{\pm 2N}=i\,h(u\pm i\epsilon)=\pm i\,h(u+i\epsilon)\,,
\end{equation}
i.e. all upper/lower jumps are the same and this property is also
inherited by the dressing part of the full discontinuity function:
\begin{equation}
[\Delta_d(u)]_{\pm 2N}=2j(u\pm i\epsilon)\,,
\label{Deltadu}
\end{equation}
where
\begin{equation}
j(u)=
\sum_k\ln\frac{\frac{1}{x(u)}-x^+_k}{x(u)-x^+_k}\,
\frac{x(u)-x^-_k}{\frac{1}{x(u)}-x^-_k}=\ln\left(\frac{B_m}{R_m}\,
\frac{R_p}{B_p}\right)=\ln\left(\frac{Y_-^{(0)}}{Y_+^{(0)}}\right).
\label{b24}
\end{equation}

Here $\Delta_d=\ln\frac{y_d^+}{(y_d^+)_*}$ using (\ref{ydC}) and we see
from (\ref{yd}) and (\ref{calD1C}) that only the dressing part ($\theta$ part)
contributes to (\ref{Deltadu}) for $N\geq1$.

Finally we note that because of the $h$ function parts added to 
$\psi^{{\rm (u)}}_{\rm m}$ and $\psi^{(\ell)}_{\rm m}$ 
these functions also have an infinite number of poles
and zeros. We find that $\Delta_d^\prime(u)$
has double zeroes at $u_k+\frac{(2N+1)i}{g}$ ($N=1,2,\dots$) and double
poles at $u_k-\frac{(2N+1)i}{g}$ ($N=-1,0,1,\dots$) for all $u_k$.

\section{The asymptotic T-system: Bethe Ansatz solution}

\subsection{Asymptotic transfer matrices}

In the asymptotic limit the $Y_{Q,0}$ functions tend to zero and the Y-system of AdS/CFT splits into
two $SU(2|2)$ Y-systems. Correspondingly two independent $SU(2|2)$ T-systems generate the asymptotic
solutions for the Y-functions. The asymptotic solution consistent with the asymptotic Bethe Ansatz
equations \cite{BS} and the multiparticle L\"uscher formulae \cite{BJ08} was given in \cite{GKV09}.

In this appendix we discuss the form and the most important analyticity properties of the solution 
of the asymptotic T-system for states when there are $N$ fundamental magnons with rapidities $u_j$ present
in the system. These solutions correspond to the eigenvalues of the fusion hierarchy of the transfer matrices 
built from the S-matrices of the centrally extended  $SU(2|2)$ algebra such that the magnon rapidities
$u_j$ play the role of the inhomogeneities of the $SU(2|2)$ vertex model.

We introduce the following functions:
\begin{equation}
R_m(u)=\prod_{j=1}^{N} \, \frac{x(u)-x_j^+}{({x_j^+})^{\frac12}}, \qquad B_m(u)=\prod_{j=1}^{N} \, 
\frac{\frac{1}{x(u)}-x_j^+}{({x_j^+})^{\frac12}},    \label{RB_m}
\end{equation}
\begin{equation}
R_p(u)=\prod_{j=1}^{N} \, \frac{x(u)-x_j^-}{({x_j^-})^{\frac12}}, \qquad B_p(u)=\prod_{j=1}^{N} \, 
\frac{\frac{1}{x(u)}-x_j^-}{({x_j^-})^{\frac12}},
\end{equation}
where $x_j^{\pm}=x_s(u_j\pm \frac{i}{g})$. These functions satisfy the relation
\begin{equation}
R_m^+(u) \, B_m^+(u)=R_p^-(u) \, B_p^-(u)=(-1)^N \, Q(u),
\end{equation} 
with $Q(u)=\prod_{j=1}^{N} (u-u_j)$.

For states outside the ${\alg{sl}(2)}$ sector auxiliary
Bethe roots appear in the formulae. To take into account their contribution as well, we
need to introduce the following functions:
\begin{equation}
R_l(u)=\prod_{j=1}^{K_l} \, \frac{x(u)-y_{l,j}}{(y_{l,j})^{\frac12}}, \qquad B_l(u)=\prod_{j=1}^{K_l} \, 
\frac{\frac{1}{x(u)}-y_{l,j}}{(y_{l,j})^{\frac12}}, \quad Q_l(u)=\prod_{j=1}^{K_l} (u-u_{l,j}), \quad l=1,2,3,
\end{equation}
where $y_{l,j}=x(u_{l,j})$ and they satisfy the relation
\begin{equation}
R_l(u) \, B_l(u)=(-1)^{K_l} \, Q_l(u), \qquad l=1,2,3.  \label{RBQl}
\end{equation}
The 3 family of Bethe roots $\{u_{l,j}\}_{l=1,2,3}$ correspond to the 3 levels of the $SU(2|2)$ nested Bethe
Ansatz.

Using the definitions above the asymptotic solution of the T-system is given by the formulae as follows.
\begin{equation}
T^{(0)}_{0,s}(u)=1, \qquad T^{(0)}_{a,0}(u)=1,    \label{T0sT0a0}
\end{equation} 
\begin{eqnarray}
T^{(0)}_{a,1}&=&(-1)^{a} \, \frac{Q_3^{[-a]} \, Q_1^{[a]}}{Q^{[a+1]}} \, \frac{B_p^{[a]}}{B_m^{[a]}} \,
\left \{  \frac{Q^{[1+a]}}{Q_1^{[a]} \, Q_3^{[a]}} \, \frac{B_m^{[a]}}{B_p^{[a]}}+ 
\frac{Q^{[1-a]}}{Q_3^{[-a]} \, Q_1^{[-a]}} \, \frac{B_m^{[-a]}}{B_p^{[-a]}}
\right. \nonumber \\
&+& \Theta(a-2) \, \sum_{n=0}^{a-2} \, \frac{Q^{[a-1-2n]}}{Q_3^{[a-2-2n]} \, Q_1^{[a-2-2n]}} \, 
\left( \frac{B_m^{[a-2-2n]}}{B_p^{[a-2-2n]}}+\frac{R_m^{[a-2-2n]}}{R_p^{[a-2-2n]}} \right) \label{Ta10} \\
&-& \Theta(a-1) \, \sum_{n=0}^{a-1}  
\left. \frac{Q^{[a-1-2n]}}{Q_3^{[a-2-2n]} \, Q_1^{[a-2n]} } \, \left(  
\frac{Q_1^{[a-2n]} \, Q_2^{[a-3-2n]}}{ Q_1^{[a-2-2n]} \, Q_2^{[a-1-2n]}}+
\frac{ Q_3^{[a-2-2n]} \, Q_2^{[a+1-2n]}}{ Q_3^{[a-2n]} \, Q_2^{[a-1-2n]}}  \right)\right\}, \nonumber
\end{eqnarray}
where $\Theta(x)$ is the unitstep function such that $\Theta(0)=1$. $T^{(0)}_{a,1}$ are the eigenvalues of 
the transfer matrices corresponding to the anti-symmetric irreducible
representations of $SU(2|2)$ in the auxiliary space. The eigenvalues belonging to the symmetric
representations are given by:

\begin{eqnarray}
T_{1,s}^{(0)}&=&\frac{1}{Q_1^{[-s]} \, Q_3^{[s]}} \, \prod_{j=1}^{s-1} \frac{R_m^{[2j-s]}}{R_p^{[2j-s]}} \,
\left \{ Q_2^{[-s-1]} \, Q_2^{[s+1]} \, \frac{R_m^{[s]}}{R_p^{[s]}} \, \sum_{k=0}^{s} \, F_{s,k}
\right. \nonumber \\
&-& 
\Theta(s-1) \, \left( Q_2^{[s-1]} \, Q_2^{[-s-1]} \, \sum_{k=0}^{s-1} \, F_{s,k} 
+Q_2^{[1-s]} \, Q_2^{[s+1]} \, \frac{B_m^{[-s]}}{B_p^{[-s]}} \, \frac{R_m^{[s]}}{R_p^{[s]}} \,
\sum_{k=1}^{s} F_{s,k} \right) \nonumber \\
&+& \left. \Theta(s-2) \, Q_2^{[1-s]} \, Q_2^{[s-1]} \, \frac{B_m^{[-s]}}{B_p^{[-s]}} \, \sum_{k=1}^{s-1} \, 
F_{s,k}
\right \},    \label{T1s0}
\end{eqnarray}
where 
\begin{equation}
F_{s,k}=\frac{Q_1^{[2k-s]} \, Q_3^{[2k-s]}}{Q_2^{[2k-1-s]} \, Q_2^{[2k+1-s]} }. \label{Fsk}
\end{equation}
Finally T-functions on the interior boundaries of the fat-hook are given by:
\begin{equation}
T^{(0)}_{a,2}=\left(  {\cal{F}}^{(0)} \, \frac{R_m^-}{R_p^-} \, \frac{1}{Q^{[-2]}} \, \frac{Q_1^-}{Q_3^-} 
\right)^{[a]} 
\, \left( {\cal{G}}^{(0)} \, \frac{R_m^+}{R_p^+} \, Q^{[2]} \, \frac{Q_3^+}{Q_1^+} \right)^{[-a]}, \qquad a 
\geq 2 
\label{Ta20}
\end{equation}
\begin{equation}
T^{(0)}_{2,s}=\left(  {\cal{F}}^{(0)} \, \frac{R_p^-}{R_m^-} \, \frac{Q_1^-}{Q_3^-} \right)^{[s]} 
\, \left( {\cal{G}}^{(0)} \, \frac{R_m^+}{R_p^+} \, \frac{Q_3^+}{Q_1^+} \right)^{[-s]}  \,
\left( \prod_{k=1}^{s-1} \frac{R_m^{[2k+1-s]}}{R_p^{[2k+1-s]}} \right)^2, \qquad s \geq 2, \label{T2s0}
\end{equation}
where ${\cal{F}}^{(0) +}$ and ${\cal{G}}^{(0) -}$ are the analytic continuations of $T_{1,1}^{(0)}$ through 
the
branch cut at $\frac{i}{g}$ and $-\frac{i}{g}$ respectively.
\begin{equation}
{\cal{F}}^{(0)}=(T_{1,1}^{(0) +})_*^- ,\qquad {\cal{G}}^{(0)}=(T_{1,1}^{(0) -})_*^+.    \label{F0G0T11}
\end{equation}
They are given explicitly by: 
\begin{equation}
{\cal{F}}^{(0)}=-\frac{Q_3^{-}}{Q_3^{+}}+\frac{R_m^+}{R_p^+} \, 
\left( \frac{Q_2^{++} \, Q_3^-}{Q_2 \, Q_3^+} + \frac{Q_2^{--} \, Q_1^+}{Q_2 \, Q_1^-} \right)-
\frac{R_m^- \, R_m^+}{R_p^- \, R_p^+} \, \frac{Q_1^+}{Q_1^-}, \label{F0}
\end{equation}
\begin{equation}
{\cal{G}}^{(0)}=-\frac{Q_3^{-}}{Q_3^{+}}+\frac{B_m^+}{B_p^+} \, 
\left( \frac{Q_2^{++} \, Q_3^-}{Q_2 \, Q_3^+} + \frac{Q_2^{--} \, Q_1^+}{Q_2 \, Q_1^-} \right)-
\frac{B_m^- \, B_m^+}{B_p^- \, B_p^+} \, \frac{Q_1^+}{Q_1^-}.     \label{G0}
\end{equation}
The Bethe Ansatz equations follow from requiring that the residues of the (would-be) poles of the 
transfer matrices at the roots of the polynomials $Q_l, \quad l=1,2,3$ vanish:
\begin{equation}
\frac{Q_2^-(u_{1,j})}{Q_2^+(u_{1,j})} \frac{B_p (u_{1,j})}{B_m (u_{1,j})}=1, \qquad j=1,.., K_1
\label{BA1}
\end{equation}
\begin{equation}
\frac{Q_2^{++}(u_{2,j})}{Q_2^{--}(u_{2,j})}  \frac{Q_1^-(u_{2,j})}{Q_1^+(u_{2,j})}
\frac{Q_3^-(u_{2,j})}{Q_3^+(u_{2,j})}=-1, \qquad j=1,.., K_2
\label{BA2}
\end{equation}
\begin{equation}
\frac{Q_2^-(u_{3,j})}{Q_2^+(u_{3,j})} \frac{R_p(u_{3,j})}{R_m(u_{3,j})}=1, \qquad j=1,.., K_3.
\label{BA3}
\end{equation}

The analyticity properties of the asymptotic T-functions can be easily read off from the formulae
above. Now we summarize their most important properties. $T^{(0)}_{a,1}$ has square root branch cuts
along the lines $\mbox{Im} u= \pm \frac{a}{g}$. ${\cal{F}}^{(0)}$ and ${\cal{G}}^{(0)}$ have square root branch 
cuts
along the lines $\mbox{Im} u= \pm \frac{1}{g}$.
From (\ref{T1s0}) it can be seen that $T_{1,s}^{(0)}$ has several square root branch cuts between the lines
$\mbox{Im} u= \pm \frac{s}{g}$, but most of these cuts are generated by a gauge transformation and are 
cancelled
from the Y-functions. Separating the gauge factor:
\begin{equation}
\tilde{T}^{(0)}_{1,s}=T^{(0)}_{1,s} \, \prod_{j=1}^{s-1} \frac{R_p^{[2j-s]}}{R_m^{[2j-s]}}, \label{T1s0tilde} 
\end{equation}
it can be seen that $\tilde{T}^{(0)}_{1,s}$ has discontinuities only along the lines $\mbox{Im} u= \pm 
\frac{s}{g}$. 

For our considerations another important analyticity information is the large $u$ behavior of the T-functions. 
Analysing the formulae above it turns out that imposing the level matching condition
(i.e. $\prod_{k=1}^N \, (x_k^+/x_k^-)=1$):
\begin{equation}
T^{(0)}_{a,1}(u) \sim 2(-1)^a\,A^{(0)} \, \frac{a}{u^2} \qquad \quad \mbox{for} \quad |u| \rightarrow \infty 
\qquad a=1,2,...,
\end{equation} 
when $|\mbox{Im} u| < \frac{a}{g}$.  The coefficient $A^{(0)}$ is real and negative. In the ${\alg{sl}(2)}$ 
sector
it is given by the simple formula  $A^{(0)}=-\frac{\mu \, (\mu+2)}{2g^2},$
where $\mu= i \, g\, \sum\limits_{j=1}^{N} \left( \frac{1}{x_j^+}-\frac{1}{x_j^-}  \right)$ is a real
and positive number.
In the region $|\mbox{Im} u| > \frac{a}{g}$ the large 
$u$ behavior of $T^{(0)}_{a,1}(u)$ is different, there the decay is only as $\frac{1}{u}$. For $a=1$ we have
\begin{equation}
T^{(0)}_{1,1}(u) \sim \pm \frac{2B^{(0)}}{u} \qquad B^{(0)}=\frac{i}{g}(N+\mu)
\end{equation} 
and the upper (lower) sign is valid for $u$ above (below) the physical strip.

In the mirror TBA equations only the zeroes and poles of $T_{a,1}(u)$ and $T_{1,s}(u)$
located in the physical strip  (i.e the strip $\mbox{Im}u \leq \frac{1}{g}$)
furthermore the zeroes and poles of ${\cal{F}}^{+}$ and ${\cal{G}}^{-}$ in the upper and lower half
planes respectively are relevant. Though the asymptotic formulae presented above are valid for a 
general state in most of this paper we considered states where $T_{1,1}$, ${\cal{F}}^+$ and ${\cal{G}}^-$
have no zeros and poles in the regions $|\mbox{Im} \, u|\leq 1/g$, $\mbox{Im} \, u \geq 0$ and $\mbox{Im} \, u 
< 0$ respectively. We think (but have not proved) that this condition is satisfied in 
the ${\alg{sl}(2)}$ sector of the theory. 
This set of states is definitely non-empty (see below) but since we are not certain if it is the entire
${\alg{sl}(2)}$ sector or only a proper subset of it we have called it in the paper
the ${\alg{sl}(2)}$ (sub-)sector. Here we list the properties of the transfer matrices that
characterize this (sub-)sector and introduce some notations and definitions.
In this (sub-)sector there is a symmetry between the left and right wing variables so the wing index of the 
variables
can be suppressed. 

In the asymptotic limit the ${\alg{sl}(2)}$ (sub-)sector is characterized by the properties:

\begin{itemize}

\item $T^{(0)}_{1,1}$ has neither zeros nor poles in the physical strip.


\item $T^{(0)}_{a,1}$ and $\tilde T^{(0)}_{1,s}$ 
have no poles in the physical strip except $T^{(0)}_{2,1}$ which 
has  poles at positions $u_j^-  \qquad j=1,..,N$.

\item $T^{(0)}_{a,1}$ has ${\cal N}_a$ zeroes in the physical strip at positions $\xi_{a,j}$ 
for $a \geq 2$, and we define $\tau_{a}(u)=\prod\limits_{j=1}^{{\cal N}_{a}} t(u-\xi_{a,j})$.
 
\item $\tilde{T}^{(0)}_{1,s}$ has $\tilde{\cal N}_s$ zeroes in the physical strip at positions 
$\tilde{\xi}_{s,j}$ 
for $s \geq 2$, and their contributions are taken into account by the factor
$\tilde{\tau}_{s}(u)=\prod\limits_{j=1}^{\tilde{\cal N}_{s}} t(u-\tilde\xi_{s,j})$.

\item The transfer matrices have no zeroes (or poles) at the boundaries of the physical strip (with imaginary 
part
$\pm 1/g$).

\item ${\cal{F}}^{(0) +}$ has neither zeros nor poles in the region $\mbox{Im} u \geq 0$.

\item ${\cal{G}}^{(0) -}$ has neither zeros nor poles in the region $\mbox{Im} u<0$.

\end{itemize}

We also define $\tau_1(u)=\prod\limits_{j=1}^{{ N}} t(u-u_j)$, $\tilde\tau_1(u)\equiv1$.
the poles located 

In the ${\alg{sl}(2)}$ (sub-)sector the source terms appearing in the TBA equations (\ref{TBAmvw}-\ref{TBA-}) 
can be expressed in terms of the $\tau_m$ and $\tilde{\tau}_{m}$ functions as follows:
\begin{eqnarray}
t_{m \vert vw}&=&\tau_m \tau_{m+2}, \qquad m=1,2,... \label{tmvw}\\
t_{m \vert w}&=&\tilde{\tau}_m \tilde{\tau}_{m+2}, \qquad m=1,2,... \label{tmw}\\
t_Q&=& \tau_Q^2, \qquad Q=2,3,... \label{tQ}\\
t_1&=&1 \label{t1}\\
t_{-}&=&\tau_2 /\tilde{\tau}_2, \label{t-}
\end{eqnarray}
where because of the left-right symmetry we have omitted the wing index $^{(\alpha)}$.

The class of minimal energy twist-two states forms an important subset within the ${\alg{sl}(2)}$ (sub-)sector. 
Because of the importance of these states we studied numerically the qualitative analyticity properties of the
$T$-functions of this class. These can be summarized as follows:
\begin{itemize}

\item $T^{(0)}_{1,1}$ has neither zeros nor poles in the physical strip,

\item $T^{(0)}_{2,1}$ has $2(N-2)$ zeros on the real axis and poles at positions $u_j^-$,

\item $T^{(0)}_{a,1}$ has $2(N-2)$ zeros on the real axis for $a \geq 3$,
 
\item $\tilde{T}^{(0)}_{1,s}$ has neither zeros nor poles in the physical strip,

\item ${\cal{F}}^{(0) +}$ has neither zeros nor poles in the region $\mbox{Im} u \geq 0$,

\item ${\cal{G}}^{(0) -}$ has neither zeros nor poles in the region $\mbox{Im} u<0$.

\end{itemize}
Near the small coupling limit all the magnon rapidities and the real zeroes of $T_{a,1}$ are larger than 2, 
because they are of order $\frac{1}{g}$.

The analyticity properties of the asymptotic T-functions summarized above determine the
analyticity properties of the asymptotic Y-functions through the formula (\ref{YT}). 
Their state independent analyticity properties can be summarized as follows.
$Y_{a,1}^{(0)}$ and $Y_{1,s}^{(0)}$ tend to constants at infinity and they have square root branch cuts
along the lines $\mbox{Im} u =  \pm \frac{(a \pm 1)}{g}$ and $\mbox{Im} u =  \pm \frac{(s \pm 1)}{g}$ 
respectively.
$Y_{1,1}^{(0)}$ and $Y_{2,2}^{(0)}$ tend to $-1$ at infinity and have square root branch cuts along the lines
$\mbox{Im} u=0,\pm \frac{2}{g}$, such that $1+Y_{1,1}^{(0)} \sim 1/u$ and $1+\frac{1}{Y_{2,2}^{(0)}} \sim 1/u$
at infinity. 
 
The state dependent analyticity properties like the distribution of zeroes and poles can be read off
from the local singularity structure of the corresponding T-functions studied above.

\subsection{Asymptotic solution for the momentum carrying nodes}

In the asymptotic limit the Y-system decouples to two $SU(2|2)$ T-systems. Correspondingly
we will put an L (left) or R (right) index on the asymptotic transfer matrices. 
Using this notation the asymptotic solution for the momentum carrying nodes takes the form \cite{GKV09}:
\begin{equation} \label{Ya00}
Y_{a,0}^{(0)}=\left( \frac{x^{[a]}}{x^{[-a]}} \right)^{J_{eff}} \, {\cal{D}}_a 
\,\frac{B_{1,L}^{[-a]}}{B_{1,L}^{[a]}} \, \frac{B_{3,L}^{[a]}}{B_{3,L}^{[-a]}}
\,\frac{B_{1,R}^{[-a]}}{B_{1,R}^{[a]}} \, \frac{B_{3,R}^{[a]}}{B_{3,R}^{[-a]}} \,
T^{(0) L}_{a,-1} \,T^{(0) R}_{a,1} \qquad a=1,2,...,  
\end{equation}
where $J_{eff}=J+\frac{K_{3,L}-K_{1,L}+K_{3,R}-K_{1,R}}{2}$ and the factor ${\cal{D}}_a $
contains the dressing phase of fundamental magnons as follows.
Let: 
\begin{equation}
{\cal{D}}_1(u)=\frac{B_m^-(u) \, R_p^+(u)}{B_p^+ (u)\, R_m^-(u)} \, 
\prod\limits_{j=1}^N e^{-2 \, i\, \theta(u,u_j)} \equiv \prod\limits_{j=1}^N S_{\alg{sl}(2)}^{1 1_*}(u,u_j),
\label{calD1C}
\end{equation}
with $\theta(u,u_j)$ given by (\ref{theta}). 
Then
\begin{equation}
{\cal{D}}_a(u)=\prod_{k=0}^{a-1} \, {\cal{D}}_1^{[a-1-2k]}(u) \equiv 
\prod\limits_{j=1}^N S_{\alg{sl}(2)}^{a 1_*}(u,u_j).
\end{equation}
In the ${\alg sl}(2)$ sector the following representation for $Y_{1,0}^{(0)}$ proved to be useful:
\begin{eqnarray}
Y_{1,0}^{(0)}=Y_{d}^{(0)} \, T_{1,-1}^{(0)L} \, T_{1,1}^{(0)R}, \qquad Y_{d}^{(0)}=y_d,
\label{ydC}
\end{eqnarray} 
where $y_d$ contains the dressing part of $Y_{1,0}^{(0)}$:
\begin{equation}
y_d=\left(\frac{x^+}{x^-}\right)^J \, {\cal D}_1.
\label{yd}
\end{equation}
The Beisert-Staudacher equations follow from the requirement that $Y^{(0)}_{1,0 (*)}(u_k)=-1$, where
$Y^{(0)}_{1,0 (*)}=\left((Y^{(0) -}_{1,0})_*\right)^+$.
In words $Y^{(0)}_{1,0}(u)$ analytically continued through the branch cut at $-\frac{i}{g}$
to the physical sheet and taken at the positions of the physical rapidities is equal to $-1$.
Starting from (\ref{Ya00}) this analytical continuation can be implemented simply by using (\ref{F0G0T11}) and 
(\ref{G0}). The equations take the form:
\begin{equation}
\left( \frac{x_s^{+}}{x_s^{-}} \right)^{J_{eff}}  S_{\alg{sl}(2)} \,
\,\frac{B_{1,L}^{-}}{R_{1,L}^{+}} \, \frac{R_{3,L}^{-}}{B_{3,L}^{+}}
\,\frac{B_{1,R}^{-}}{R_{1,R}^{+}} \, \frac{R_{3,R}^{-}}{B_{3,R}^{+}}
\bigg|_{u_k} =-1, \qquad k=1,..,N,
\label{Beisert}
\end{equation}
where we introduced the notation 
\begin{equation}
S_{\alg{sl}(2)}(u)
=\prod\limits_{j=1}^N S_{\alg{sl}(2)}^{1_* 1_*}(u,u_j)
=\left(\prod\limits_{j=1}^N S_{\alg{sl}(2)}^{1 1_*}(u,u_j)\right)_{(*)}
\end{equation}
for short. For states outside the ${\alg{sl}(2)}$ sector there are 6 other Bethe equations that can be obtained 
by
putting $L$ and $R$ indices on the 3 auxiliary Bethe equations (\ref{BA1})-(\ref{BA3}).

We close this appendix by the description the analyticity properties of $Y_{a,0}^{(0)}$.
 They have square root branch cuts along the lines $\mbox{Im} u =\pm \frac{a}{g}$. 
They do not have any zeros in the complex plane and the distribution of their poles is as follows.
\begin{itemize}
\item $Y_{1,0}^{(0)}$ has poles at positions $u_j^{[\pm 2]}$,
\item $Y_{a,0}^{(0)}$ has poles at positions $u_j^{[\pm(a-1)]}$ and $u_j^{[\pm(a+1)]} \qquad a=2,3...$.
\end{itemize}

For large $u$ the $Y_{a,0}^{(0)}$ behaves as:
\begin{eqnarray}
Y_{a,0}^{(0)}(u) \sim  
{1 \over u^{2(J+2+N+\mu)}}, \qquad \mbox{for} \qquad \mbox{Im}\, u \leq \frac{a}{g}. 
\nonumber
\end{eqnarray}


The qualitative analyticity information which can be obtained from the formulae given
in this appendix was used in the main text in the derivation of the TBA equations
for the exact energies of these states.

\section{On the refined interpretation of the discontinuity 
relations}


This appendix is devoted to the detailed explanation of the refined interpretation of the discontinuity 
relations (\ref{Y1cut}-\ref{Yvw1cut}) when local singularities lying exactly on the cut lines are present.
We will explain the formula (\ref{CutInterpret}) in detail. To this end we will invoke the asymptotic solution
and in order to get rid of the logarithmic type discontinuities we will examine the discontinuity relations
for the derivative of $\Delta$. We will show that (\ref{CutInterpret}) is valid for all states of the model
and not only for states from the ${\alg{sl}(2)}$ (sub-)sector discussed mostly in the main text. 

 From (\ref{Ya00}), which gives the asymptotic form
of $Y_1^{(0)}$ for a general state it follows that asymptotically $\Delta$ takes the form:
\begin{equation}
\Delta^{(0)}=\Delta_R^{(0)}+\Delta_L^{(0)}+\Delta_d+\Delta_4, 
\end{equation}
where
\begin{equation}
\Delta_R^{(0)}=\left[ \ln T_{1,1}^{R (0)} \right]_{+1}, \qquad 
\Delta_L^{(0)}=\left[ \ln T_{1,-1}^{L (0)} \right]_{+1}, \qquad
\Delta_d=\left[ \ln {\cal D}_1 \right]_{+1}
\end{equation}
and
\begin{equation}
\Delta_4=-2J_{eff}\ln x\,+\,\ln\left(\frac{B_{1,L}}{R_{1,L}}\,\frac{R_{3,L}}{B_{3,L}}\,
\frac{B_{1,R}}{R_{1,R}}\,\frac{R_{3,R}}{B_{3,R}}\right).
\end{equation}
This last contribution has cuts along the real axis only and thus plays no role in the
discontinuity relations at $\pm\frac{2Ni}{g}$ for $N\not=0$.

First we recall that the main result (\ref{Deltadu}-\ref{b24}) of appendix B is that 
\begin{equation} \label{A0}
\left[ \Delta_d \right]_{\pm 2N}=2 \, \ln \frac{Y_-^{(0)}}{Y_+^{(0)}}=\sum_{\alpha} \,
\ln \frac{Y_-^{(\alpha) (0)}}{Y_+^{(\alpha) (0)}}.
\end{equation}

Using (\ref{Ta10}) of appendix C we will show that if we forget about the local singularities
of the asymptotic $T$- and $Y$-functions 
we get the naive equality for the derivatives:
\begin{equation} \label{naive}
\left[ \Delta_R^{(0) \prime} \right]_{\pm 2N}= D_{12R}^{(\pm 2N) (0) \prime},
\end{equation}
where 
\begin{equation}
D_{12R}^{(\pm 2N) (0)}=
\pm  \left(
\left[\ln \!  \left(1-\frac{1}{Y_{\mp}^{(+) (0)}}   \right)\right]_{\pm 2N}   \!
+\sum\limits_{m=1}^N \, \left[\ln  \! \left(1+\frac{1}{Y_{m|vw}^{(+) (0)}}\right)\right]_{\pm (2N-m)}
\right)
\end{equation}
and similarly we get analogous expressions for $\Delta_L^{(0)}$.
This complicated expression of $Y$-functions becomes much simpler in the language
of $T$-functions. For the sake of simplicity let us consider the discontinuity relation in the upper half plane.
Similar considerations that lead to (\ref{6.11}) in the main text give
\begin{equation} \label{D120}
D_{12R}^{(2N) (0) \prime}=\left[ \ln ^{\prime} T_{1,1}^{R (0) +} \right]_{+2N}+
\left[ \ln^{\prime} T_{N+1,1}^{R (0)} \right]_{N-1}-\left[ \ln^{\prime} T_{N+2,1}^{R (0)} \right]_{N},
\qquad N \geq 1.
\end{equation} 
Using (\ref{F0G0T11}) we write
\begin{equation}
\left[ \Delta_R^{(0) \prime} \right]_{\pm 2N}=\left[ \ln ^{\prime} T_{1,1}^{R (0) +} \right]_{+2N}-
\left[ \ln ^{\prime} {\cal F}^{R (0) +} \right]_{+2N}.
\end{equation}
From the expressions of appendix C it can be seen that ${\cal F}^{R (0) +}$ and $T_{N,1}^{(0)}$ 
has no square-root discontinuities in the upper half plane and in the strip $|\mbox{Im} u |< \frac{N}{g}$ 
respectively, and it follows that
\begin{equation}
\left[ \ln ^{\prime} {\cal F}^{R (0) +} \right]_{+2N}=\left[ \ln^{\prime} T_{N+1,1}^{R (0)} \right]_{N-1}=\left[ 
\ln^{\prime} T_{N+2,1}^{R (0)} \right]_{N}=0,
\end{equation}
which implies  (\ref{naive}). This together with (\ref{A0}) would justify formula (\ref{Deltacut})
for the discontinuities of $\Delta$. 

Now let us examine whether local singularities of $T$-functions lying exactly on the cut lines might modify
 (\ref{Deltacut})? To investigate this point some information on the local singularities of the $T$-functions
is necessary. Since we require the $Y$-functions to be real analytic functions of $u$ their local singularities
can be either real or come in complex conjugate pairs. Due to the T-representation of the $Y$-functions, the
local singularities of the $T$-functions generate this structure. Thus, $T$-functions have the same 
local singularity structure apart from some irrelevant modification coming from the fact that the asymptotic
form of the $T$-functions given in appendix C corresponds to a gauge where the $T$-functions are not real 
analytic. 

From this discussion it is clear that the real zeroes are the most important objects for the question we are
investigating here since they preserve their imaginary parts, while the other singularities move in the complex 
plane if we tune
the coupling $g$ from low to higher values. In principle it can happen that at certain values of $g$ some complex
roots lie exactly on one of the cut lines, but since their imaginary parts are not restricted by any symmetry 
they move off the cut line if $g$ is changed a little.

For the sake of simplicity we exclude this zero measure set of the possible values of $g$ from our discussion
and then we have to consider whether the real zeroes of the $T$-functions modify (\ref{Deltacut}).
Formula (\ref{D120}) implies that the real zeroes of the $T$-functions can lie on the cut lines only in the term
 $\left[ \ln^{\prime} T_{N+1,1}^{R (0)} \right]_{N-1}$  and only when $N=1$.
This means that
\begin{equation} \label{refined}
\left[ \Delta_R^{(0) \prime} (u)\right]_{+2 } \neq D_{12R}^{(+2) (0) \prime}(u), \qquad
\mbox{if} \qquad u=u_{2,k}^{(+)},
\end{equation}
where the set $\{ u_{2,k}^{(+)}\}_{k=1}^{N_{2}^{(+)}}$ denotes the real zeroes of $T_{2,1}^{R(0)}$ 
with absolute value larger than 2.
The equality can be restored if we remove the contribution of these local singularities from the right hand 
side.
For this purpose we introduce the polynomial $ p_2^{(\alpha)}(u)=\prod\limits_{k=1}^{N_2^{(\alpha)}} \, 
(u-u_k^{(\alpha)})$ and the modified equality which holds for all $|u| \geq 2$ takes the form: 
\begin{equation} \label{modifiedD}
\left[ \Delta_R^{(0) \prime} (u)\right]_{+2 } = D_{12R}^{(+2) (0) \prime}(u)-\ln^{\prime} p_2^{(+)}.
\end{equation}
In order to rephrase this modification in the language of gauge invariant $Y$-functions, let us recall the
naive discontinuity relation (\ref{naive}) for $N=1$:
\begin{equation}
\left[ \Delta_R^{(0) \prime} \right]_{+ 2}=  
\left[\ln^\prime \!  \left(1-\frac{1}{Y_{-}^{(+) (0)}}   \right)\right]_{+ 2}   \!
+ \, \left[\ln^\prime  \! \left(1+\frac{1}{Y_{1|vw}^{(+) (0)}}\right)\right]_{+ 1}.
\end{equation}
Since $1+\frac{1}{Y_{1|vw}^{(+) }} \sim T_{2,1}^+ \, T_{2,1}^-$, it is clear that this term
is responsible for the neccessary modifications. On the other hand $Y_{-}^{(+)} \sim T_{2,1}$, 
thus the zeroes of $p_2^{(+)}$ can also be defined as the zeroes of $Y_{-}^{(+)}$ lying on the real cut line.

Similar considerations can be applied for the discontinuity relations corresponding to the lower half
plane and to the left part $\Delta_L$.
Putting together the results of these considerations it turns out that the form of the proposal (\ref{Deltacut}) 
is unaffected by the local singularities for $N \geq 2$, but must be modified when $N=1$ as follows:
\begin{equation}   \label{CutInterpretD}
[\Delta]_{\pm 2}= \pm \sum_{\alpha=\pm} \left(
\left[\ln \!  \left(1-\frac{1}{Y_{\mp}^{(\alpha)}}   \right)\right]_{\pm 2}   \!
+ \, \left[\ln  \! \left\{ \left(1+\frac{1}{Y_{1|vw}^{(\alpha)}}\right)
\,\frac{1}{p_2^{(\alpha) \mp}}  \right\} \right]_{\pm 1} 
\! \! \! \! \! \! \!  \!+ \! \ln \!  \left( \frac{Y_{-}^{(\alpha)}}{Y_{+}^{(\alpha)}} \right)  
\right),
\end{equation}
where $p_2^{(\alpha)}(u)$ is the polynomial having zeroes at the positions of the real zeroes 
of $Y_{-}^{(\alpha)}$ with absolute values larger than $2$.

Similar arguments as we applied for $\Delta$ shows that the other discontinuity relations 
(\ref{Ypmcut}) and (\ref{Yvw1cut}) are not to be modified.
We note that the problem of local singularities lying on the cut lines is not only of academic
interest, but also occurs in realistic cases, for example 
 for the minimal energy twist-two states in the ${\alg{sl}(2)}$ sector of the model.

A final remark on the modification of (\ref{Deltacut}) is in order. If in (\ref{Deltacut}) the term 
discontinuity is
not defined by (\ref{[f]_Z}), but is defined as that part of the discontinuity which comes from the square-root
behavior of the functions, then (\ref{Deltacut}) is correct for $N=1$, too. The subtlety is associated with
the $\log$
type branch points lying on the cut lines which becomes important when (\ref{Deltacut}) is translated to 
dispersion relations.

\section{Regularity and the exact Bethe equation}
In this appendix we clarify the consequences of imposing the exact
Bethe equations and compare this to what is known in analogous relativistic
integrable models.

In simple relativistic integrable models we have only one massive node
(with corresponding Y-system function denoted $Y_0$) and, like for the 
Sine-Gordon model, the XXX model, or (with some small modification) for
the O(4) nonlinear sigma model, the only Y-system equation containing
the factor $1+Y_0$ is for the neighboring Y-function $Y_1$ and is of the form
\begin{equation}
Y_1^+\,Y_1^-=(1+Y_0)\,(1+Y_2).
\label{rela}
\end{equation}
For the asymptotic Y-system we have to drop the $(1+Y_0)$ factor since
$Y_0=0$ asymptotically. 

The zeroes of $Y_1$ are the physical rapidities $\left\{\theta_j\right\}$
and the elements of the set $\left\{r_k\right\}$. Both sides of (\ref{rela})
vanish at the points $r^\pm_k$. For the right hand side of the equation
this vanishing is due to the factor $(1+Y_2)$. The above conclusion is
the same whether we consider the exact problem or the asymptotic limit.
In contrast, the behavior of the two sides of the equation at singular points
associated to the physical rapidities $\theta_j$
is very different asymptotically and for the exact solution.
Asymptotically the right hand side is regular at $\theta_j^\pm$ and since
one of the factors vanishes on the left hand side, the other factor has to 
have a pole at $\theta_j^{[\pm2]}$ to compensate. However, for the exact 
solution the analog of (\ref{exactBethe}) 
\begin{equation}
1+Y_0(\theta^\pm_j)=0
\label{exactBetherela}
\end{equation}
is imposed and now both sides vanish at $\theta_j^\pm$. There is no need for
compensation and $Y_1$ remains regular at $\theta^{[\pm2]}_j$. Thus as
a consequence of imposing the exact Bethe equation (\ref{exactBetherela})
the neighboring Y-function becomes more regular. This occurs two steps
away from the real line.

The consequences of the exact Bethe equations (\ref{exactBethe}) are very
similar, but we have to take into account that in the AdS/CFT case these
quantization conditions are imposed after analytical continuation
to a different sheet, the string sheet.

The relevant equation is (\ref{Y-}). We rewrite its inverse after the analytic
continuation: 
\begin{equation}
\frac{1}{Y_+^{(\alpha)+}}\,\xi_{(*)}=(1+Y_{1(*)})\,{\cal M}_{(*)},
\label{Y-mod}
\end{equation}
where
\begin{equation}
\xi=\frac{1}{Y_-^{(\alpha)-}}\,\, ,\qquad\quad
{\cal M}=\frac{1+Y^{(\alpha)}_{1\vert w}}{1+Y^{(\alpha)}_{1\vert vw}}
\end{equation}
and we recall that for any function $f$ we have
\begin{equation}
f_{(*)}=((f^-)_*)^+.
\end{equation}
Again, as for the relativistic case, the first factor on the right hand
side of (\ref{Y-mod}) is absent in the asymptotic limit.

Using the explicit formulae given in appendix C, we can convince ourselves that
the factor ${\cal M}_{(*)}$ is regular at singular points associated to the
physical rapidities $u_j$. We also see that the first factor on the left hand
side of (\ref{Y-mod}) has zeroes at $\left\{u_j\right\}$. (For the asymptotic 
case this can be seen using the explicit formulae of appendix C again and
for the exact problem it follows from the discussion of subsection 5.2).

The consequences of (\ref{Y-mod}) are very similar to what we have seen above.
Asymptotically $\xi_{(*)}$ has to have poles at $u_j$ to compensate the
zeroes of the first factor. In the exact case because of (\ref{exactBethe})
the right hand side also vanishes and $\xi_{(*)}$ remains regular. Thus
also in our case the exact solution is more regular than the asymptotic limit.
We note that this occurs at two TBA steps away from the real line in the 
physical (string) sheet: while asymptotically $(Y_-^{(\alpha)--})_*$ has
zeroes at $u^+_j$, these zeroes are absent for the exact solution.

\end{document}